\documentclass[12pt]{iopart}
\usepackage{iopams}
\usepackage{graphicx}
\usepackage{dcolumn}                    
\usepackage{bm}                         
\newcommand {\ba} {\begin{eqnarray}}
\newcommand {\ea} {\end{eqnarray}}
\newcommand {\be} {\begin{equation}}
\newcommand {\ee} {\end{equation}}
\usepackage{times}

\begin{document}

\title[Superconducting Properties of the $s^{\pm}$-wave state]
{Superconducting Properties of the $s^{\pm}$-wave state: Fe-based superconductors}

\author{Yunkyu Bang$^{1}$ and G. R. Stewart$^{2}$}

\address{$^{1}$Department of Physics, Chonnam National University, Kwangju
500-757, Republic of Korea \\
$^{2}$ Physics Department, University of Florida, Gainesville, FL
32611-8440, USA}

\begin{abstract}
Although the pairing mechanism of the Fe-based superconductors (FeSCs) has not yet been settled with a consensus, as to the pairing symmetry and the superconducting (SC) gap function, the abundant majority of experiments are supporting for the spin-singlet sign-changing s-wave SC gaps on multibands ($s^{\pm}$-wave state). This multiband $s^{\pm}$-wave state is a very unique gap state {\it per se} and displays numerous unexpected novel SC properties such as a strong reduction of the coherence peak, non-trivial impurity effects, nodal-gap-like nuclear magnetic resonance (NMR) signals, various Volovik effects in the specific heat (SH) and thermal conductivity, and anomalous scaling behaviors with the SH jump and the condensation energy vs. $T_c$, etc. In particular, many of these non-trivial SC properties can be easily mistaken as evidence for a nodal gap state such as a d-wave gap.
In this review, we provide detailed explanations of theoretical principles for the various non-trivial SC properties of the $s^{\pm}$-wave pairing state, and then critically compare the theoretical predictions with the experiments of the FeSCs. This will provide a pedagogical overview of how much we can coherently understand the wide range of different experiments of the FeSCs within the $s^{\pm}$-wave gap model.
\end{abstract}

\pacs{74.20.-z,74.20.Rp,74.25.-q,74.70.-b}

\date{\today}
\maketitle
\tableofcontents

\section{Scope and Introduction}
\subsection{Scope}

There exist already many good review articles\cite{paglione2010high,johnston2010puzzle,stewart2011superconductivity,hirschfeld2011gap,
wen2011materials,chubukov2012pairing,chubukov2015iron,hosono2015iron,hirschfeld2016using,hosono2016exploration} for various issues of the Fe-based superconductors (FeSCs) with different aims and focuses since its discovery\cite{kamihara2008iron} and following explosion of the world-wide research activity on these superconducting (SC) compounds in the last several years. The justification for writing yet another review on this subject is as follows.
This review has a particularly narrow scope and focused aim. The whole exposition and discussions in this article are dedicated to one particular SC pairing gap model, the $s^{\pm}$-wave pairing state, as a pairing state of the FeSCs. We will not discuss much about its underlying pairing mechanism except some general concept and plausibility arguments to yield this pairing state. Assuming the $s^{\pm}$-wave pairing state as the SC ground state of the real Fe-based SC compounds, we then examine its compatibility, as well as its failures, with available experiments. In doing so, we also intentionally use a minimal two band $s^{\pm}$-wave pairing model to understand experiments.

Introducing more tuning parameters like the gap anisotropy, more degrees of freedom like orbitals and more bands, and more realistic coupling matrix elements, etc can better fit the experimental data, and it is also true that these details do exist in real Fe-based SC materials and can play important roles to completely understand various aspects of these materials.
However, it is not our purpose to fit better the experimental data, and we would like to show the proof of concepts and emphasize mainly the generic features, but not the parameter dependent tuning ability, of the $s^{\pm}$-wave pairing state.
This is because of two reasons: (1) the $s^{\pm}$-wave pairing state itself is an interesting new SC state, having many unexpected interesting SC properties regardless of its realization in the FeSCs, hence it is worthy of study by itself; (2) the unrealistically simplified -- in some sense --  minimal two band $s^{\pm}$-wave model, ignoring the apparent details mentioned above, is surprisingly good at explaining almost all, often either peculiar or anomalous, experimental data.

Another important purpose of this article is to provide a pedagogical detailed exposition how to understand the experimental data of the representative SC properties of the materials and how to understand them theoretically, side by side. We hope this second purpose serves as a useful guideline, in particular, for young researchers in the field. Needless to say, if we omit or miss some important references for the relevant issues dealt with in this paper, it is not intentional. We tried our best to give a fair treatment to all research papers.

\subsection{Brief Summary of the Fe-based Superconductor Theories}

\subsubsection{random phase approximation (RPA) type theories}
Immediately after the discovery of La(O$_{1-x}$F$_x$)FeAs ($x= 0.05-0.12$)  superconductor with $T_c \sim 26K$ in 2008 \cite{kamihara2008iron}, several theorists -- Mazin et al.\cite{Mazin} and Kuroki et al.\cite{Kuroki} among others -- have carried out a weak coupling BCS calculations combining the essential band structure and the antiferromagnetic (AFM) spin fluctuations arising from the local interactions between the $d$-orbital electrons of the Fe atoms, and found that the leading SC pairing solution is the sign-changing $s$-wave state: $s$-wave order parameters (OPs) formed on the hole Fermi surfaces (FSs) around $\Gamma$ point and the electron FSs around $M$ point in the Brillouin zone (BZ) (in this paper we use the two Fe/cell BZ if not otherwise specified) with opposite signs from each other, therefore conveniently called as the $s^{\pm}$-wave state. These early theories are RPA theories, where the theory constructs the low energy effective pairing interaction by calculating a dynamic spin susceptibility $\chi_s(q,\omega)$ using the RPA method and solves Eliashberg gap equation with the effective interaction. This approach is pretty standard weak coupling theory and still faces objections because the Fe-based SC compounds are believed to be a strongly correlated electron system (SCES) like the high-$T_c$ cuprates and heavy fermion systems.  Thus, there is the belief that the description of the superconductivity in the SCES should be something beyond the weak coupling BCS-Eliashberg type theory.

After these early RPA theories, more extensions and elaborations of the RPA type approaches\cite{graser2009near,kuroki2009pnictogen,graser2010spin,kemper2010sensitivity} have been applied on more realistic models of Fe-based SC compounds for a wider parameter space of $U, U'$ (on-site Coulomb repulsions between intra- and inter orbitals, respectively) and $J, J'$ (on-site Hund coupling and pair hopping, respectively), changing dopings and pnictogen height\cite{kuroki2009pnictogen}, etc. It was found that in most of cases the spin susceptibility $\chi_s(q,\omega)$ is dominant at a large momentum $q=(\pi,0)/(0,\pi)$. The gap solutions from the multiband Eliashberg equations can obtain more complicated structures than the original RPA solution: it is quite natural to have a strong anisotropy in the $s$-wave gap function $\Delta_a(k)$ around each FS \cite{graser2009near,kuroki2009pnictogen,kemper2010sensitivity,maier2009origin} and three dimensional warping (along $c$-direction), and can even develop vertical line nodes in the gap function with some parameters but still remains in the $A_{1g}$ symmetry\cite{graser2010spin}. With a particular (unphysical) choice of parameter (e.g. large values of $J$ and $J'$), the $d_{x^2-y^2}$-wave solution can also be a dominant solution\cite{graser2009near}.
Therefore, we can say that the dominant solution of the RPA approaches for the FeSCs in most of the parameter space is basically the $s^{\pm}$-wave state.

Along this line of development, Kontani et al. \cite{kontani2010orbital,onari2012self} have extended it toward a charge instability (there are many motivations to this direction such as $C_2$ structural phase transition, stripe AFM order, many signals of nematic order/fluctuations, etc), searching for the optimal conditions for the dominant charge/orbital fluctuations instead of the spin fluctuations as a pairing glue. These authors found two routes to enhance the charge/orbital fluctuations: (1) coupling with in-plane Fe phonon, and (2) vertex correction (a usual RPA theory ignored it). Once the charge/orbital fluctuations $\chi_c$ are found to be dominant, they mediate an attractive interaction for singlet channel (AFM spin fluctuations mediate a repulsive interaction for singlet channel). Therefore if a dominant $\chi_c (q,\omega)$ occurs at around $q=(0,0)$, it helps any kind of SC pairing\cite{hosono2015iron}, and if a dominant $\chi_c (q,\omega)$ occurs at around $q=(\pi,0)/(0,\pi)$, it will promote $s^{++}$-wave pairing and compete against $s^{\pm}$-wave pairing\cite{kontani2010orbital,onari2012self}. Whether this scenario of pairing glue is relevant with the FeSCs mainly depends on the judgement on whether the choice of interaction parameters necessary for dominant charge/orbital fluctuations is physically relevant with the real compounds. Although it is still an open issue, theoretically it seems to require a rather unphysical parameter choice to obtain the dominant charge/orbital fluctuations $\chi_c$, in particular, at around $q=(\pi,0)/(0,\pi)$, and there are not so strong experimental supports for the $s^{++}$-wave pairing state. For more in-depth discussions on the $s^{++}$ scenario in FeSCs, see \cite{hosono2015iron,hirschfeld2011gap}.

A more elaborate extension of the RPA type approach is called fluctuation-exchange (FLEX) method\cite{bickers1989conserving}, which basically add a self energy correction to one particle propagators and the self-consistent vertex corrections to the standard RPA type calculations, satisfying so-called conserving approximation. This method is theoretically better justified, but empirically it has been known that the results are not necessarily better than the RPA results when the system is a strong correlated (or strong coupling) one. Nevertheless, the FLEX studies of the Fe-based SC systems\cite{ikeda2008pseudogap,ikeda2010phase,zhang2009orbital} basically produced qualitatively similar results to the RPA results.

\subsubsection{functional renormalization group (fRG) technique}
Another quite powerful technique is the fRG technique. This theoretical technique is supposed to be unbiased, starting from the high energy Hamiltonians with local interactions $U, U'$,$J$, $J'$, etc., and without approximation traces down, through RG process, several instabilities simultaneously on a equal footing -- superconductivity(S), spin density wave (SDW), charge density wave (CDW), etc -- of a given Hamiltonian\cite{wang2009functional,wang2009antiferromagnetic,wang2010nodes,thomale2011exotic}. Hence this technique can reveal a close competition between S and SDW, for example, with changes of doping and other interaction parameters. For superconductivity itself, it can also trace the competition  among the different Cooper channels with the SC order parameter (OP) decomposition using lattice harmonics: which would correspond to $s$-, $d$-, $s^{\pm}$-wave, etc. in the band picture. At the moment, it is fair to say that this is the most unbiased theoretical tool to study the ground state of the interacting many body systems, but the weak point is that being a numerical technique, there is a limitation to analyze the underlying physics for the particular ground state.
Nonetheless, most of the results from the fRG method\cite{wang2009functional,wang2009antiferromagnetic,wang2010nodes,thomale2011exotic} are qualitatively similar to the ones from the RPA type theories, and the $s^{\pm}$-wave state was found a leading SC instability in the major region of the parameters $U, U'$,$J$, $J'$.

\subsubsection{local pairing approach: strong coupling theories}
Then there is a local pairing approach\cite{si2008strong,seo2008pairing,fang2008theory,daghofer2008model}. This approach starts with the local magnetic Hamiltonian, $J-J'$ model, added with a itinerant part $"t"$-term, hence called $t-J-J'$ model (the $t-J$ model is the counterpart model for the high-$T_c$ cuprates). The motivation comes from the experimental fact that the magnetism of the Fe-based SC compounds has a strong local character as well as the itinerant magnetism. The model itself is appealing and can be theoretically justified to some extent\cite{si2008strong,daghofer2008model}, however its solutions are not controlled. The simplest of all methods is
to decouple the local magnetic part Hamiltonian ($J-J'$ term) by mean field method into all possible SC pairing channels and determine the dominant pairing channel by diagonalizing the total $H$. At this level, this approach is technically nothing but BCS theory but these authors argued that it better incorporates the physics of the high energy local interactions like Hund and exchange couplings which should be important in the strongly correlated electron systems. Interestingly, the resulting pairing solutions compare quiet well with the results of other methods like fRG and RPA, namely the dominant pairing solution is again found to be the $s^{\pm}$-wave gap in the large parameter space.

With all theories mentioned above, the so-called $s^{\pm}$-wave state or its continuously modified pairing state have been consistently found as the dominant pairing solution in most of the parameter space. As to the pairing mechanism, RPA and FLEX belong to the BCS mechanism. fRG and local pairing theory also belong to the BCS framework in the sense of finding Cooper pair instability, but these methods contain unknown parts about how pairing glues arise (fRG) or how to solve this pairing interactions beyond a mean field method (local pairing theory).
On the other hand, people with more extreme viewpoint suspect that there should be fundamentally different pairing mechanism beyond a BCS paradigm, with which the strong correlation effects such as quantum criticality (QC) plays an active role.
In this review, we do not discuss this issue of the pairing mechanism, but will mainly focus on examining and testing the consistency of the $s^{\pm}$-wave pairing state in comparison with available experiments. Only in the last section 11, we will discuss "Experimental hints for pairing mechanism."

\begin{figure}
\noindent
\includegraphics[width=180mm]{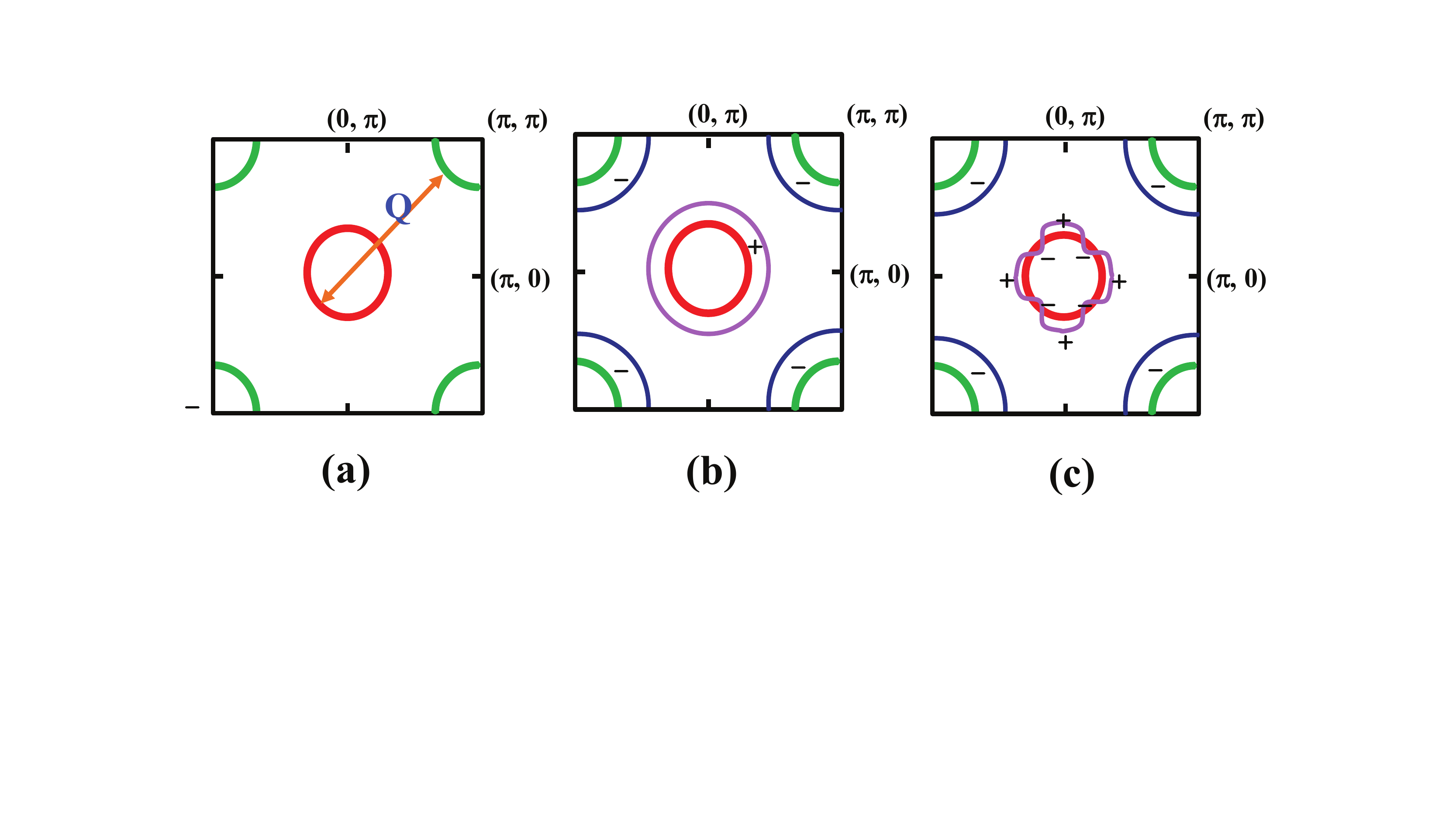}
\vspace{-4.5cm}
\caption{(Color online) (A) A typical FSs of $\epsilon_{h} (k)$ (red) and
$\epsilon_{e} (k)$ (green) band of the two band model. The AFM
wave vector ${\bf Q}$ spans between two bands. Here we use the BZ for 2 Fe/unit cell. (B) A sketch of the $s^{\pm}$-wave gap solution. (C) A sketch of the $s+g$-wave gap solution which can be continuously evolved from a solution of (B) without changing the gap symmetry of $A_{1g}$. \label{fig1}}
\end{figure}

\section{The $s^{\pm}$-wave pairing state}

\subsection{Phenomenological two band model}
In this paper, we use a minimal two band model for the $s^{\pm}$-wave pairing state to emphasize the proof of concept and to clarify the generic properties of the $s^{\pm}$-wave state for understanding the experimental data. By doing this, we can test the pairing symmetry and the structure of gap function of the FeSC without introducing ad-hoc assumptions and material specific fine tuning. The minimal two band model consists of one hole band and one electron band representing the generic $s^{\pm}$-pairing state. In real Fe-based SC compounds, there exists more than one hole bands around $\Gamma$ point $(0,0)$ and more than one electron bands around the $M$ point $(\pi,\pi)$ in the Brillouin zone (BZ) -- in this paper, we use the two dimensional BZ for two Fe/cell as depicted in Fig.1, therefore the hole (electron) band in our two band model should represent the thermodynamic average of a group of hole (electron) bands. The model is described with the Hamiltonian consisting of two bands,

\begin{eqnarray}
H &=& \sum_{k \sigma} \epsilon_h (k) h^{\dag}_{k \sigma} h_{k
\sigma} + \sum_{k \sigma} \epsilon_e (k) e^{\dag}_{k \sigma} e_{k
\sigma} \nonumber \\
&+& \sum_{k k^{'} \uparrow \downarrow} V(k,k^{'}) h^{\dag}_{k
\uparrow} h^{\dag}_{-k \downarrow}
h_{k^{'} \downarrow}h_{-k^{'} \uparrow}
+\sum_{k k^{'} \uparrow \downarrow} V(k,k^{'}) e^{\dag}_{k
\uparrow} e^{\dag}_{-k \downarrow} e_{k^{'} \downarrow}e_{-k^{'}
\uparrow} \nonumber \\
&+& \sum_{k k^{'} \uparrow \downarrow} V (k,k^{'}) h^{\dag}_{k
\uparrow} h^{\dag}_{-k \downarrow} e_{k^{'} \downarrow}e_{-k^{'}
\uparrow}
+\sum_{k k^{'} \uparrow \downarrow} V(k,k^{'}) e^{\dag}_{k
\uparrow} e^{\dag}_{-k \downarrow} h_{k^{'} \downarrow}h_{-k^{'}
\uparrow},
\end{eqnarray}

\noindent where $h^{\dag}_{k \sigma}$ and $e^{\dag}_{k \sigma}$ are the electron creation operators on the hole and the electron
bands, respectively. $\epsilon_{h,e} (k)$ are the dispersions of the hole band and electron bands in the two dimensional BZ, respectively.
The band dispersions $\epsilon_{h,e} (k)$ need not be specified for the purpose of this paper, but the generic Fermi surfaces and the BZ of
the model is depicted in Fig.1

The microscopic origin of the pairing interaction $V(k,k^{'})$ could be AFM fluctuations of the magnetic moment of the Fe 3$d$-electrons and theoretically connected to the dynamic spin susceptibility $\chi_s(\omega, q)$. Many authors have calculated $\chi_s(\omega, q)$, mostly using generalized RPA methods\cite{Kuroki,graser2009near,kuroki2009pnictogen,graser2010spin,kemper2010sensitivity,kontani2010orbital,onari2012self,
ikeda2008pseudogap,ikeda2010phase,zhang2009orbital} starting from more microscopic Hamiltonians. When only spin degrees of freedom are considered, several theoretical results produced a common feature, i.e., $\chi_s(\omega, {\bf q})$ is strongly peaked at ${\bf q}=(\pi,\pi)$ at low energy, indicating a nearby AFM instability, which is also qualitatively in accord with the early inelastic neutron experiments\cite{christianson2008resonant,qiu2008neutron} (see section 5 for more discussions and references). This kind of $\chi_s(\omega, {\bf q})$ with the AFM spin fluctuations is well known to lead repulsive interactions between electrons in the singlet Cooper channel over all momentum exchanges. However, there is also a large discrepancy between theory and experiment for the prediction of the size of magnetic moment when $\chi_s(\omega, {\bf q})$ is ordered. This issue is related to the fundamental question of how localized or itinerant the $d$-electrons are inside the Fe-based SC materials.
On the other hand, presumably more elaborate theory, which includes both orbital as well as spin degrees of freedom\cite{kontani2010orbital,onari2012self}, yields the result of $\chi_{charge}(\omega, {\bf q})$ in which the orbital fluctuations are strongly peaked at ${\bf q}=(0,0)$ and lead to an attractive interaction for small momentum exchanges. There are some indirect evidences of the strong orbital/charge fluctuations such as structural and nematic instabilities but no clear evidence exists for strong fluctuations in the small momentum sector ${\bf q} \approx (0,0)$ coming from $\chi_{charge}(\omega, {\bf q})$ with inelastic neutron scattering.

With this much discussion about the possible origin of the pairing interactions, in the above model, the pairing interaction $V(k,k^{'})$ is phenomenologically defined and assumed coming from an AFM spin fluctuations.
Therefore, it is all repulsive in momentum space for singlet Cooper channel and strongly peaked around ${\bf q}=\vec{k}-\vec{k^{'}}=(\pi,\pi)$ as

\begin{equation}
V(k,k^{'}) = V_M \frac{\kappa^2}{|(\vec{k}-\vec{k^{'}})-\vec{Q}|^2
+\kappa^2}
\end{equation}

\noindent where $\vec{k}$ and $\vec{k^{'}}$ are momenta in the two dimensional BZ and the parameter $\kappa$ controls the magnetic correlation length as $\xi_{AFM} = 2 \pi a/ \kappa$ ($a$ is the unit-cell dimension). This interaction
mediates the strongest repulsion when two momenta $\vec{k}$ and $\vec{k^{'}}$ are spanned by the ordering wave vector $\vec{Q}$. This condition is better fulfilled when the two momenta $\vec{k}$ and $\vec{k^{'}}$ reside each on hole band and electron band, respectively, as shown in the model FS structure (see Fig.1).

The SC ground state of the Hamiltonian Eq.(1) is solved using the BCS approximation and the two bands need two SC order parameters (OPs) as

\begin{eqnarray}
\Delta_{h} (k) & = &  \sum_{k^{'} } V(k,k^{'}) <h_{k^{'} \downarrow} h_{-k^{'} \uparrow}> , \\
\Delta_{e} (k) & = &  \sum_{k^{'} } V(k,k^{'}) <e_{k^{'}
\downarrow} e_{-k^{'} \uparrow}>.
\end{eqnarray}

\noindent After decoupling the interaction terms of Eq.(1) using the above OPs, the self-consistent mean field conditions lead to the following two coupled gap equations.

\begin{eqnarray}\label{gap_eq1}
\Delta_h (k)  &=&   - \sum_{k^{'} }  [V_{hh}
(k,k^{'}) \chi_h (k^{'}) + V_{he}
(k,k^{'}) \chi_e (k^{'})],  \nonumber \\
\Delta_e (k)  &=&    - \sum_{k^{'} } [V_{eh}
(k,k^{'}) \chi_h (k^{'}) + V_{ee}
(k,k^{'}) \chi_e (k^{'})].
\end{eqnarray}
\noindent where $V_{hh} (k,k^{'})$, $V_{he} (k,k^{'})$, etc are the interactions defined in Eq.(2) and the subscripts are
written to clarify the meaning of $V_{hh} (k,k^{'})$ =$V (k_h,k^{'} _h)$, $V_{he} (k,k^{'})$ =$V (k_h ,k^{'}_e)$, etc., and
$k_h$  and $k_e$ specify the momentum $k$  located on the hole and electron bands, respectively.
The pair susceptibilities are defined as
\begin{equation}\label{eq.chi}
\chi_{h,e}(k) = T \sum_{n}  \frac{\Delta_{h,e}}{\omega_n^2 + \epsilon_{h,e}^2 + \Delta_{h,e}^2(k)},
\end{equation}

\noindent where $\omega_n=\pi T (2n+1)$ are Matsubara frequencies.
For our purpose in this paper -- which is the demonstration of principle rather than a better fitting experimental data, one more simplification makes our discussions clearer without loss of essential features. Namely, we will assume constant isotropic $s$-wave gaps $\Delta_{h,e}(k)=\Delta_{h,e}$ on each band and the Fermi surface averaged pairing interactions between bands (interband) and within each band (intraband) such as $<V_{he}(k_h,k'_e)> = V_{he}$, $<V_{hh}(k_h,k'_h)>= V_{hh}$, and $<V_{ee}(k_e,k'_e)>= V_{ee}$, etc. Then the coupled gap equations (Eq.[\ref{gap_eq1}]) can be written as

\begin{eqnarray}\label{gap_eq2}
\Delta_h (T) &=&   -  [V_{hh} \chi_h (T) + V_{he}   \chi_e (T) ],  \nonumber \\
\Delta_e (T) &=&  - [V_{eh}  \chi_h (T) + V_{ee}  \chi_e (T)].
\end{eqnarray}

\noindent with the momentum integrated pair susceptibilities
\begin{eqnarray}
\chi_{h,e}(T) &=&  T \sum_{n} N(0)_{h,e} \int _{-\omega_{AFM}} ^{\omega_{AFM}} d \epsilon \frac{\Delta_{h,e}}{\omega_n^2 + \epsilon_{h,e}^2 + \Delta_{h,e}^2}  \\
&=& N(0)_{h,e} \int _{-\omega_{AFM}} ^{\omega_{AFM}} d \epsilon \frac{\Delta_{h,e}(T)}{2 E_{h,e}}
\tanh (\frac{E_{h,e}}{2 T}),
\end{eqnarray}
where $E_{h,e} =\sqrt{\epsilon_{h,e}^2 + \Delta_{h, e}^2}$ and $N(0)_{h,e} $ are the quasiparticle excitations and the DOS of
the hole and electron bands, respectively, and $\omega_{AFM}$ is the cutoff energy of the pairing potential $V(q)$.

Assuming all repulsive pairing potentials $V_{ab} > 0 ;(a,b=h,e)$ and with a dominance of the interband potentials as $V_{he} (= V_{eh})  > V_{hh}, V_{ee}$, the above gap equations produce the $s^{\pm}$-wave gap solution as proposed and reconfirmed by many authors\cite{Mazin,Kuroki,graser2009near,seo2008pairing,chubukov2008magnetism,DHLee,bang2008possible}. The schematic picture of the $s^{\pm}$-wave state is drawn in Fig.1(b).
And considering more realistic pairing potentials $V(k,k')$ with detailed coupling matrix element $M_{\alpha, \beta}(k,k')$ including the orbital degrees of freedom ($\alpha, \beta = d_{xz}, d_{yz}, d_{xy}$ on the bands), the isotropic $s$-wave OP on each band (Fig.1(b)) can develop an anisotropy\cite{kuroki2009pnictogen,graser2010spin,kemper2010sensitivity,maier2009origin} and in its extreme case even a nodal gap is possible\cite{graser2010spin,kemper2010sensitivity} as depicted in Fig.1(c).
However, in this case, this nodal gap doesn't break any additional symmetry from the case of Fig.1(b) but continuously keeps the same lattice symmetry of $A_{1g}$. As a result, the number of nodal points of Fig.1(c) are 8 instead of the 4 nodal points as in a $d$-wave gap.
This $A_{1g}$ nodal gap can occur either on the hole band or on the electron band depending on the details of the real Fe-based SC compounds. There have been several theoretical predictions for a nodal gap on the electron FS around the $M$ point\cite{graser2009near,maier2009origin,chubukov2009momentum,wang2010nodes} which recently has a supporting ARPES measurement\cite{xu2013possible}. The $A_{1g}$ nodal gap structure with 8 nodes depicted in Fig.1(c) was indeed confirmed in the heavily K-doped (Ba,K)Fe$_2$As$_2$ by ARPES experiments\cite{shinsik_node,okazaki2012octet} and theoretically explained\cite{bang2014shadow}.
There is also possibility to have $d$-wave nodal/nodeless gap solution\cite{graser2009near,seo2008pairing,DHLee,chubukov2009momentum} with the model of Eq.(1) in the parameter space of interactions nearby from the $s^{\pm}$-wave solution because the Fe-based SC systems are now well known to have several instabilities closely competing\cite{graser2010spin,chubukov2009momentum}.

Here we would like to emphasize that (1) the genuine $s^{\pm}$-wave state can develop a nodal gap without changing the gap symmetry or introducing any new pairing mechanism. (2) Surely, having nodes or not in the SC gap introduces distinctively different features in the SC properties. However, the type of the nodal gap in Fig.1(c) does not mean anything new or novel physics; they are just accidental nodes.

\subsection{Similarity to the $d$-wave solution}
\begin{figure}[h]
\centering
\noindent
\vspace{0cm}
\includegraphics[width=110mm]{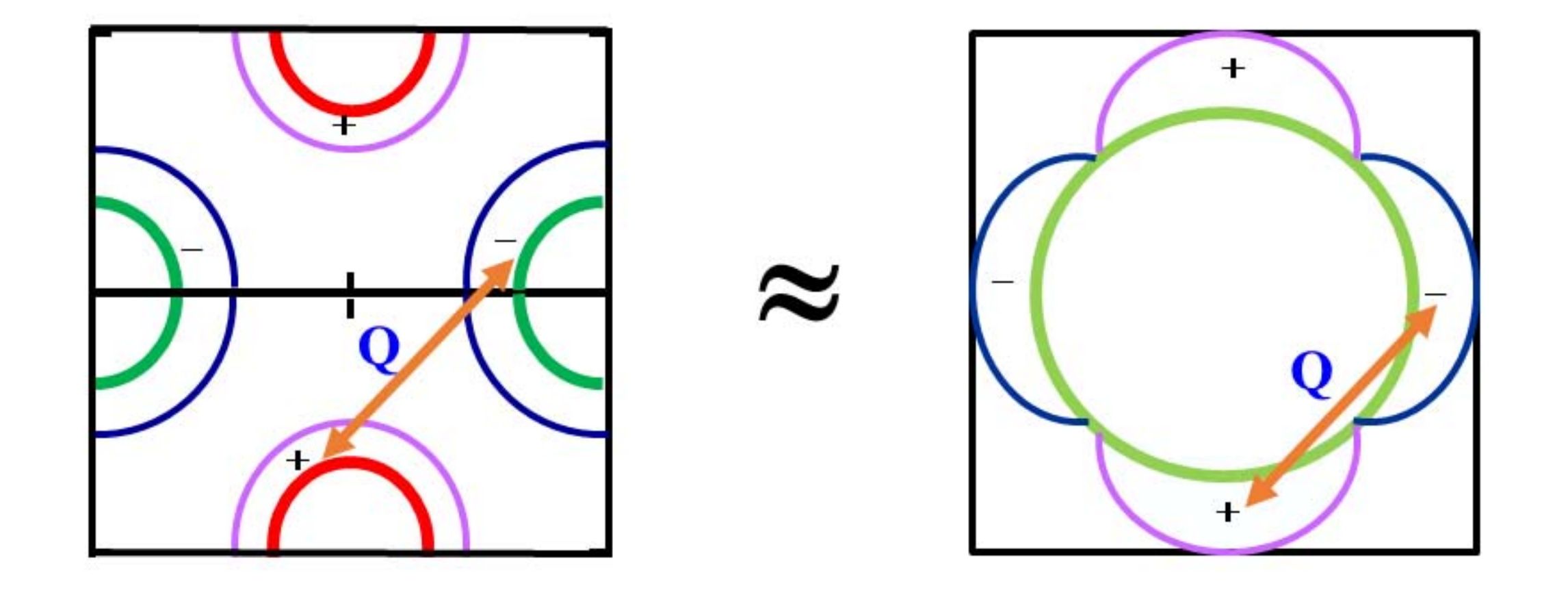}
\caption{(Color online) The illustrations of the $s^{\pm}$-wave gap solution in different cutting of the BZ and a typical $d$-wave gap solution. The comparison clearly shows the similarity of the pairing symmetry between two pairing solutions. In the weak coupling theory, it also strongly suggests a common pairing mechanism. \label{fig2}}
\end{figure}

On the other hand, by shifting the BZ by a half unit cell distance along either $x$- or $y$-direction (shifting the BZ by $(\pm \pi,0)$ or $(0, \pm \pi)$ as shown in Fig.2, we can see that the genuine $s^{\pm}$-wave state appears to have the same pairing symmetry as the $d$-wave state: more precisely the $s^{\pm}$-wave state has the gliding+$C_2$ symmetry and the $d$-wave state has only $C_2$ symmetry. More importantly, Fig.2 also suggests a possible common pairing mechanism for both SC gap states if both SC states are described within the weak coupling BCS theory although many researchers believe that both high-$T_c$ cuprate and Fe-based superconductivities should be governed by theories beyond a standard BCS theory.

\subsection{Some unique features of the $s^{\pm}$-wave solution}
Although it is genuinely a BCS theory, the two band $s^{\pm}$-wave superconductor described by the coupled gap equation (Eq.(\ref{gap_eq2})) has several novel features that are not shared with a standard single band BCS theory, and therefore are often and easily mistaken as evidences for a non-BCS superconductivity. It is the main objective of this review to provide a pedagogical overview of these new SC features of the $s^{\pm}$-wave superconductor and clarify possible confusions. To list those main novel features beforehand:

1) The gap sizes of $|\Delta_h|$ and $|\Delta_e|$ are not equal in general. Approximately they are inversely related to the DOSs $N_{h,e}$ as $\frac{|\Delta_h|}{|\Delta_e|} \approx \frac{N_e}{N_h}$ when $T \rightarrow 0$, and $\frac{|\Delta_h|}{|\Delta_e|} \approx \sqrt{\frac{N_e}{N_h}}$ when $T \rightarrow T_c$ (this second relation becomes exact when $V_{he,eh} \gg V_{hh,ee}$.

2) As a result, the gap-to-$T_c$ ratio can be much larger or smaller than the BCS prediction \cite{BCS} depending on which gap size is used such as $2 \Delta_L / T_c > 3.5$ and  $2\Delta_S / T_c < 3.5$, where $\Delta_{L,S}$ are the gaps of the larger/smaller of $\Delta_{h,e}$. The proper ratio can be calculated with the thermodynamically averaged gap value $\Delta_{ave}=(N_h |\Delta_h| + N_e |\Delta_e|)/(N_h+N_e)$, and then the ratio $2 \Delta_{ave}/T_c$  can be compared to the BCS value to judge whether the given Fe-based SC compound is in weak coupling limit or in strong coupling limit.

3) Because the two OPs $\Delta_{h,e}$ are coupled and induce each other, the small gap OP only cannot be destroyed by some perturbations and the larger gap OP still remains. This feature produces the unexpected Volovik effect in the vortex state with magnetic field.

4) The opposite signs of the two OPs produces similar effects as in $d$-wave superconductors such as the resonant impurity scattering, suppression of the coherence peaks in nuclear magnetic resonance (NMR). On the other hand, these effects are not as perfect as in $d$-wave because of the inequivalent size of the opposite-signed OP as $|\Delta^{+}| \neq |\Delta^{-}|$.

5) Finally, different combinations of the above properties produce many interesting and exotic SC properties in the $s^{\pm}$-wave state.

\subsection{Experimental tests for Pairing Symmetries: Density of States $N_s(\omega)$}
\begin{figure}[h]
\centering
\includegraphics[width=140mm]{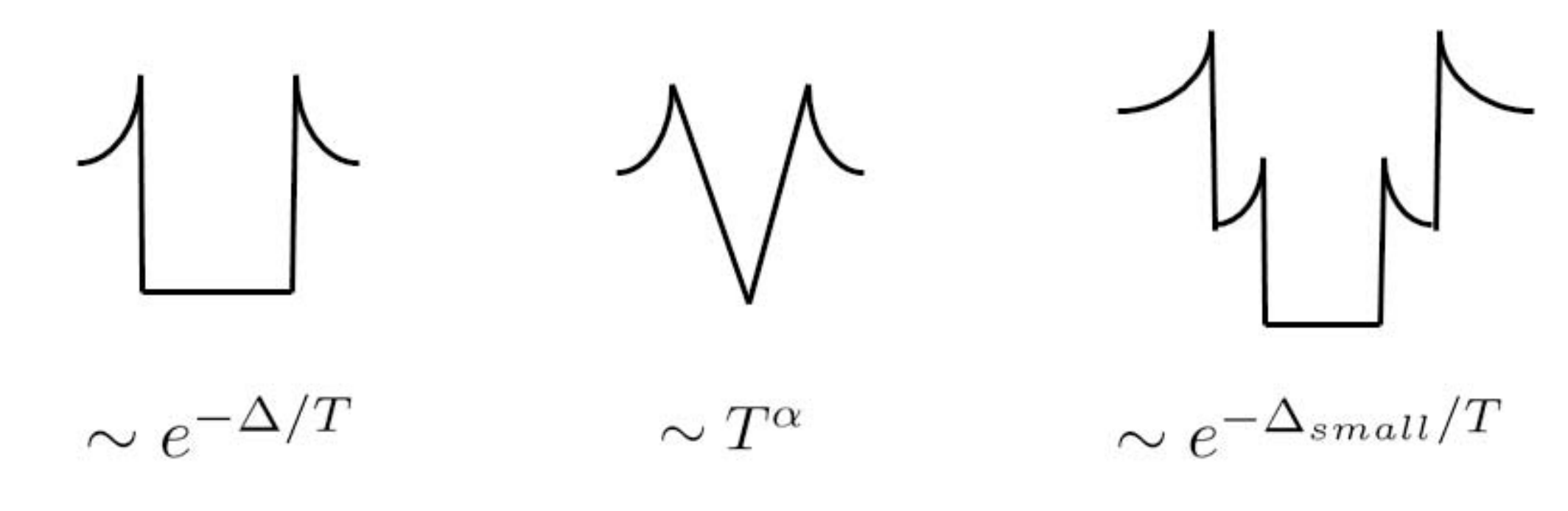}
\caption{(Color online) Schematic shapes of DOS $N(\omega)$ of three representative SC gap states: single $s$-wave, $d$-wave, and $s^{\pm}$-wave gaps, respectively. At bottom, typical low temperature behaviors of each SC state are written.
\label{DOS_power}}
\end{figure}

Most of experiments for testing the gap symmetry and gap function are basically probing the shape of DOS (see Fig.\ref{DOS_power}) by measuring various transport, thermodynamic, electro-magnetic, and optical properties in the SC state: e.g. angle resolved photoemission spectroscopy (ARPES), specific heat (SH), thermal conductivity, penetration depth, NMR, optical conductivity, Raman spectroscopy, etc.
Therefore, we expect that the clean $s^{\pm}$-wave superconductors should display the full gap behaviors just like a standard $s$-wave superconductor as shown in Fig.3(c), and indeed many FeSCs show various full gap SC behaviors, for example, the exponentially flat temperature dependence of the penetration depth, $\lambda(T) \sim \exp{[-\Delta/T]}$, for SmFeAsO$_{0.8}$F$_{0.2}$\cite{malone2009magnetic}, PrFeAsO$_{1-y}$\cite{hashimoto2009microwave1111}, and  (Ba,K)Fe$_2$As$_2$ \cite{hashimoto2009microwave122}, etc.

As to the sign-changing OP nature of the $s^{\pm}$-wave state, the most direct observation would be the Josephson tunnel junction experiment as done with the $d$-wave cuprate superconductor \cite{van1995phase}. Unfortunately, however,  the $s^{\pm}$-wave state in the FeSCs does not allow to make any real space contacts preferentially to each one of two different OPs because both OPs $\Delta_1^{+}$ and $\Delta_2^{-}$ are isotropic in $ab$-plane, so that the Josephson tunnel junction experiment is not so useful to distinguish the gap symmetry, or it has produced only a limited evidence\cite{chen2010integer}.

On the other hand, if only the DOS is probed, another two gap SC state, the $s^{++}$-wave state (e.g. MgB$_2$), which has the same signs on the two $s$-wave OPs as $\Delta_1^{+}$ and $\Delta_2^{+}$, should have the same SC features as in the $s^{\pm}$-wave superconductor.
Indeed, in the clean limit SC state, many SC properties  -- such as the SH, thermal conductivity, penetration depth, Knight shift, etc --  of these two SC states should be identical and cannot be distinguished.
However, in the last several years, it was found by many researchers that the sign-changing OP feature in the $s^{\pm}$-wave SC state can produce some unique and distinct SC properties compared to the $s^{++}$-wave superconductor both in clean limit and, in particular, more in impure case. Theoretical investigations of these new SC properties of the sign-changing $s$-wave superconductor have been challenging {\it per se} and their experimental comparisons with the numerous FeSCs have been very successful. As a result, now the $s^{\pm}$-wave state is mostly accepted as a standard pairing state of the FeSCs, possibly with a few exceptions.

The basic principle which enables us to probe this sign-changing OPs is to utilize processes involving a large momentum exchange, ${\bf q}\sim O(\pi)$, so that it connects two OPs located on the separate Fermi pockets in the BZ. There are three possibilities:

(1) Impurity scattering: local impurities (both magnetic and non-magnetic impurities) scatter quasiparticles with all momentum exchanges so that it connects the OPs between the same signs as well as between the opposite signs. Therefore, the impurity scattering will change all SC properties of the $s^{\pm}$-wave state very differently from the $s^{++}$-wave state. This is the main process to detect the $s^{\pm}$-wave state because impurities always exist in real materials, often inevitably but also controlled to some extent.

(2) Some selected SC properties intrinsically contain large momentum process so that they intrinsically probe the sign-changing OP nature. Examples are: NMR relaxation rate 1/T$_1$, and dynamic structure function $\chi_s(Q,\omega)$ by inelastic neutron scattering measurement.

(3) Combination of the above two processes also appears in various SC properties and results are often very interesting in unexpected ways.

\section{Impurity Effects on the $s^{\pm}$-wave state: the  $\mathcal{T}$-matrix theory}

The study of the impurity effects on the $s^{\pm}$-wave state is an interesting and also important subject. Theoretically, it is interesting because the $s^{\pm}$-wave state is a new pairing gap state and was not studied for its impurity effects before, and several novel impurity effects were indeed found with it. It has also practical importance in order to understand the SC properties of the FeSCs and identify the pairing symmetry of it.

Real materials, including SC materials, always contain impurities at some level, as well as defects where the sub-lattice site occupancy is not thermodynamically perfect.  Historically, the seminal paper by Abrikosov and Gorkov\cite{AG1960} has provided the theoretical ground to study this important effect. Using the Green's function method, these authors have shown that non-magnetic impurities don't affect the $s$-wave superconductors confirming Anderson's theorem\cite{anderson1959}, which was argued on the ground of the time-reversal symmetry of the singlet $s$-wave superconductor. This Green's function method for impurity scattering is very powerful to study various realistic cases such as magnetic/non-magnetic impurities, $s$- and non-$s$-wave superconductors, etc.

The work of Abrikosov and Gorkov was a Born approximation study of the impurity scattering and this theory was soon extended to a $\mathcal{T}$-matrix theory by many authors\cite{hirschfeld1988,balatsky2006imp} which includes a certain set of the multiple scattering process to infinite order at low density limit of impurity concentration.  Being a low impurity density expansion but not a coupling constant expansion, the $\mathcal{T}$-matrix theory can continuously describe from the Born (weak coupling) limit to the unitary (strong coupling) limit, hence can capture  a phenomena like the impurity resonance which is not possible with the Born approximation. This $\mathcal{T}$-matrix theory of impurity scattering in superconductors has been successfully applied to the unconventional superconductors such as the heavy fermion and high-$T_c$ cuprate superconductors. There is an extensive amount of literature on it, but we refer to two representative review papers on this subjects\cite{hirschfeld1988,balatsky2006imp} and readers can find more literature therein.

\subsection{Formalism}

\begin{figure}[t]
\centering
\includegraphics[width=130mm]{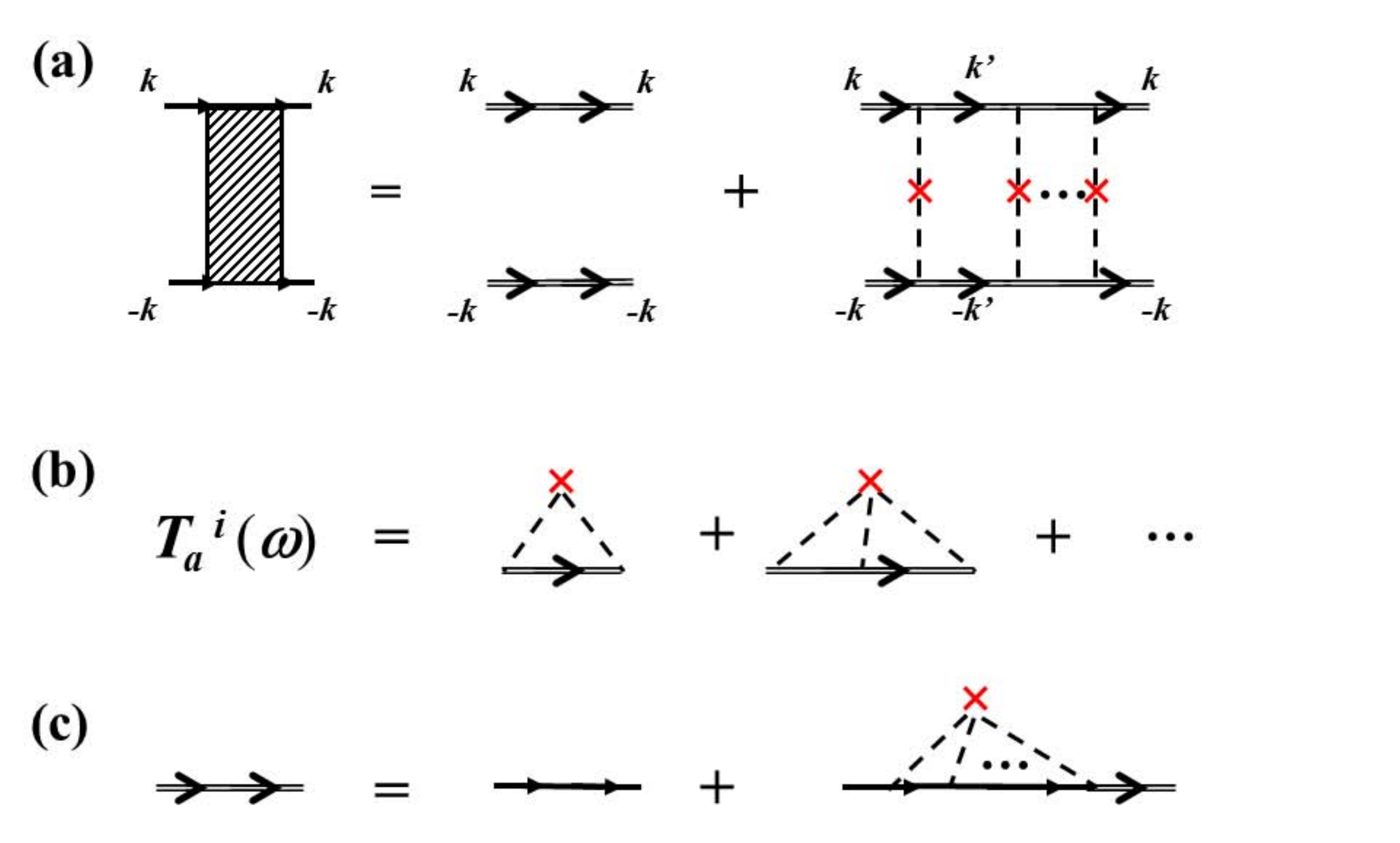}
\vspace{0cm}
\caption{(Color online) (A) Renormalized pair susceptibility $\chi_{h,e}(T)$ including normal self energy correction and vertex correction. (B) Definition of the $\mathcal{T}$-matrix. Notice that our $\mathcal{T}$-matrix begins with the second order in impurity  interaction. The first order diagrams are absorbed to chemical potential. (C) Dressed one-particle Green function with $\mathcal{T}$-matrix self energy correction. In all diagram, the red cross symbol indicate a single impurity.
\label{T-mtx_diagram}}
\end{figure}

This standard $\mathcal{T}$-matrix theory was generalized to the two band $s^{\pm}$-wave superconductors by one of us\cite{bang2009imp} and we briefly explain its essence here. All impurity scattering effects in SC state enter the pair susceptibility of the SC state Eq.(10)
\begin{equation}
\chi_{h,e}(T) = T \sum _n N(0)_{h,e} \int _{-\omega_{AFM}}
^{\omega_{AFM}} d \epsilon \frac{ \tilde{\Delta}_{h,e}(k) } {
\tilde{\omega}_n^2 +\xi^2 + \tilde{\Delta}_{h,e} ^2 (k)},
\end{equation}
with the self energy corrections to renormalize $\tilde{\omega}_n$ and $\tilde{\Delta}_{h,e}$ as follows
\ba \tilde{\omega}_n =\omega_n + \Sigma^{0}
 _h(\omega_n) + \Sigma^{0} _e(\omega_n),  \\
\tilde{\Delta}_{h,e} = \Delta_{h,e} + \Sigma^1 _{h} (\omega_n) +
\Sigma^1 _{e} (\omega_n).
\ea
and the selfenergies are calculated with the $\mathcal{T}$-matrices as
\begin{equation}
 \Sigma_{h,e} ^{0,1} (\omega_n)  = \Gamma \cdot \mathcal{T}^{0,1}
_{h,e} (\omega_n), ~~ \Gamma= \frac{n_{imp}}{\pi N_{tot}},
\end{equation}
where $\omega_n= T \pi (2n +1)$ is the Matsubara frequency,
$n_{imp}$ the impurity concentration, and $N_{tot}=N_h(0) +N_e(0)$
is the total DOS. Since it is in SC state, the selfenergies $\Sigma_{a}^{0,1}$ are defined for normal ($\Sigma_{a}^{0}$) and anomalous ($\Sigma_{a}^{1}$) parts on each band $"a=h,e"$.
The corresponding $\mathcal{T}$-matrices $\mathcal{T}^{0,1}_a$
are the Pauli matrix $\tau^{0,1}$ components in the Nambu space and are calculated as follows
\ba
\mathcal{T}^{i} _{a} (\omega_n) &=& \frac{G^{i} _{a} (\omega_n)}{D} ~~~~~(i=0,1; ~~a=h,e), \\
D &=& c^2 +[G^0 _h + G^0 _e]^2 + [G^1 _h + G^1 _e]^2,\\
G^0 _a (\omega_n) &=& \frac{N_a}{N_{tot}}  \Bigg\langle
\frac{\tilde{\omega}_n} {\sqrt{\tilde{\omega}_n^2 + \tilde{\Delta}_{a} ^2 (k) }} \Bigg\rangle,\\
G^1 _a (\omega_n) &=& \frac{N_a}{N_{tot}} \Bigg\langle
\frac{\tilde{\Delta}_{a}} {\sqrt{\tilde{\omega}_n^2 + \tilde{\Delta}_{a} ^2 (k) }} \Bigg\rangle,
\ea
where $c=\cot \delta_0$ is a convenient measure of scattering strength, with $c=0$ for the unitary limit and $c > 1$ for the Born limit. $\Big\langle ...\Big\rangle$ denotes the Fermi surface average. In Fig.\ref{T-mtx_diagram}, we show the schematic Feynman graphs of the above formulas: Fig.\ref{T-mtx_diagram}(a) the renormalized pair susceptibility $\chi_{h,e}(T)$,  Fig.\ref{T-mtx_diagram}(b) the $\mathcal{T}$-matrix $\mathcal{T}^{i} _{a} (\omega_n)$, and Fig.\ref{T-mtx_diagram}(c) the dressed one-particle Green function $G^{i} _{a} (\omega_n)$.

The most unique point of the impurity effects in the $s^{\pm}$-wave state is the term $[G^1 _h + G^1 _e]$ in the denominator $D$ (Eq.(15)). Because of the opposite signs of $\Delta_h$ and $\Delta_e$,  this term becomes almost zero (it becomes exactly zero in the case of the $d$-wave state, for example). When $[G^1 _h + G^1 _e] \rightarrow 0$ and $c \rightarrow 0$, the $\mathcal{T}$-matrices $\mathcal{T}^{0,1}_a$ can develop a resonant impurity band inside the gap at $\omega=0$, as happens in the $d$-wave state by the same mechanism. What is more interesting with the $s^{\pm}$-wave state is that this cancelation is not perfect, hence the term $[G^1 _h + G^1 _e]$ remains tiny but still finite. As a result, the impurity resonance is not located at zero energy as in the $d$-wave state but located at finite energies $\omega_{res} \ll \Delta_s$, symmetrically around the zero energy.

\begin{figure}
\hspace{2cm}
\includegraphics[width=130mm]{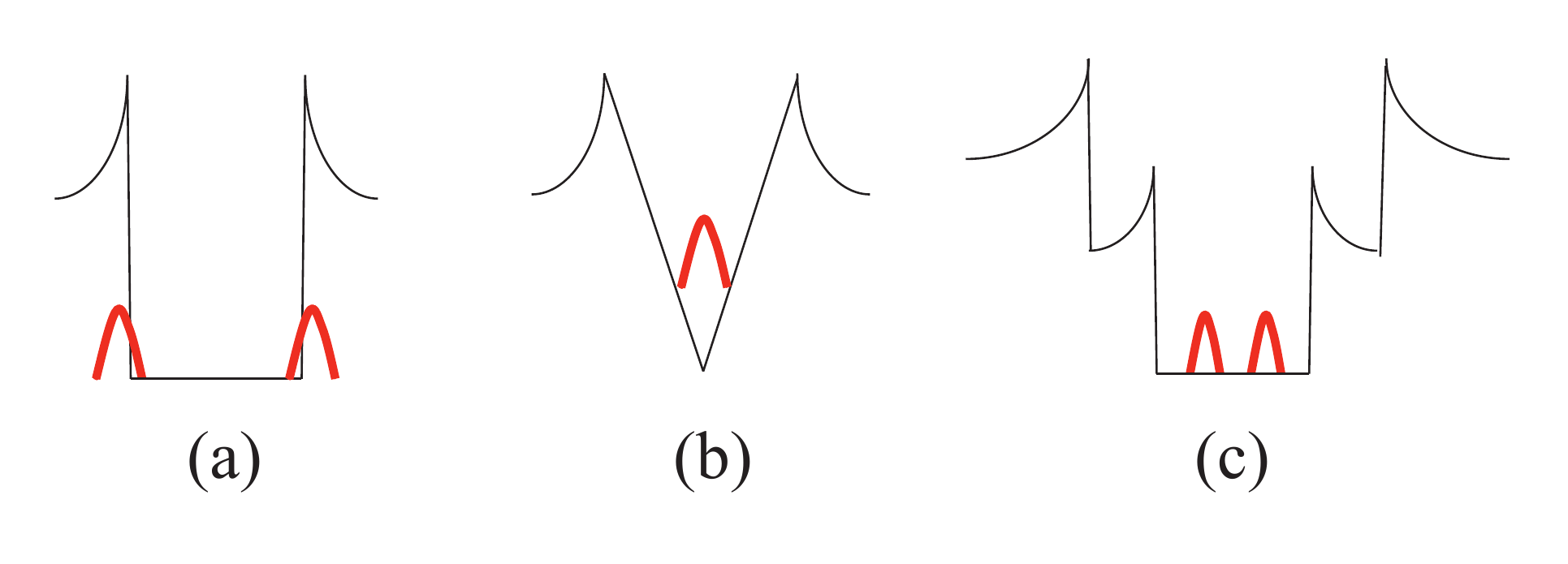}
\vspace{-0.5cm}
\caption{(Color online) Schematic illustrations of the impurity bound states (red lines) in (A) $s$-wave state;  (B) $d$-wave state; and (C) $s^{\pm}$-wave state. All cases are with non-magnetic impurities in the unitary (strong coupling) limit.
\label{sc_bound_imp}}
\end{figure}

\subsection{Impurity Resonance and In-gap states}

In Fig.\ref{sc_bound_imp}, the schematic pictures of the impurity bound states with the non-magnetic unitary impurities for different SC states, (A) $s$-wave, (B) $d$-wave, and (C) $s^{\pm}$-wave state, respectively, are depicted.
First, Fig.\ref{sc_bound_imp}(a) shows the case of the standard $s$-wave superconductors where the non-magnetic impurities do not form an in-gap state, hence do not induce any significant changes for the SC properties.
In contrast, Fig.\ref{sc_bound_imp}(b) shows the case of the $d$-wave superconductor where the non-magnetic unitary impurities induces a bound state at zero energy inside the SC gap. With a finite impurity concentration, this in-gap state forms a impurity band with a finite DOS around zero energy.
This so-called in-gap state induced by impurities in $d$-wave SC states has been well studied in connection with the heavy fermion and high-$T_c$ cuprate superconductors \cite{hirschfeld1988,balatsky2006imp}. The presence of the in-gap state significantly changes all SC properties of the $d$-wave superconductor and the accurate theoretical predictions of the systematic changes of these SC properties with impurities have played a crucial role to understand many puzzling experiments and finally to identify the $d$-wave gap symmetry.
Finally, Fig.\ref{sc_bound_imp}(c) shows the case of the $\pm s$-wave superconductor. Qualitatively and even physically we can understand it as an intermediate case between the $s$-wave and $d$-wave superconductors.

\begin{figure}
\hspace{2cm}
\includegraphics[width=130mm]{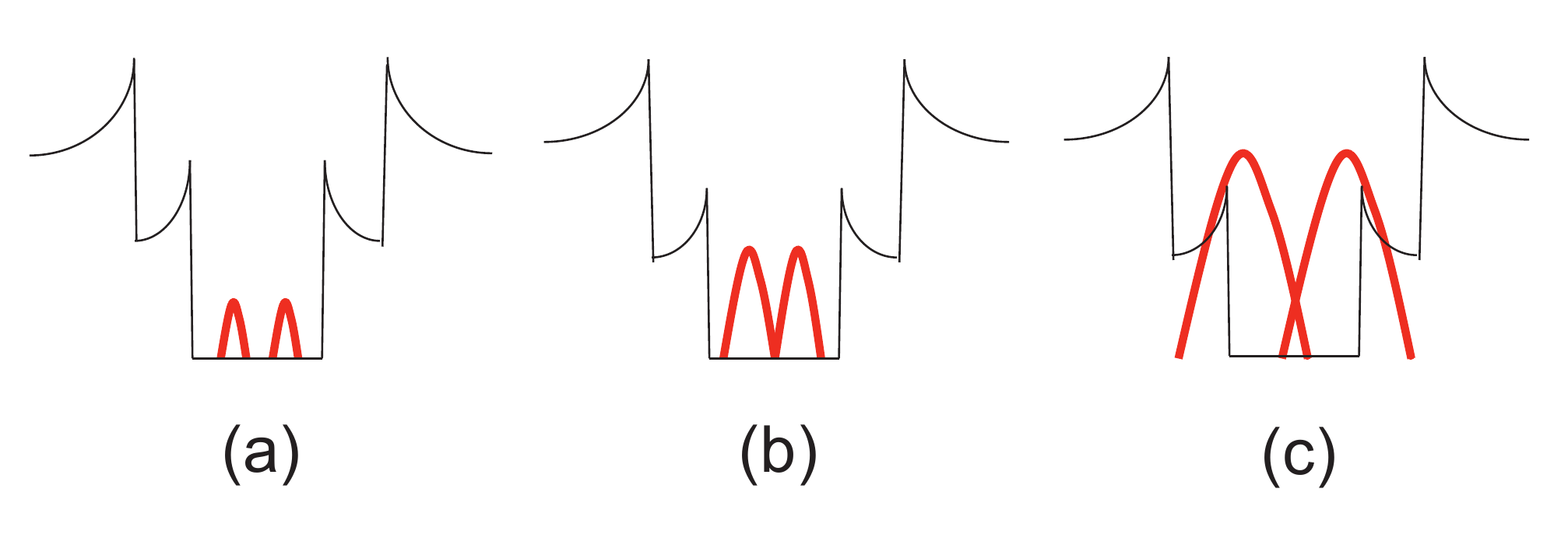}
\vspace{-0.5cm}
\caption{(Color online) Systematic evolution of the impurity bound states (red lines) in the $s^{\pm}$-wave state with increasing impurity concentration $n_{imp}$.
(A) $n_{imp} < n_{imp}^{crit}$, (B) $n_{imp} = n_{imp}^{crit}$, and (C) $n_{imp} > n_{imp}^{crit}$.
\label{spm_bound_imp}}
\end{figure}

In Fig.\ref{spm_bound_imp},  we illustrate how the impurity band systematically evolves inside of the gap in the $s^{\pm }$-wave superconductor as the impurity concentration increases. The low energy DOS $N(\omega)$ is still gapped at very low impurity concentration (Fig.\ref{spm_bound_imp}(a)), then it becomes a $V$-shape DOS (thermodynamically the same as the clean $d$-wave DOS) at the critical impurity  concentration $n_{imp}=n_c$ (Fig.\ref{spm_bound_imp}(b)), and finally evolves to the finite $N_{imp}(0)+V$-shape DOS with higher impurity concentration $n_{imp} > n_c$ (Fig.\ref{spm_bound_imp}(c)). This kind of a systematic evolution does't occur with a $d$-wave superconductor. From the clean limit to the very low impurity concentration of $n_{imp}$, the system should show a gapped $s$-wave SC behaviors. And for a finite range of concentration of $n_{imp}$ around $n_{imp}^{crit}$, the system shows $d$-wave-like SC properties unless the experimental probes go to a very low energy scale with $T$, or $\omega$, or fields $H$. And for higher concentrations of $n_{imp} > n_c$, it shows a dirty  $d$-wave-like behaviors, yet with some differences from it. In this case, the low energy $N(\omega)$ becomes $N_{imp}(0)+V$-shape DOS, and the sharp $V$-shape DOS continues to exist on top of a finite DOS at $N(0)$. This type of DOS (Fig.\ref{spm_bound_imp}(c)) looks different from a dirty $d$-wave DOS (Fig.\ref{sc_bound_imp}(b)) where the $V$-shape DOS becomes immediately flattened with a finite DOS at $N(0)$. This difference between the dirty $s^{\pm}$-wave and the dirty $d$-wave superconductors can easily be discerned by measuring the specific heat $C(T)$ and Knight shift $K(T)$, for example. In the next sections, we will show more details how this systematic evolution of the DOS at $N(\omega \approx 0)$ in the $s^{\pm}$-wave superconductor can show up as various non-trivial behaviors in different SC properties like NMR, specific heat, thermal conductivity, etc

\subsection{Examples of Impure SC DOS: $d$-wave and $s^{\pm}$-wave states.}

\begin{figure}[h]
\hspace{2cm}
\includegraphics[width=140mm]{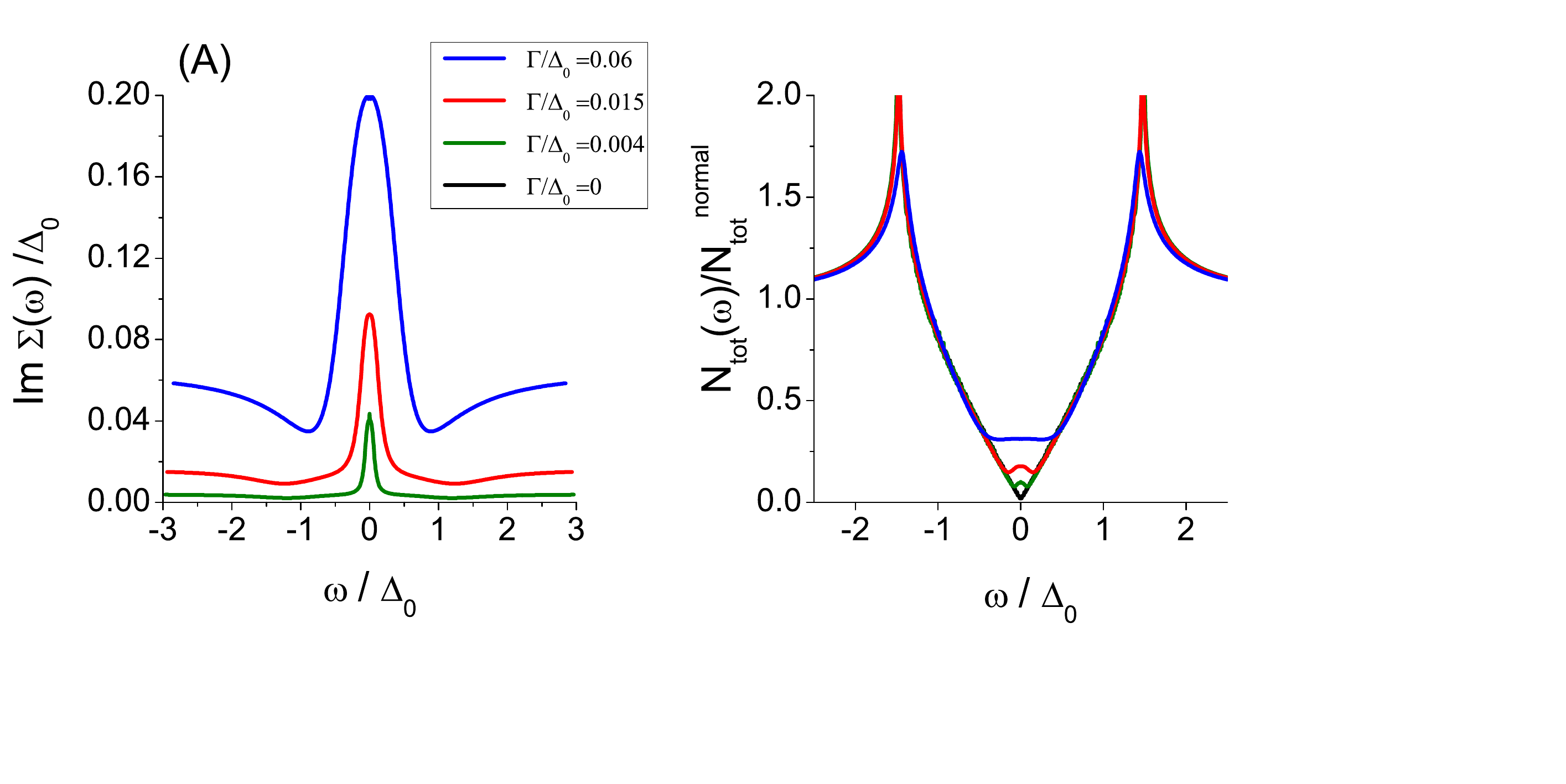}
\vspace{-1.5cm}
\caption{(Color online) The $d$-wave impurity band formation. (A) Impurity induced selfenergies $Im \Sigma^0 (\omega)$ for
different impurity concentrations, $\Gamma / \Delta_0= 0.0, 0.004, 0.015$, and, $0.06$, respectively.   (B) Systematic evolution of the DOS $N(\omega)$ of the $d$-wave state with the self energy corrections of (A).
\label{DOS_d_wave_imp}}
\end{figure}
%
\begin{figure}[h]
\hspace{2cm}
\includegraphics[width=140mm]{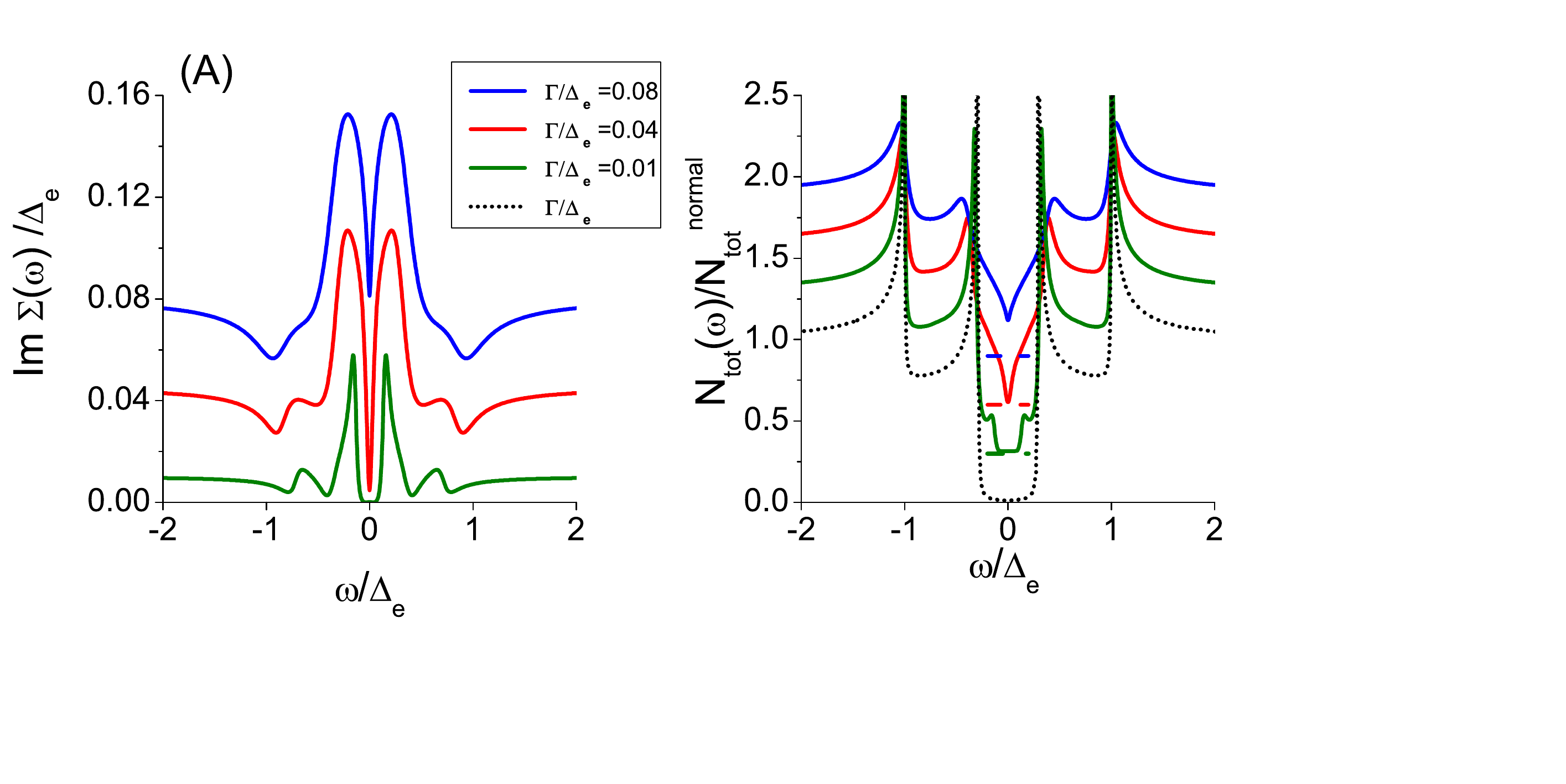}
\vspace{-2.0cm}
\caption{(Color online) The $s^{\pm}$-wave impurity band formation. (A) Impurity induced selfenergies $Im \Sigma^0 _{tot} (\omega) = Im \Sigma^0 _h + Im \Sigma^0 _e$ for different impurity concentrations, $\Gamma / \Delta_e= 0.0, 0.01, 0.04$, and, $0.08$, respectively.   (B) Systematic evolution of the DOS $N_{tot}(\omega)$ of the $s^{\pm}$-wave state with the self energy corrections of (A).  Each DOS are offset for clarity (the zero baselines of the offset are marked by the horizontal bars with the corresponding colors). From \cite{bang2009imp}.
\label{DOS_spm_wave_imp}}
\end{figure}

Figure \ref{DOS_d_wave_imp} and Figure \ref{DOS_spm_wave_imp} show the impurity induced selfenergies $Im \Sigma^0 (\omega)$ and the corresponding DOSs $N(\omega)$ with the self energy corrections for the $d$-wave and $s^{\pm}$-wave states, respectively. The results are self explaining by themselves as the key points were explained in the previous sections. The imaginary part of the impurity self energy $Im \Sigma^0 (\omega)$ for the $d$-wave state (Fig.\ref{DOS_d_wave_imp}(A)) clearly shows the zero energy resonance peak, and it induces the zero energy in-gap states in the total DOS $N(\omega)$. In the case of the $s^{\pm}$-wave state, the resonance peaks shown with $Im \Sigma^0 (\omega)$ (Fig.\ref{DOS_spm_wave_imp}(A)) are split symmetrically into four peaks $\pm \omega_{1} \ll \Delta_s$ and $\pm \omega_{2} \ll \Delta_L$, where $\Delta_s$ and $\Delta_L$ are small gap and large gap, respectively. The corresponding total DOS $N(\omega)$ shown in Fig.\ref{DOS_spm_wave_imp}(B) display the systematic evolution with increasing impurity concentration. More realistic calculations with a five orbital model also produced qualitatively similar results\cite{onari2009violation,kariyado2010single}

\section{Angle Resolved PhotoEmission Spectroscopy}

\subsection{Superconducting Gaps measured by ARPES}
\begin{figure}[h]
\centering
\includegraphics[width=130mm]{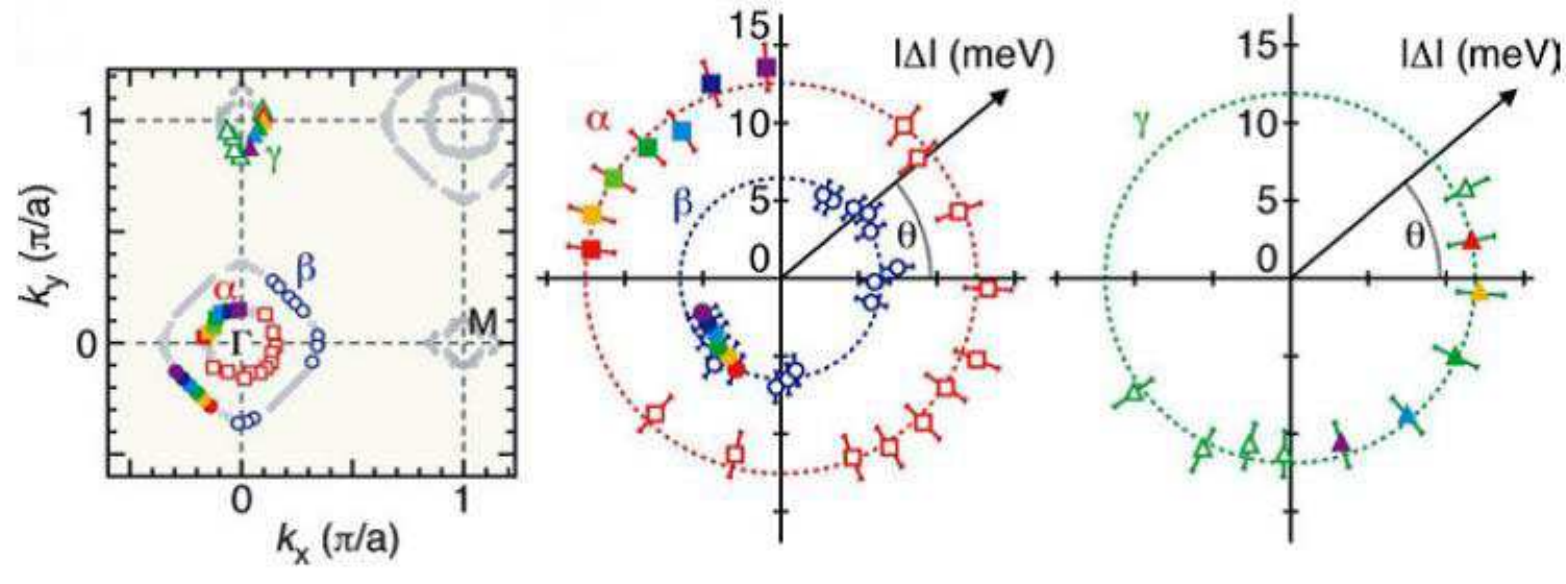}
\caption{(Color online) (Left) Fermi Surface mapping of Ba$_{0.6}$K$_{0.4}$Fe$_2$As$_2$. (Middle, Right) SC gap values at 15K  for the $\alpha, \beta,$ and $\gamma$ bands respectively, in polar coordinate \cite{ding2008}.
\label{ARPES_ding}}
\end{figure}

The angle resolved photoemission spectroscopy measures the quasiparticle dispersion and its spectra with energies and momenta resolved.
Nowadays the best energy resolution of the leading group is $\sim 1-2 meV$  with Synchrotron Radiation light\cite{damascelli2003angle} and can be  much better
with laser lights (laser ARPES). With this level of resolution, the ARPES is the most powerful and versatile experimental tool for studying the electronic properties of solids. For  the correlated metals and superconductors, it can measure the spectra of the bands near the Fermi level and provide the fundamental information of FS shapes/topology of the system. By comparing with the band calculations of Density Functional Theory (DFT), these ARPES results can also provide the information of the strength of renormalization (or effective masses $m^{\ast}$) of each band.
These are the most important information to begin with for any theoretical investigations of the correlated metal systems. Using polarizations (either linear or circular) of light, it can also provide an information of orbital degrees of freedom of the bands in the multi-orbital compounds like $d$-band metals, which is another very valuable information to understand these correlated metals\cite{lee2012orbital,yoshida2014orbital,yi2015observation,liu2015electronic,richard2015arpes}.
Most importantly, by changing temperature, the ARPES spectra also deliver information about how the electronic properties evolve with temperature; hence it reveals not only SC transition but also various magnetic and orbital transitions. We refer the readers for these interesting issues to two review papers\cite{seo2014review,zi2013angle}, among many.

In this review, since our main focus is limited to examining the consistency of the $s^{\pm}$-wave pairing state for the FeSCs, we will only briefly touch upon a small part of the ARPES experiments about the SC gap.
Concerning the SC gap symmetry, the ARPES experiment would measure the SC gap magnitude $|\Delta(k)|$ around the FSs and its interpretation is straightforward. Ideally it measures, after the thermal factor is subtracted, the one particle spectral density in the SC state \cite{damascelli2003angle,schrieffer1963theory} defined as

\begin{equation}
A(k,\omega) = - \frac{1}{\pi} Im G(k,\omega) = - \frac{1}{\pi} Im \frac{(\omega+\Sigma(k,\omega)) + \epsilon(k)}{(\omega+\Sigma(k,\omega))^2 - E^2(k)}
\label{ARPES}\end{equation}
with $E^2(k) = \epsilon^2(k)+\Delta^2(k)$. Therefore, tracking the FS ($\epsilon(k)=0$), ARPES can measure the momentum dependent SC gap size $|\Delta(k)|$ around the FSs, but not the sign of the gap.

\begin{figure}
\noindent
\centering
\hspace{1cm}
\includegraphics[width=150mm]{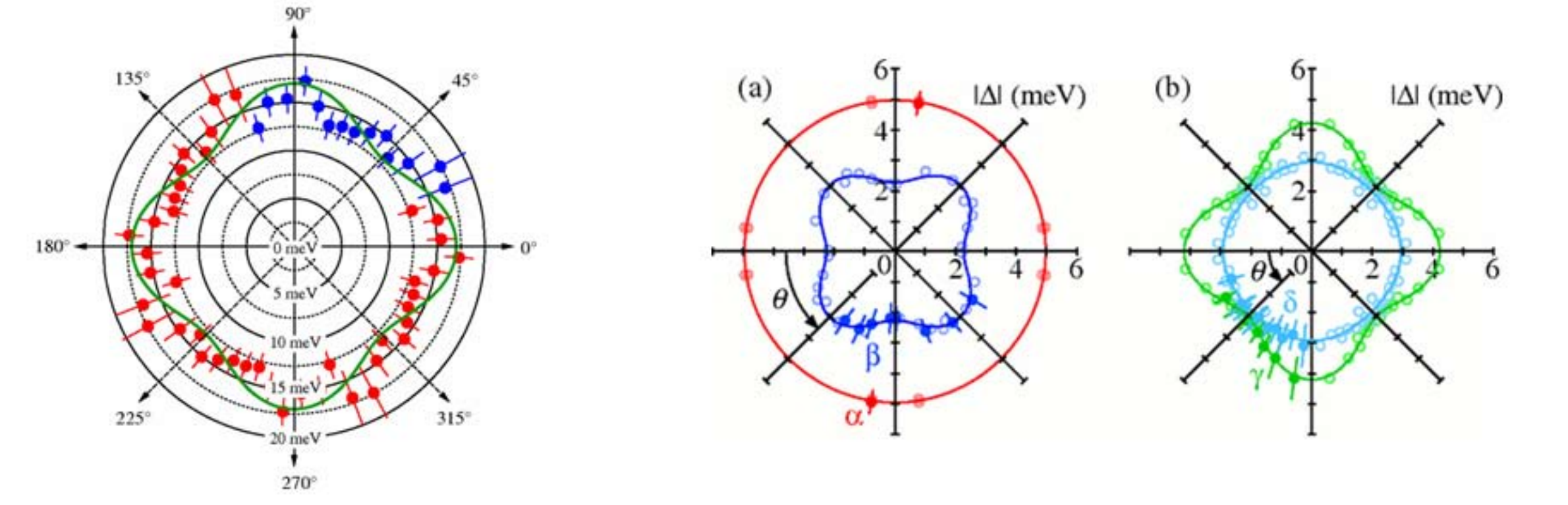}
\caption{(Color online) (Left) The ARPES data of $|\Delta(k)|$ on a hole pocket at $\Gamma$ of a NdO$_{0.9}$F$_{0.1}$FeAs single crystal ($T_c\sim 53$K) at $T=$ 20K \cite{kondo2008momentum}. (Right) The ARPES data of $|\Delta(k)|$s on the hole pockets at $\Gamma$ (a), and on the electron pockets at $M$ (b) of LiFeAs single crystal ($T_c\sim18$K) at $T=$8K \cite{umezawa2012unconventional}.
\label{ARPES_kondo}}
\end{figure}

A typical ARPES data of $|\Delta(k)|$ of the FeSC, Ba$_{0.6}$K$_{0.4}$Fe$_2$As$_2$\cite{ding2008}, are shown in Fig.\ref{ARPES_ding}. Shown in the left panel are the BZ (one Fe per unit cell) and measured FSs: the two hole band FSs ($\alpha$ and $\beta$) around $\Gamma$ point and one electron band FS ($\gamma$) around $M$ point. And in the right two panels, the measured SC gaps $|\Delta_a (k)|, (a=\alpha, \beta, \gamma)$, around each FS are displayed in polar coordinate. As seen, the gaps $|\Delta_a (k)|$ are quite isotropic and fully opened around each FS. This was an undeniable evidence for the $s$-wave superconductor.

Soon after, more systematic ARPES measurements for various Fe-based SC compounds with different dopings were carried out. The main findings of ARPES results for the FeSCs are:
(1) most majority of the FeSCs show almost isotropic full gaps around both the hole pockets and the electron pockets as shown in Fig.\ref{ARPES_ding}; (2) however, many FeSCs also show varying degree of anisotropy in $|\Delta_a (k)|$ as shown in Fig.\ref{ARPES_kondo} for NdO$_{0.9}$F$_{0.1}$FeAs\cite{kondo2008momentum}, and LiFeAs \cite{umezawa2012unconventional}; (3) for a small number of Fe-based SC compounds, the ARPES data of $|\Delta_a (k)|$ also show a strong evidences for possible nodal gaps either in the hole pockets\cite{shinsik_node,okazaki2012octet} or in the electron pockets\cite{xu2013possible}.

Although the ARPES cannot detect the sign of the gap function $\Delta_a (k)$, the item (1) and (2) above are consistent with the $s^{\pm}$-wave pairing gap scenario. Regarding the item (3) of possible nodal gaps, although some researchers tend to interpret the presence of nodal gap itself as a signature for a distinct novel pairing mechanism, we think it can still be very naturally accommodated within the $s^{\pm}$-wave pairing scenario (see Fig.1(c)). In particular, all the reported possible nodal gap structures\cite{xu2013possible,shinsik_node,okazaki2012octet} either on hole pockets or on electronic pockets didn't break $A_{1g}$ symmetry of the compounds, therefore they all belong to the same pairing symmetry class as the standard $s^{\pm}$-wave gap.

For more detailed analysis, the effects and consequences of the self energy correction $\Sigma(k,\omega)$ in Eq.(\ref{ARPES}) need to be included, which contains the renormalization and correlation effects from inelastic scattering as well as the impurity scattering effects. In particular, the impurity scattering induced self energy correction plays an important role in the SC phase as discussed in section 3. The non-magnetic impurities induce in-gap bound states (see Fig.6)  (Fig.8(A)) in the $s^{\pm}$-wave state, which would substantially change the shape of quasiparticle spectra $A(k,\omega)$ of Eq.(18) and consequently the total DOS, $N_{tot}(\omega)= \sum_k A(k,\omega)$ (Fig.8(B)).
In particular, with the impurity density higher than a critical amount $n_{imp}^{crit}$, the total DOS, $N_{tot}(\omega)= \sum_k A(k,\omega)$, becomes a $V$-shape DOS just like a $d$-wave superconductor as seen in Fig.8(B). However, an important distinction from a $d$-wave superconductor comes from the ARPES spectra. Namely, although the total DOS looks like a $d$-wave gap,  the individual q.p. spectra $A(k,\omega)$, which is measured by ARPES experiment, shows an isotropic non-zero gap $|\Delta_a (k)| \neq 0$ everywhere around the whole FSs. This is what has been observed with numerous ARPES experiments.

\subsection{Summary}
The main message from the ARPES experiments regarding the SC gap in the FeSCs is simple: except a few compounds or for exceptional dopings, most of the ARPES experiments with the FeSCs have been showing fully opened $s$-wave gaps with some degree of anisotropy around the FSs, which is consistent with the $s^{\pm}$-wave gap scenario. This probe itself, however, cannot tell the sign-changing nature of the $s^{\pm}$-gap function.
On the other hand, the interesting and challenging issue is that even when the ARPES experiments measured isotropic full gaps, various other experimental probes -- NMR, specific heat, thermal conductivity, penetration depth, etc --  have shown strong nodal gap features (various power law behaviors) in the SC state with the basically same (nominally) compounds whence the ARPES experiments saw full $s$-wave gaps.  Resolving this contradictory dilemma is the main subject of the remaining sections.

\section{Inelastic Neutron Scattering (INS)}

\subsection{Neutron resonance in the $s^{\pm}$-wave state}
INS measures the dynamic spin susceptibility $\chi_s(q,\omega)$, and it is well known that the pairing symmetry and the gap function can be probed utilizing the coherence factor of the spin susceptibility in SC state. The coherence factor of the non-interacting spin susceptibility $\chi_s ^0(q,\omega) \sim [G G + F F]$  is a case II type and defined  as follows for  $\omega > (|\Delta_k| + |\Delta_{k+q}|)$,

\be
\chi_s ^0 \sim \frac{1}{2} \Big( 1 - \frac{\epsilon_k \epsilon_{k+q} + \Delta_k \Delta_{k+q}}{E_k E_{k+q}} \Big) \sim \frac{1}{2} \Big( 1 - \frac{\Delta_k \Delta_{k+q}}{E_k E_{k+q}} \Big).
\label{coherence}
\ee
This coherence factor becomes $\sim 0$ when $\Delta_k \Delta_{k+q} >0$, or $\sim 1$ when $\Delta_k \Delta_{k+q} <0$. Therefore, depending on the SC gap function, the INS experiments can scan over the momentum $q$ and frequency $\omega$ space to find a constructive or destructive effect from the above coherence factor.

Then it was first noticed that this coherence factor can be utilized to identify the $d$-wave pairing state of the high-$T_c$ cuprate superconductors\cite{monthoux1994self,bulut1996neutron,liu1995theory,mazin1995neutron,abanov1999relation,morr1998resonance}, because choosing $q={\bf Q}$ -- which connects the $"+"$ part and the $"-"$ part of the gap function $\Delta_{d-wave}(k)$ (see Fig.2) -- the coherence factor is enhanced as $\sim 1$ for $\omega \sim 2 \Delta_{max}$. This enhanced non-interacting spin susceptibility $\chi_s ^0(q,\omega)$ can have a more dramatic effect in the interacting susceptibility as, for example, using an RPA approximation,

\be
\chi_s (q,\omega) = \frac{\chi_s ^0(q,\omega)}{1 - U \chi_s ^0(q,\omega)}
\ee
where $U$ is a local interaction. As $T$ goes to zero, the real part of $\chi_s ^0(q,\omega)$ has so-called logarithmic divergence at $q={\bf Q}$ and $\omega=2 \Delta_{max}$\cite{morr1998resonance} -- this singularity will be mitigated over frequencies and momenta because of the distribution of $d$-wave gap $\Delta(k)$ -- due to the constructive coherence factor of Eq.(\ref{coherence}), then the denominator $[1 - U \chi_s ^0(q,\omega)]$ can approach zero near $\omega \sim 2 \Delta_{max}$, for a wide range of values of $U$. As a result, the imaginary part of dynamic spin susceptibility $Im \chi_s (q,\omega)$ can form a "resonance" peak  which was detected by numerous INS experiments with cuprate as well as heavy fermion superconductors \cite{dai1999magnetic,fong1999neutron,stock2008spin,sato2001strong} confirming the $d$-wave pairing state in these materials. In this spin exciton (or resonance) mechanism, it is important to notice that the value of the RPA interaction $U$ needs not a fine tuning due to the logarithmic divergence of $Re \chi_s ^0({\bf Q},\omega=2 \Delta_{max})$. In reality, you need a minimum strength of interaction, but there should be a wide window of strength to form a resonance, so that this spin exciton resonance mechanism should be quite universal for $d$-wave superconductors as well as $s^{\pm}$-wave superconductors.

\begin{figure}[t]
\hspace{1cm}
\includegraphics[width=150mm]{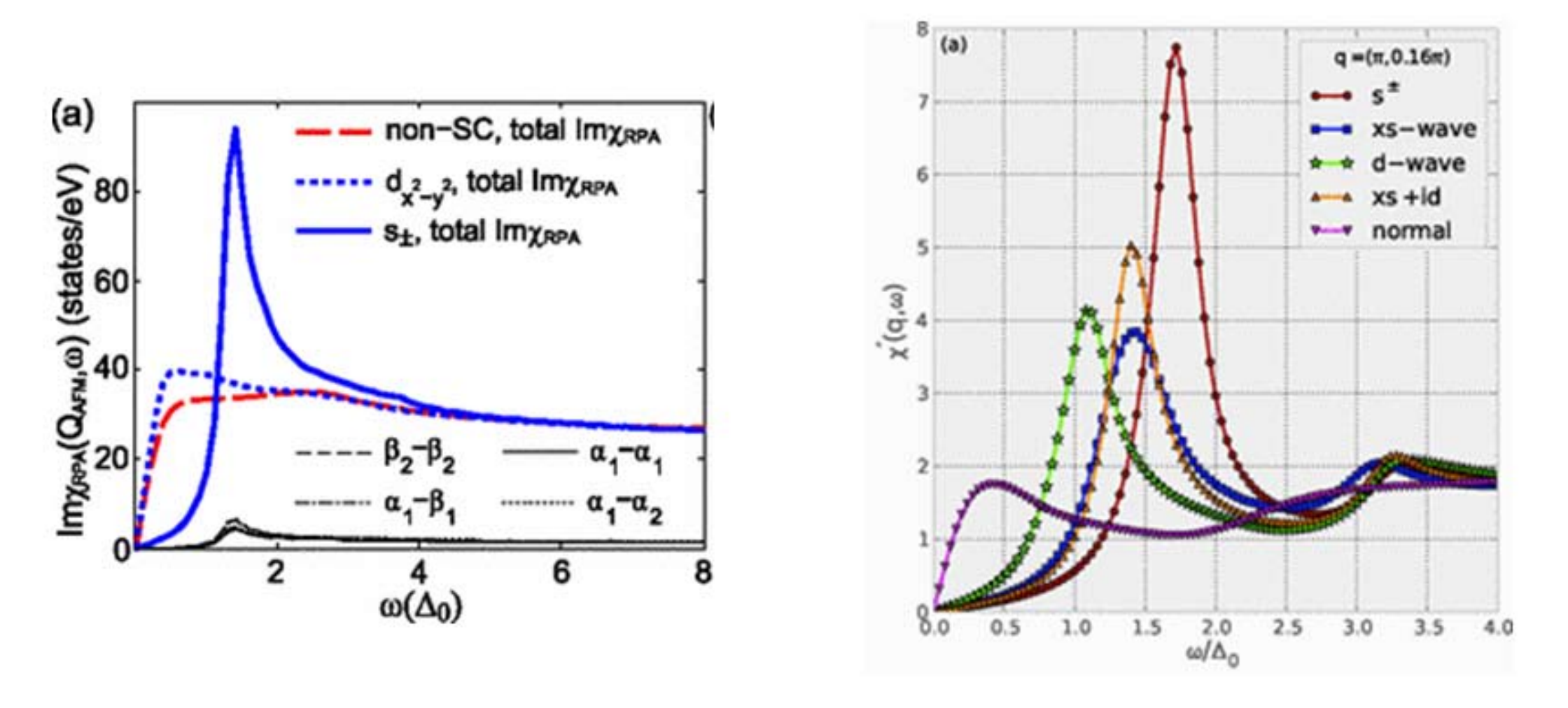}
\vspace{-1.0cm}
\caption{(Color online)  (Left) The imaginary part of RPA spin susceptibility $Im \chi_s({\bf Q_{AFM}},\omega)$ for the normal and SC states calculated with four band model. The sharpest spectrum (sold blue) is for the $s^{\pm}$-wave state. From \cite{korshunov2008theory}.  (Right) The imaginary part of RPA spin susceptibility $Im \chi_s({\bf q^{\ast}},\omega)$ (here ${\bf q^{\ast}}$ is the best nesting vector for a chosen band structures) for the normal and SC states calculated with five orbital model. The sharpest spectrum (red line $+$ circles) is for the $s^{\pm}$-wave state. From \cite{maier2012evolution}.
\label{INS_theory}}
\end{figure}

\begin{figure}[h]
\hspace{1cm}
\includegraphics[width=150mm]{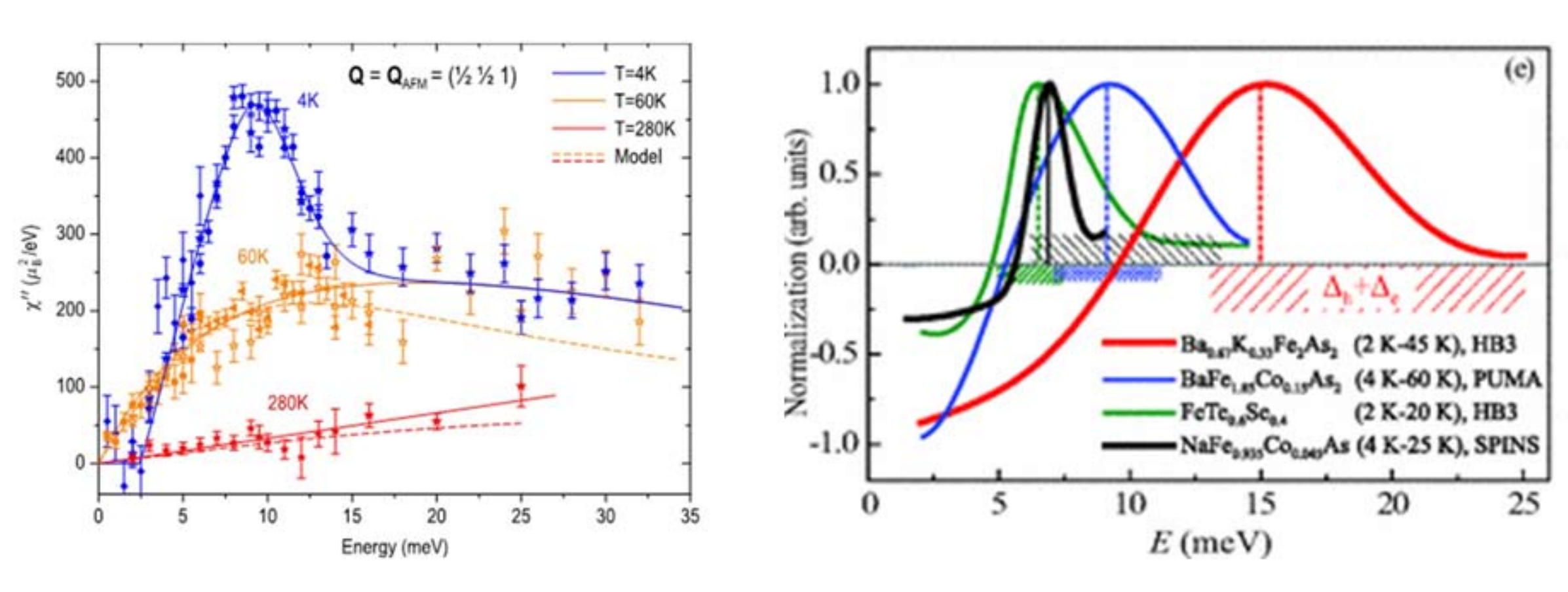}
\caption{(Color online) (Left) Imaginary part of the spin susceptibility $Im \chi_s({\bf Q_{AFM}},\omega)$ in the superconducting (T = 4 K) and the normal state (T = 60 and 280 K), obtained from the INS data $S({\bf Q},\omega)$ of BaFe$_{1.85}$Co$_{0.15}$As$_2$ ($T_c$=25K) by correcting for the thermal population factor. The solid lines are guides to the eye and the dashed lines represent fits with a standard theoretical formula of the normal state spin susceptibility with a short range AFM correlation. From \cite{inosov2010normal}.
(Right) The schematics of neutron spin resonance for various iron pnictide superconductors. The red, blue, black, and green dashed regions show the range of $2\Delta$ as determined from ARPES and other experiments. From \cite{zhang2013distinguishing}.
\label{INS_exp}}
\end{figure}

The exact same mechanism for the neutron resonance can occur with the $s^{\pm}$-wave state because $\Delta_k \Delta_{k+q} <0$ if the momentum $q$ is selected to satisfy $\Delta_k \Delta_{k+q} \sim \Delta_h \Delta_{e}$. Here this particular momentum $q={\bf Q}$ is the nesting vector or near from it $q={\bf q^{\ast}} \approx {\bf Q}$ which best connects the hole band and the electron band in FeSCs (see Fig.2). This possible neutron resonance peak in the FeSCs was theoretically \cite{korshunov2008theory,maier2008theory,maier2009neutron} suggested as a proving signature of the $s^{\pm}$-wave state, and almost simultaneously detected by INS experiment with optimal K-doped Ba-122\cite{christianson2008resonant}
and  La-1111\cite{qiu2008neutron} in accord with the theoretical prediction. Soon numerous INS experiments with K-doped Ba-122\cite{castellan2011effect,lee2011incommensurate}, Co-doped Ba-122\cite{lumsden2009two,christianson2009static,inosov2010normal}, Co-doped Na-122 \cite {zhang2013distinguishing}, Fe(SeTe) \cite{qiu2009spin,iikubo2009antiferromagnetic,babkevich2010magnetic,argyriou2010incommensurate} have reported the occurrence of the resonance peak below $T_c$ and disappearance of it above $T_c$, confirming that this resonance is related to the superconductivity and the most natural explanation would be with the $s^{\pm}$-wave pairing state.

Figure \ref{INS_theory} show two representative theory (RPA) calculations with different choices of bands and interactions\cite{korshunov2008theory,maier2012evolution} for the FeSCs, and shared the main feature in common: the resonance peak appears below the SC gap edge ($\omega_{res} < |\Delta_h| + |\Delta_e|$)  at the nesting vector $q={\bf Q}$ or near, and the sharpest for the $s^{\pm}$-wave state. These results are well compared to Figure \ref{INS_exp} which show the representative INS experiments for various FeSCs reporting the resonance peak appearing below $T_c$. Therefore, theories and experiments seem to be quite consistent each other and support the $s^{\pm}$-wave pairing state for the FeSCs.

\subsection{Some questions for the neutron resonance in the $s^{\pm}$-wave state}

Despite the very natural explanation of the neutron peak, some questions were raised for the $s^{\pm}$-wave state scenario. The main question was that the INS experiments show much too broad resonance peak, compared to the very sharp peak from the RPA calculations (see Fig.\ref{INS_theory} and Fig.\ref{INS_exp}. However, this sharpness of the resonance peak with the RPA theory calculations can be improved considering many realistic reasons such as impurity scattering, gap anisotropy\cite{hirschfeld2011gap}. Also the experimental data of the resonance peak shape is not always broad and can be rather sharp for some Fe-based SC compounds (see the right panel of Fig.\ref{INS_exp}).
Nevertheless, based on this critique, Kontani and coworkers\cite{onari2010structure,onari2011neutron} rejected the $s^{\pm}$-wave scenario and proposed the $s^{++}$-wave state to explain the broad neutron peak. In this model with the $s^{++}$-wave state,  the coherence factor of Eq(\ref{coherence}) is destructive, hence there is no logarithmic divergence in $Re \chi_s^0$ and therefore no "resonance" below $2 \Delta$ ($|\Delta_h| \approx |\Delta_e|$) possible. Instead, these authors claimed that the quasiparticle (q.p.) damping -- which should be sufficiently strong because of strong correlation -- should drop in the SC state but only for $\omega < 3 \Delta$. Then, because of this sudden drop of the q.p. damping, the dynamic spin susceptibility $\chi_s (q,\omega)$ in the SC state can have a hump like enhancement in the region of $2 \Delta <\omega <  3 \Delta$. While their numerical calculations of $Im \chi_s (q,\omega)$ in Ref.\cite{onari2011neutron} appear consistent with the broad peak of neutron experiments, this scenario requires a fine tuning of parameters like damping rates in normal state and SC state, and the RPA interaction strength to produce the sizable hump structure. Considering almost universal observation of the neutron resonance peak in the various FeSCs\cite{christianson2008resonant,qiu2008neutron,castellan2011effect,lee2011incommensurate,lumsden2009two,christianson2009static,
inosov2010normal,qiu2009spin,iikubo2009antiferromagnetic,babkevich2010magnetic,argyriou2010incommensurate}, this $s^{++}$-wave scenario seems to be too artificial.
However, this point is still under debate\cite{nagai2011determination,onari2011reply,nagai2011comment}.

The second question is about the temperature dependence of the resonance energy $\omega_{res}(T)$. Because the spin resonance is a particle-hole exciton in spin channel in the SC state, the constraint of the resonance peak position should be $\omega_{res}(T) < |\Delta_h| + |\Delta_e|$. Therefore, increasing temperature as $T \rightarrow T_c$, it is expected that $\omega_{res}(T)$ should decrease. Indeed, the neutron peaks of BaFe$_{1.85}$Co$_{0.15}$As$_2$\cite{inosov2010normal} showed the expected temperature variation, while the data of FeTe$_{0.6}$Se$_{0.4}$ shows that $\omega_{res}(T)$ is almost temperature independent up to very close $T_c$ but only the peak height decreases\cite{harriger2012temperature}. This needs an explanation.

\subsection{Summary}
The INS resonance peak in SC state observed in numerous Fe-based SC compounds\cite{christianson2008resonant,qiu2008neutron,castellan2011effect,lee2011incommensurate,lumsden2009two,christianson2009static,
inosov2010normal,qiu2009spin,iikubo2009antiferromagnetic,babkevich2010magnetic,argyriou2010incommensurate} is absolutely consistent with the $s^{\pm}$-wave state. The underlying mechanism of this phenomena is the constructive coherence factor of the $s^{\pm}$-wave state and identically operating and confirmed with the $d$-wave cuprate superconductors. Although there are a few details -- shape of peak spectra, temperature dependence of the peak frequency, etc --  needing improvement to fit the experimental neutron spectra, the overall consistency between theories and experiments is excellent.

\section{Nuclear Magnetic Resonance}
The nuclear magnetic resonance experiments consist of measurements of three major quantities: Knight shift $K(T)$, $T_1$, and $T_2$ relaxation times, respectively. Among these three quantities, Knight shift $K(T)$, and $1/T_1$ directly measure the DOS of metals below and above $T_c$, so that they can provide the valuable information about the SC gap functions $\Delta(k)$.

\subsection{Knight shift}
\subsubsection{Clean limit}

Knight shift $K(T)$ is the relative shift of the NMR resonant frequencies between the Zeeman split energy levels of the nuclear spin of the specific ions inside material. Zeeman energy is proportional to the total magnetic field $H_{eff}$ at the nuclear spin, which is defined as $H_{eff}=(1+K(T))H_{ext}$. While there are several sources for $K(T)$, in metallic systems, the main contribution for $K(T)$ is the paramagnetic uniform spin susceptibility times the hyperfine coupling.
Therefore, in the case of the singlet pairing superconductors, it basically measures the change of the DOS $N(0)$ at Fermi level of metal as temperature varies above and below $T_c$.  For the $s^{\pm}$-wave superconductors, the theoretical formula of Knight shift is given as
\begin{eqnarray}
\label{spin_susceptibility}
K(T) & \sim&  Re \chi_S(q=0, \omega \rightarrow 0)  \\
&=&  -\int_0 ^{\infty} d \omega
\frac{\partial f_{FD} (\omega)}{\partial \omega}
\sum_{a=h,e} \Big[ N_a(0) \left\langle Re \frac{\omega}{\sqrt{\omega^2-\Delta_a ^2(k,T)}}
\right\rangle_{k} \Big] \\
&=&  -\int_0 ^{\infty} d \omega \frac{\partial f_{FD} (\omega)}{\partial \omega}
\Big[ N_h(\omega,T) + N_e(\omega,T)  \Big] \label{knight}
\end{eqnarray}
where $\chi_S(q, \omega)$ is the dynamic spin susceptibility of the conduction electrons and $f_{FD}(\omega)=(1+ \e^{\omega/T} )^{-1}$ is Fermi-Dirac distribution function. $\left\langle ...\right\rangle$  means a FS average and the inside expression is nothing but the normalized DOS $N(\omega)$ in the SC state. The coherence factor for the static uniform spin susceptibility becomes $"1"$ and the uniform susceptibility limit ($q=0$ ) doesn't allow the inter-band scattering, hence the Knight shift of the $s^{\pm}$-pairing state is just summation of two $s$-wave Knight shifts from the hole band and the electron band, respectively.

We expect, therefore, a typical temperature dependence of an ordinary $s$-wave superconductor for Knight shift $K(T)$ of FeSCs: for a singlet $s$-wave superconductor, a rapid drop below $T_c$ and an exponentially flat behavior at low temperatures for $T < T_c /3$. However, being an two band model, the gap-to-$T_c$ ratio of the $s^{\pm}$-wave superconductor can be very different from the standard BCS value of $2 \Delta_{BCS} /T_c \approx 3.5$, such as $2 \Delta_L /T_c \gg 3.5 $ and $2 \Delta_S /T_c \ll 3.5$, where $\Delta_{L,S}$ are the larger and smaller gaps from the hole and electron bands. Each band has their own DOS $N(0)_{h,e}$ and it has been shown that in general the inverse relation $\frac{\Delta_h}{\Delta_e} \sim  \frac{N_e}{N_h}$ holds for $s^{\pm}$-wave model when the interband repulsion is the dominant pairing interaction\cite{bang2008possible}.
Therefore depending on the relative ratio between $N_h$ and $N_e$, and the choice of the gap-to-$T_c$ ratio $2 \Delta_{h,e} /T_c$, the shape of the temperature dependence of $K(T)$ over the wide range below $T_c$ can be very different from the standard single band BCS behavior. Choosing relatively larger values of $\Delta _{h,e}/ T_c$, we can phenomenologically simulate the effect of the strong coupling superconductivity.  For overall temperature dependence of the gaps $\Delta_{h,e}(T)$, we use a phenomenological BCS formula, $\Delta_{h,e}(T)=\Delta_{h,e}(T=0) \tanh (1.74 \sqrt{T_{c}/T-1})$.

\begin{figure}[h]
\hspace{2cm}
\includegraphics[width=150mm]{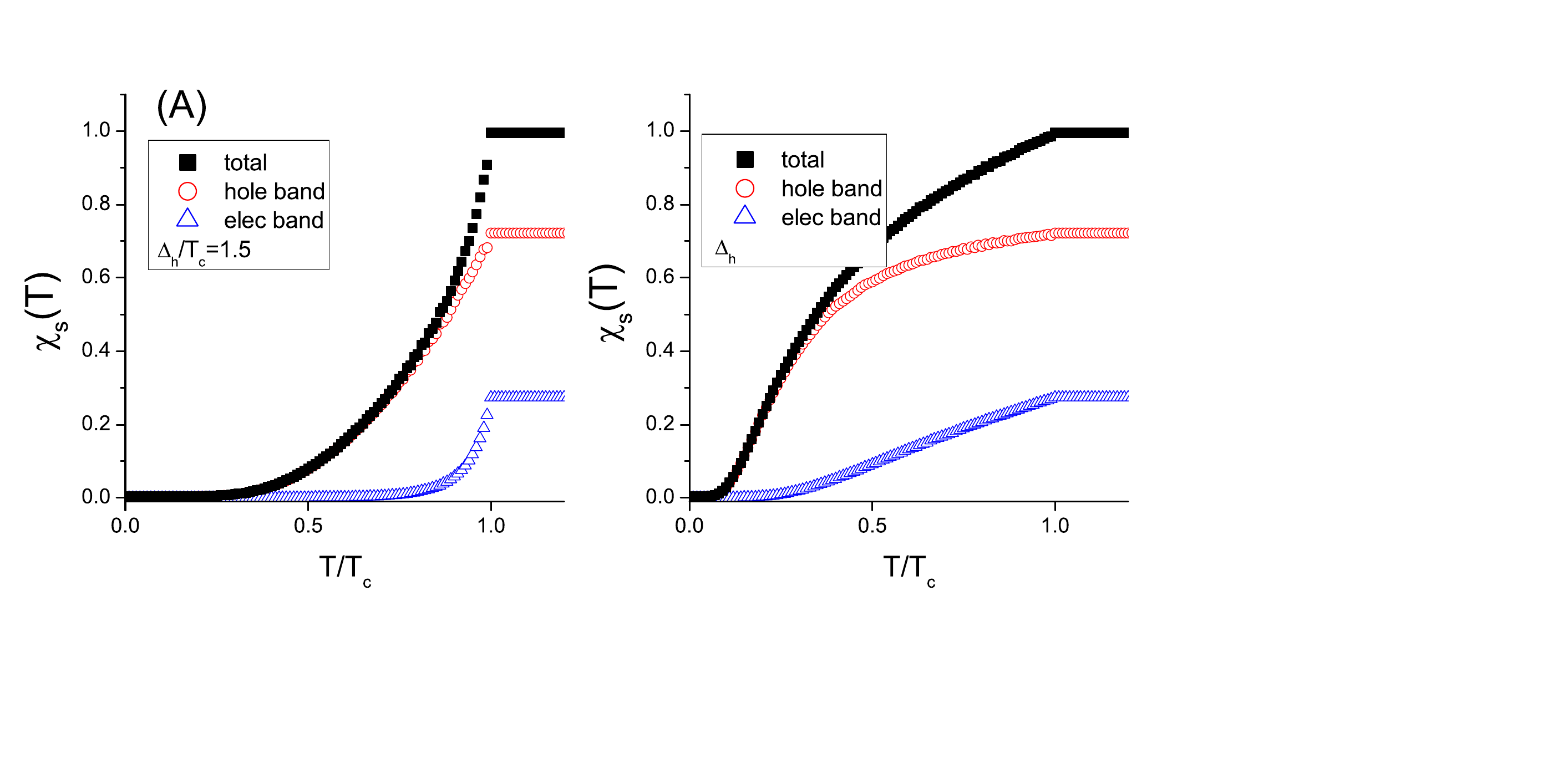}
\vspace{-2.5cm}
\caption{(Color online)  Normalized Knight shift (uniform spin susceptibility) of $s^{\pm}$-wave gap. The total (solid black
square), hole band (open red circle), and electron band (open blue triangle) contributions are shown separately.
(A) With $2\Delta_h/ T_c =3$ ($2\Delta_e/ T_c =7.5$), it shows a typical BCS $s$-wave behavior,
and (B) with $2 \Delta_h/ T_c$ =1.0 ($2\Delta_e/ T_c =2.5$), it shows a much slower reduction of Knight shift below $T_c$ because of the smaller gap-to-$T_c$ ratios. From \cite{bang2008possible}
\label{knight_theory}}
\end{figure}

\begin{figure}[h]
\hspace{1cm}
\includegraphics[width=140mm]{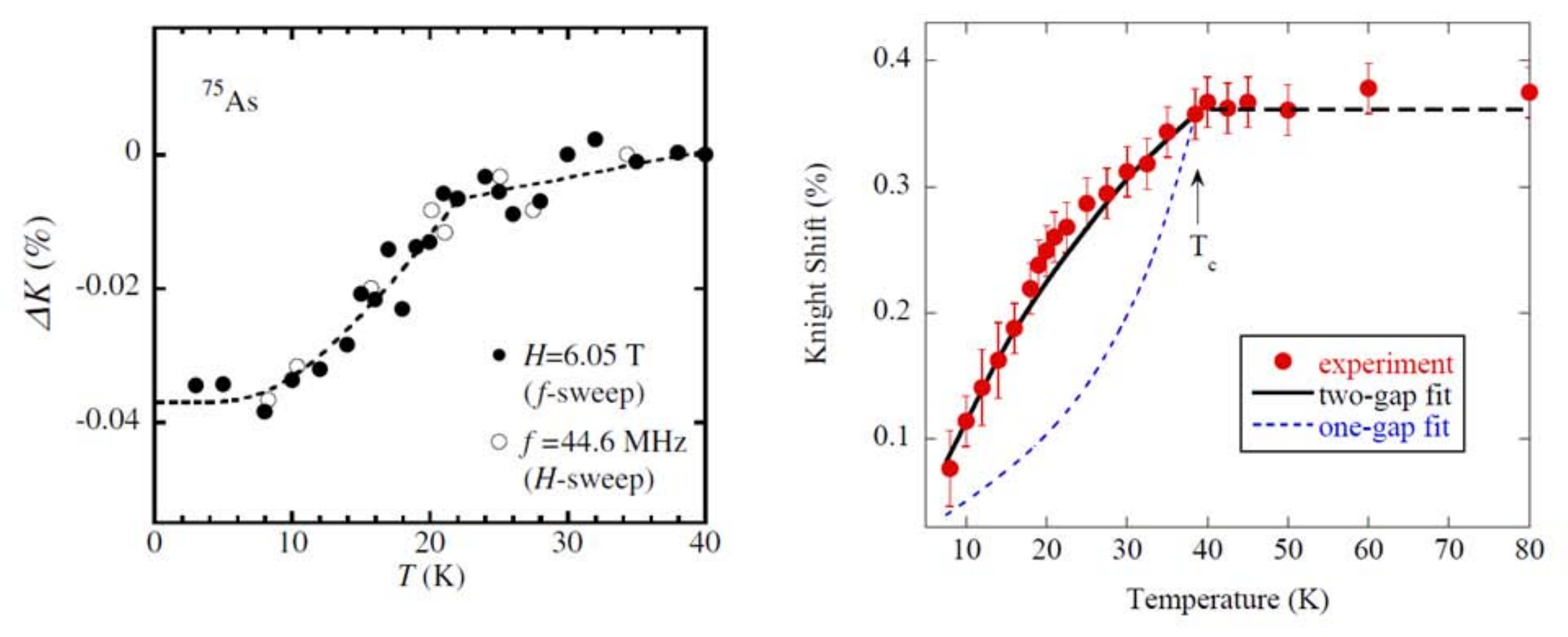}
\vspace{0cm}
\caption{(Color online)  (Left) Experimental data of $^{75}$As Knight shift (circle symbols) of LaFe$_{1-y}$Co$_{y}$AsO \cite{kawabata2008superconductivity}. (Right) Experimental data of $^{75}$As Knight shift (red symbols) of PrO$_{0.89}$F$_{0.11}$FeAs. From \cite{matano2008spin}.
\label{knight_exp}}
\end{figure}

Figure \ref{knight_theory} shows theoretical calculations of the representative cases of $K(T)$ of the $s^{\pm}$-pairing model in clean limit. For demonstration purpose, we chose the hole band as the main band ($N_h = 2 N_e$) and arbitrarily chose the gap-to-$T_c$ ratio of  $2 \Delta_h / T_c$; the other parameters are then automatically determined. The case (A) with $2 \Delta_h / T_c =3.0$ shows a typical BCS behavior: a rapid drop below $T_c$ and the exponentially flat behavior at low temperatures indicating the presence of a full gap due to a $s$-wave pairing. The case (B) with $2 \Delta_h / T_c =1.0$ shows a much slower reduction below $T_c$ because of the smaller gap-to-$T_c$ ratios, but it eventually shows the exponentially flat behavior at very low temperatures indicating a $s$-wave full gap. In both cases, (1) the clear drop of $K(T)$ immediately below $T_c$ indicates a "singlet" pairing superconductor; (2) the exponentially flat behavior at low temperatures ($T \ll T_c$) indicates an $s$-wave (full gap) superconductor. However, as demonstrated in (A) and (B), the convexity (down or up) of $K(T)$ below $T_c$ can be anything due to the two band (or multi band, in general) nature of superconductivity. These genuine behavior of the $s^{\pm}$-wave superconductor and its variations with different FeSCs are well confirmed with experiments shown in Fig.\ref{knight_exp}.

However, these genuine clean limit behaviors should be modified with impurity scattering. As explained in section 3, the $s^{\pm}$-wave state easily -- almost intrinsically -- creates in-gap states with non-magnetic impurities which modifies the typical full-gap ("U"-shape) DOS into the "V"-shape DOS. As a result, Knight shift, probing the low energy DOS $N(\omega)$, the "$s$"-wave pairing evidence of the exponentially flat behavior in $K(T)$ at low temperatures should disappear with impurities. This will be discussed in next subsection.

\subsubsection{With Impurities}
In section 3, we explained that the impurity selfenergies $\Sigma_{imp}^{0,1}(\omega)$ can form resonance states inside the SC gap (see Fig.8(A)) in the $s^{\pm}$-wave state with non-magnetic impurities. Once $\Sigma_{imp}^{0,1}(\omega)$ are calculated, we include these impurity self energy corrections into the Knight shift formula Eq.(\ref{knight}) as $\omega \rightarrow \tilde{\omega}=\omega+\Sigma_h^0(\omega)+\Sigma_e^0(\omega)$ and $\Delta_{h,e} \rightarrow \tilde{\Delta}_{h,e}=\Delta_{h,e}+\Sigma_h^1(\omega)+\Sigma_e^1(\omega)$ (Eqs.(13)-(16)). The results are basically the renormalization of DOS $N_{tot}(\omega)$ as shown in Fig.8(B), and Knight shift $K(T)$ will probe this renormalized DOS $N_{tot}(\omega)$.

\begin{figure}
\noindent
\hspace{2cm}
\includegraphics[width=150mm]{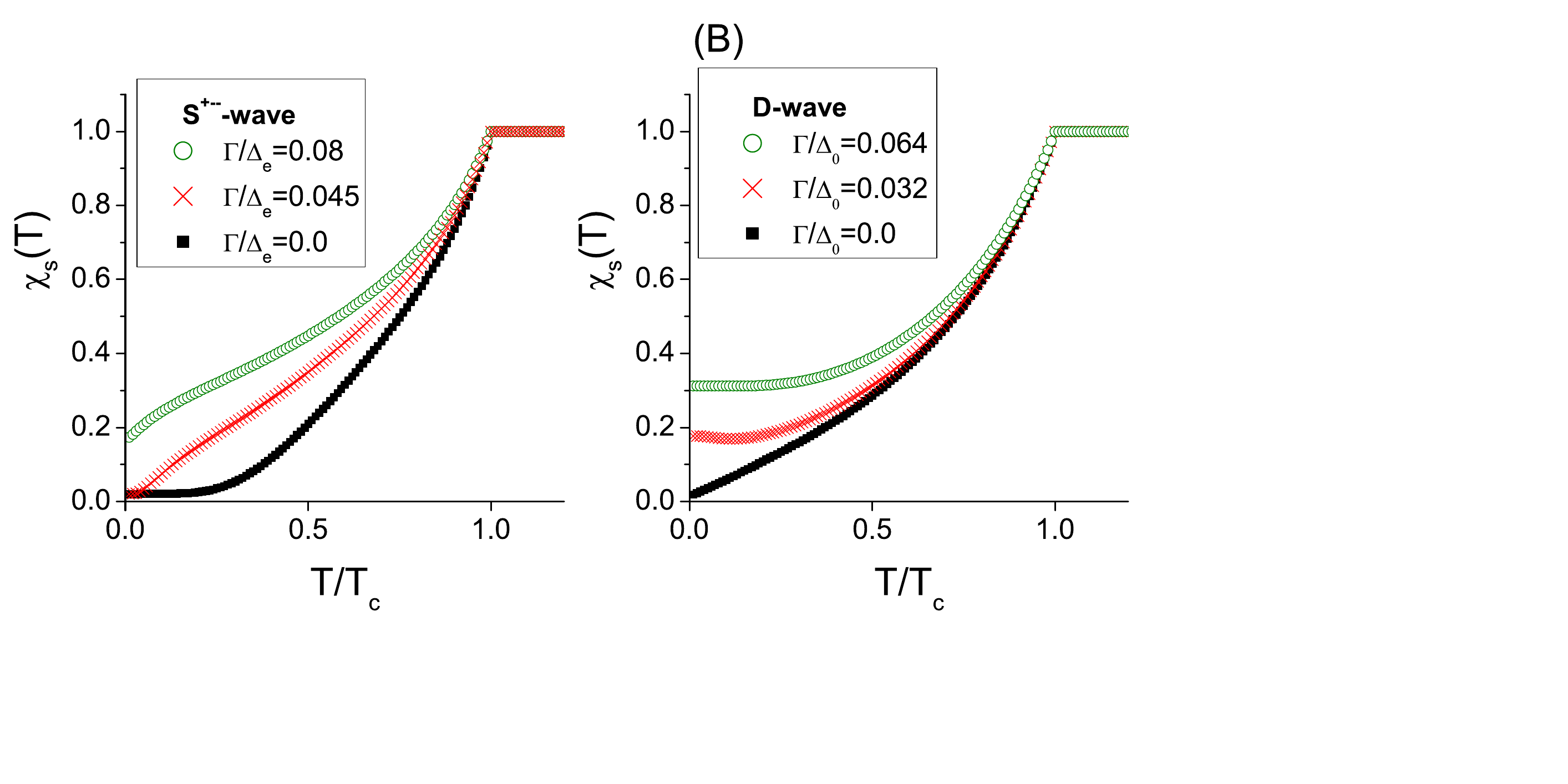}
\vspace{-2cm}
\caption{(Color online) Normalized Knight shift (uniform spin
susceptibility) with non-magnetic impurity scattering. (A) $s^{\pm}$-wave case for $\Gamma/\Delta_e=$0.0, 0.045, and 0.08, respectively, with $2 \Delta_{h}/T_c =3.0$ and $|\Delta_{e} / \Delta_{h}|=2.5$.
(B) $d$-wave case for $\Gamma/\Delta_{0}= 0.0, 0.032,$ and 0.064, respectively, with $2 \Delta_{0}/T_c =5$.
\label{knight_theor_imp}}
\end{figure}

Figure \ref{knight_theor_imp}(A) shows calculation results of $K(T)$ for the same model as in Fig.\ref{knight_theory}(A), but now including impurity scattering. In clean case ($\Gamma_{imp}=0.0$), $K(T)$ shows the typical $s$-wave Knight shift behavior, i.e. the exponentially flat at low temperatures. But with impurity scattering rate $\Gamma_{imp}=0.045 \Delta_e$, the low temperature part of $K(T)$ changes to the $T$-linear behavior just like a clean $d$-wave superconductor. This is because of the "V"-shape DOS at the critical impurity concentration $n_{imp}^{crit}$ (see Fig.6(B) and Fig.8(B)). With a higher impurity concentration, $\Gamma_{imp}=0.08 \Delta_e$,
$K(T)$ still continues to show the $T$-linear behavior but now on top of a constant shift $K_0$. These behaviors are contrasted with the Knight shift  $K_{d-wave}(T)$ in the $d$-wave (or any line-nodal) superconductor shown in Fig.\ref{knight_theor_imp}(B). There, $K_{d-wave}(T)$ in the clean $d$-wave case shows the $T$-linear behavior as expected. But with impurities (non-magnetic, unitary scatterer), the $K_{d-wave}(T)$ becomes flat at low temperatures similar as in the clean $s$-wave superconductor: however, the important difference is the constant part $K_0$. Therefore, the interpretation of Knight shift data $K(T)$ to identify the gap symmetry should not be judged only by the temperature dependence; the determination of the constant part $K_0 = K(T \rightarrow 0)$ at low temperatures is essential before analyzing the temperature dependence.

\begin{figure}[h]
\noindent
\hspace{3cm}
\includegraphics[width=70mm]{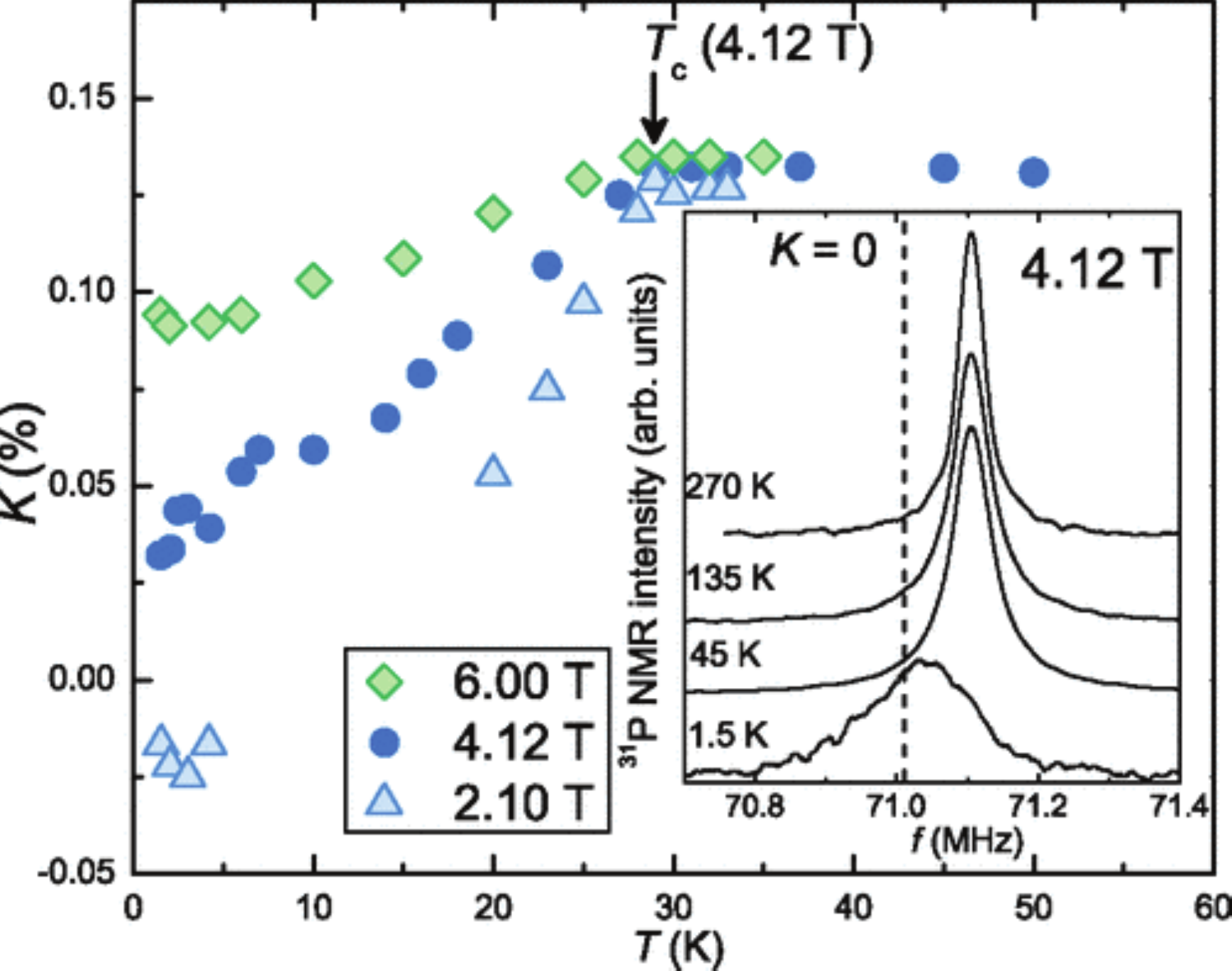}
\vspace{0cm}
\caption{(Color online) Experimental data of $^{31}$P Knight shift of BaFe$_2$(As$_{0.67}$P$_{0.33}$)$_2$ \cite{nakai2010p}.
\label{knight_nakai}}
\end{figure}

Figure \ref{knight_nakai} shows the $^{31}$P Knight shift of BaFe$_2$(As$_{0.67}$P$_{0.33}$)$_2$ \cite{nakai2010p} which shows the $T$-linear behavior at low temperatures. Judging from the temperature dependence of this Knight shift data itself, whether the SC state of BaFe$_2$(As$_{0.67}$P$_{0.33}$)$_2$ compound is a clean nodal gap superconductor or a dirty $s^{\pm}$-wave superconductor cannot be determined for certain. We need to cross check with other experimental probes of the SC properties to determine the most consistent pairing state. Incidently, the authors of \cite{nakai2010p} showed in the same paper that the BaFe$_2$(As$_{0.67}$P$_{0.33}$)$_2$ compound has a substantial amount of residual DOS $N_{res} / N_0 =0.34$ from the measurement of $1/T_1$ spin-lattice relaxation rate. Having this much of the residual DOS in a nodal gap (e.g. $d$-wave) superconductor, the low temperature part of Knight shift $K(T)$ should be flat up to at least $1/3$ of $T_c$ as demonstrated in Fig\ref{knight_theor_imp}(B). Therefore, judging from the combined data of Knight shift and $1/T_1$ spin-lattice relaxation rate, more consistent SC gap state of BaFe$_2$(As$_{0.67}$P$_{0.33}$)$_2$ should be a dirty $s^{\pm}$-wave state, rather than a clean nodal gap (or $d$-wave) state. However, in order to really pin down the correct pairing gap, it is always better to analyze more data of SC properties from various other probes such as penetration depth $\lambda(T)$, thermal conductivity $\kappa(T, H)$, etc. At the moment, the correct SC gap of BaFe$_2$(As$_{1-x}$P$_{x}$)$_2$ is still under debate.

\subsubsection{Summary}
Knight shift $K(T)$ measures the thermal average of the DOS $N(\omega)$ from normal to SC states, therefore its variation with temperature, in particular, at low temperatures, is an excellent probe for the SC gap structure. In the clean limit, it is straightforward to distinguish an $s$-wave full gap (exponentially flat in $T$) and a nodal gap (linear in $T$) superconductors. However, with a tiny amount of impurities in the cases of the $s^{\pm}$-wave state and $d$-wave (or any nodal gap) state, their typical temperature dependencies of $K(T)$ are exchanged with each other: the $s^{\pm}$-wave gap (linear in $T$) and a nodal gap  (exponentially flat in $T$) superconductors. Therefore, it is important to first determine whether the sample is in clean limit or in dirty limit before analyzing the temperature dependence of $\delta K(T)$. Here this definition of the dirty limit is not the same as the standard definition like $l_{mfp} \approx \xi_{coh}$ or $\Gamma_{imp} \approx ( T_c, \Delta_0 )$. In fact, Fig.\ref{knight_theor_imp} show that the impurity scattering rate as tiny as $\Gamma_{imp}/\Delta_0 \approx0.05$ is sufficient to see this dramatic change of the impurity effect on $K(T)$.  Most of the Knight shift $K(T)$ experimental data of FeSCs up to now appear consistent with the $s^{\pm}$-wave SC state.

\subsection{Spin-lattice relaxation rate: $1/T_1$ }

\subsubsection{Clean limit}
$T_1$ relaxation time is the longitudinal relaxation time of the nuclear spin returning back to the equilibrium direction after flipped to the $90$ degree rotated direction by a pulsed field. The relaxation process needs the angular momentum and energy dissipation to the surrounding environment of the nucleus.  The main source of the dissipation in metal is conduction electrons in contact with the each nucleus through a hyperfine coupling, therefore it is a local probe (interaction) and can detect the change of the DOS of the conduction bands from above to below $T_c$. Theoretically it is written as
\begin{eqnarray}
\frac{1}{T_1}  &\sim& \lim_{\omega_0 \rightarrow 0} \sum_q A_{hf}^2(q) \frac{Im \chi_S(q, \omega_0)}{\omega_0}  \\
&=&  - T \int_0 ^{\infty} d\omega
\frac{\partial f_{FD} (\omega)}{\partial \omega} \Biggl \lbrace
\Biggl[ N_h ^2(0)\left\langle Re
\frac{\omega}{\sqrt{\omega^2-\Delta_h ^2(k)}}
\right\rangle_{k}^2
\nonumber \\   &+& 2 N_h (0) N_e (0) \left\langle Re
\frac{\omega}{\sqrt{\omega^2-\Delta_h ^2(k)}}
\right\rangle_{k} \left\langle Re
\frac{\omega}{\sqrt{\omega^2-\Delta_e ^2(k^{'})}}
\right\rangle_{k^{'}} \Biggr] \nonumber \\
&+& \Biggl[
 N_h ^2(0)\left\langle Re
   \frac{\Delta_h(k)}{\sqrt{\omega^2-\Delta_h ^2(k)}}
  \right\rangle_{k}^2
  \nonumber \\   &+& 2 N_h (0) N_e (0) \left\langle Re
     \frac{\Delta_h(k)}{\sqrt{\omega^2-\Delta_h ^2(k)}}
    \right\rangle_{k} \left\langle Re
     \frac{\Delta_e(k^{'})}{\sqrt{\omega^2-\Delta_e ^2(k^{'})}}
    \right\rangle_{k^{'}}
 \nonumber \\   &+&  N_e ^2(0) \left\langle Re
     \frac{\Delta_e(k)}{\sqrt{\omega^2-\Delta_e ^2(k)}}
    \right\rangle_{k}^2
\Biggr]\Biggr \rbrace.
\label{1overT1}
\end{eqnarray}
where $\omega_0$ is the NMR resonance frequency and can be taken to be zero since its energy scale is much smaller than the SC gap energy as $\hbar \omega_0 \ll \Delta_{sc}$, and $A_{hf}(q)$ is a hyperfine coupling between the nuclear moment and the surrounding conduction electrons.

A key difference from the Knight shift is that while Knight shift measures the real part of uniform spin susceptibility $Re \chi_S(q=0, \omega)$, the $1/T_1$ spin-lattice relaxation rate is a local probe, hence momenta  $k$ and $k^{'}$ of each band are independently summed. This momentum integration over whole BZ leads to several important results. First, it allows the inter-band scattering process as $... N_h(0) N_e(0)  \left\langle \frac{\Delta_h}{ ... } \right\rangle_k \left\langle \frac{\Delta_e}{ ... }  \right\rangle_{k'}$ as in the above Eq.(\ref{1overT1}) which leads to a destructive coherence factor for $1/T_1$ in the $s^{\pm}$-pairing state. In ordinary $s$-wave superconductor, $1/T_1$ has the constructive coherence factor, $\sim (1 + \Delta_k \Delta_{k'})$, which produces a coherence peak (also called Hebel-Slichter peak) in $1/T_1(T)$ just below $T_c$, because $\Delta_{sc}^2(T)$ rapidly grows below $T_c$. But, the $1/T_1(T)$ expression in Eq.(\ref{1overT1}) has mixed coherence terms like $(1 + \Delta_{h(e)} \Delta_{h(e)})$, $(1 + \Delta_h \Delta_e)$, etc, where the first term $(1 + \Delta_{h(e)} \Delta_{h(e)})$ is a usual constructive (hence induce a peak structure) coherence factor, but the second term $(1 + \Delta_h \Delta_e)$ becomes a destructive coherence factor because of the opposite signs of $\Delta_h$ and $\Delta_e$ (hence induces a dip structure instead of a peak). As a result, we can expect that the Hebel-Slichter peak of the ordinary  $s$-wave superconductors will be largely suppressed in the $s^{\pm}$-wave SC state.
How much it is suppressed depends on the material specific parameters of $N_{h,e}$ and $\Delta_{h,e}(T)$. The numerical calculations found that this Hebel-Slichter peak in the $s^{\pm}$-pairing state is almost but not completely suppressed in clean limit; however only a small amount of impurities is sufficient to completely erase this peak.

\begin{figure}[t]
\hspace{3cm}
\includegraphics[width=120mm]{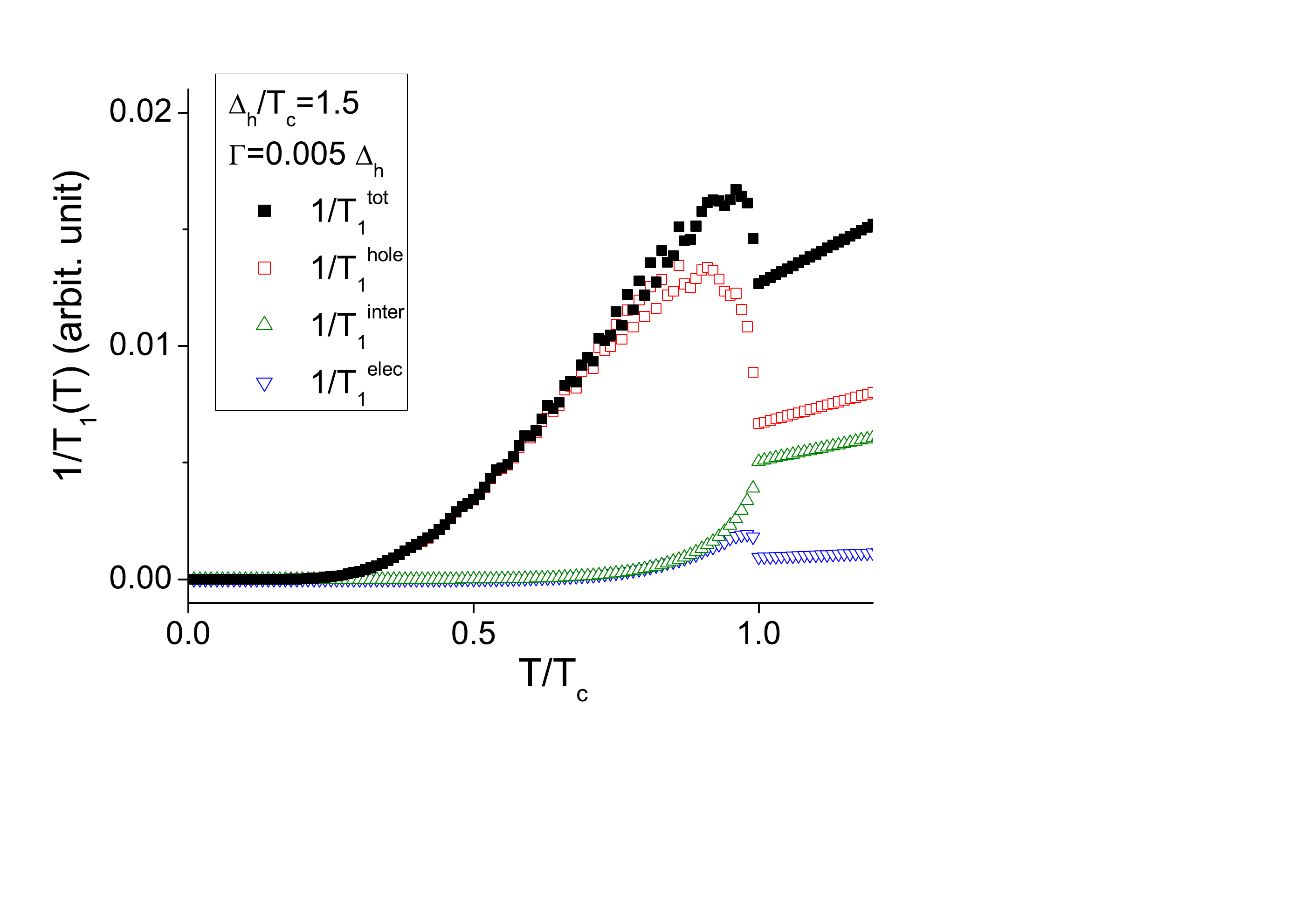}
\vspace{-2cm}
\caption{(Color online) Numerical results of $1/T_1(T)$ of the $s^{\pm}$-wave state in clean limit with $2\Delta_h / T_c$ =3.0  (with $\Delta_e / \Delta_h$ = $N_h /N_e =2.5$). (A) Displayed are the contributions of the separate terms in Eq.(\ref{1overT1}) for $1/T_1(T)$: total (solid black square), hole band (open red square),
electron band (open blue triangle), and interband term (open green triangles). The total $1/T_1(T)$ shows the Hebel-Slichter peak below $T_c$ which is much reduced compared to the BCS value. From \cite{bang2008possible}.  \label{T1_clean}}
\end{figure}

Figure \ref{T1_clean} shows a representative theoretical calculations of $1/T_1$ in clean limit of the $s^{\pm}$-pairing state with $2\Delta_h/T_c = 3.0$. It displays the separate contributions of each term in the two band model: two intra-band terms (hole and electron bands, respectively), and one inter-band term. Two intra-band contributions (red squares and blue inverted triangles) to $1/T_1$ show the typical Hebel-Slichter peaks, respectively (their jump sizes are comparable to their normal state $1/T_1(T_c)$ at $T_c$). However, the interband contribution (green triangles) shows a dip instead of a peak. As a result, the total $1/T_1$ shows a much reduced Hebel-Slichter peak, but still with a visible size.
Compared to this theoretical prediction of $1/T_1$ in clean limit of the $s^{\pm}$-pairing state, in early days, several NMR $1/T_1$ experiments with the Fe-based SC compounds,  in particular, LaOFeAs (so-called 1111) compound, \cite{grafe200875,nakai2008evolution,kawasaki2008two,terasaki2008spin,mukuda2009doping}
have reported common peculiar features:  (1) no Hebel-Slichter peak, and (2) $1/T_1 \sim T^3$ over all measured temperatures below $T_c$. These features were surprising and it was immediately taken as strong evidences for a nodal gap state, like a $d$-wave state, in Fe-based SC materials. However, this $d$-wave or a nodal gap superconductor claim was in contradiction with the other experiments (e.g. ARPES experiments \cite{ding2008}) which indicated an isotropic  $s$-wave gap. In this context, it was a challenging task to explain the $1/T_1$ experiments with the $s^{\pm}$-wave model.
The results of Fig.\ref{T1_clean} demonstrated that the $s^{\pm}$-wave SC state in clean limit is not quite consistent with these
$1/T_1$ experiments of early days although the Hebel-Slichter peak is strongly reduced by the sign-changing OPs.
In particular, there is no intrinsic mechanism to explain the $T^3$ dependence with the $s^{\pm}$-wave SC state having no nodes. In an effort to improve the reduction of Hebel-Slichter peak as well as the power law, it was attempted to add the impurity damping by hand\cite{bang2008possible}.

\begin{figure}
\hspace{1.0cm}
\includegraphics[width=160mm]{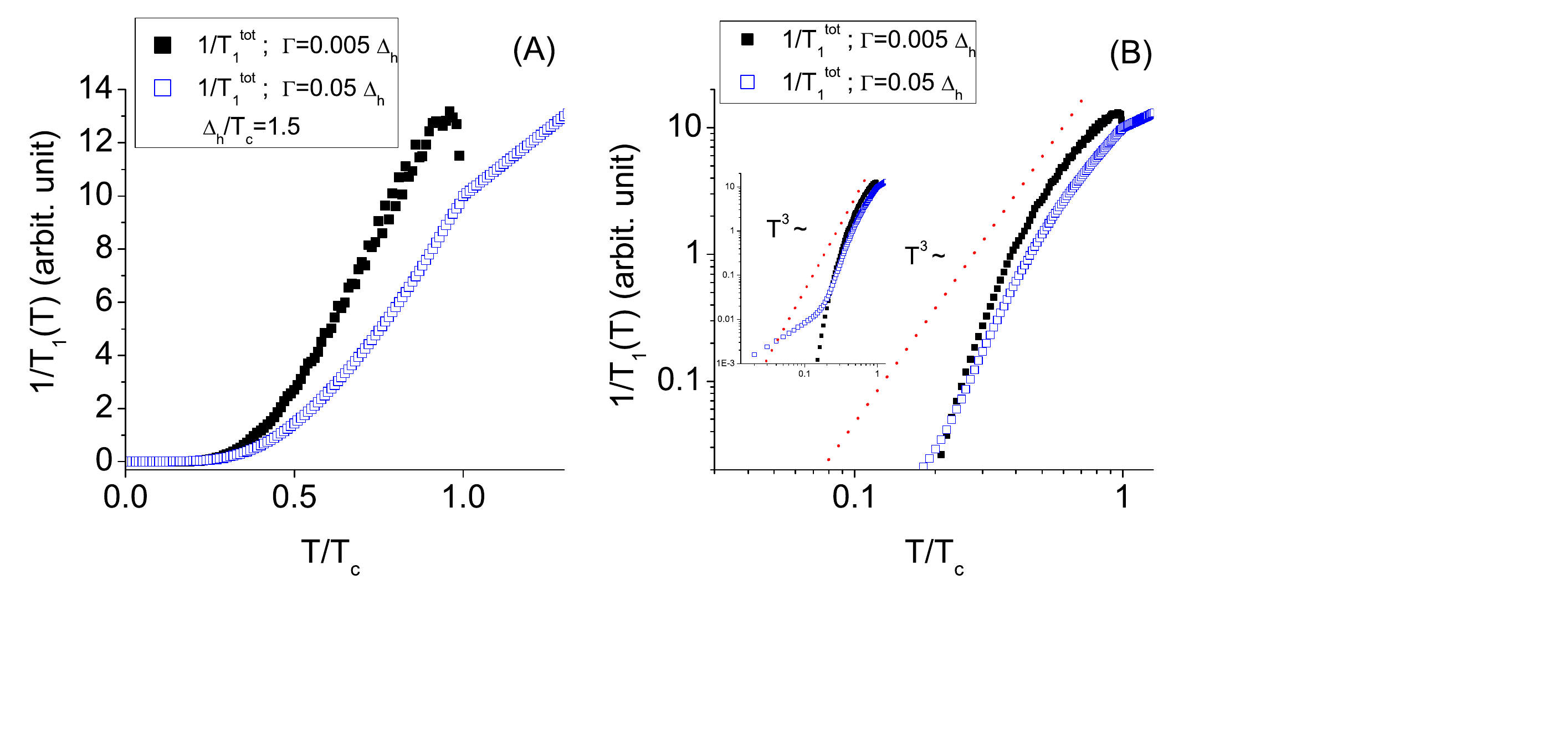}
\vspace{-2.5cm}
\caption{(Color online) Numerical results of $1/T_1$(T) of the $s^{\pm}$-wave gap including a constant impurity damping $\Gamma$, with the same parameters as in Fig.\ref{T1_clean}.
(A) Total $1/T_1$(T): in clean limit (solid black square) and with a constant damping (open blue squares).
(B) The log-log plot of the same data in (A). The inset is a wide view. From \cite{bang2008possible}. \label{T1_imp}}
\end{figure}

\begin{figure}
\hspace{0cm}
\includegraphics[width=160mm]{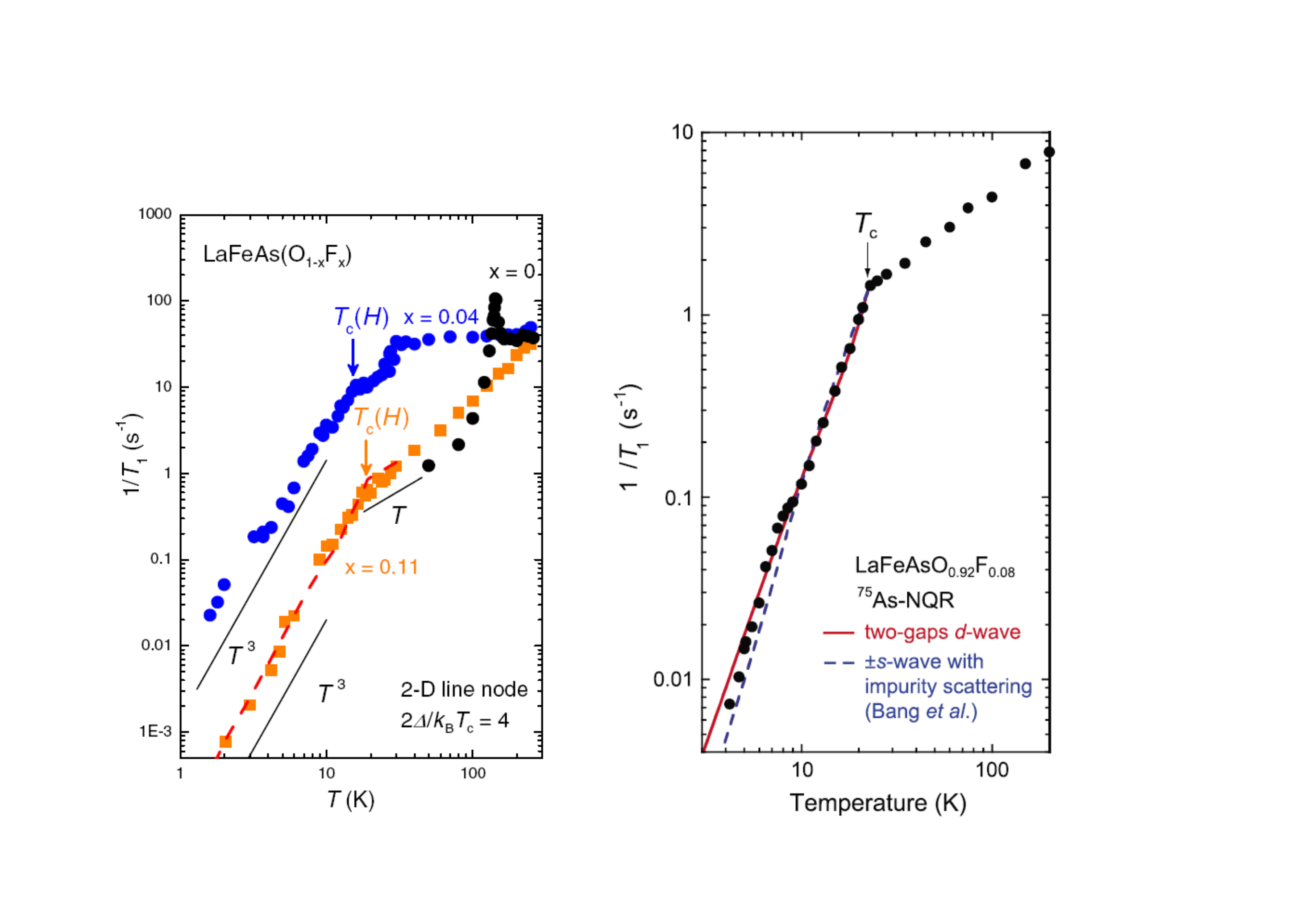}
\vspace{-2cm}
\caption{(Color online) (Left) $1/T_1$ of $^{75}$As of LaFeAs(O$_{1-x}$F$_x$) for $x=0.04$ and $0.11$. Solid lines are $d$-wave fittings. From \cite{nakai2008evolution}. (Right) $1/T_1$ of $^{75}$As of LaFeAsO$_{0.92}$F$_{0.08}$. Solid line is a clean $d$-wave fitting and the dotted line is a fitting with $s^{\pm}$-wave with impurity. From \cite{kawasaki2008two}.
\label{T3_exp}}
\end{figure}

Figure \ref{T1_imp}(A) replot the same calculations of  total  $1/T_1$ in Figure \ref{T1_clean}, but with an artificial constant damping introduced into Eq.(\ref{1overT1}) by $\omega \rightarrow \omega+ i \Gamma$. It shows that a tiny amount of damping, $\Gamma=0.05 \Delta_h$ (blue squares), completely erases the Hebel-Slichter peak of the clean limit result (black solid squares).
In Fig.\ref{T1_imp}(B),  the same data of Fig.\ref{T1_imp}(A) are plotted in log-log plot to examine an overall power law behavior of $1/T_1(T)$ below $T_c$. It shows that the $s^{\pm}$-wave state with a constant damping is only partially successful to fit the early $1/T_1$ experiments: no Hebel-Slichter peak and only an approximate power law of $1/T_1 \sim T^{3}$.
However, as shown in Fig.\ref{T3_exp}, almost all early $1/T_1(T)$ data of Fe-based SC compounds \cite{grafe200875,nakai2008evolution,kawasaki2008two,terasaki2008spin,mukuda2009doping} were not just approximately but almost perfectly $\sim T^3$ down to measured lowest temperatures while the results in Fig.\ref{T1_imp}(B) of the $s^{\pm}$-wave gap model is far from this $T^3$ behavior. To resolve this discrepancy between experimental $1/T_{1,exp}(T)$ and the $s^{\pm}$-wave gap model, we need to study the impurity effect more seriously\cite{chubukov2008magnetism,bang2009imp,parker2008extended}.

\begin{figure}
\hspace{1cm}
\includegraphics[width=60mm]{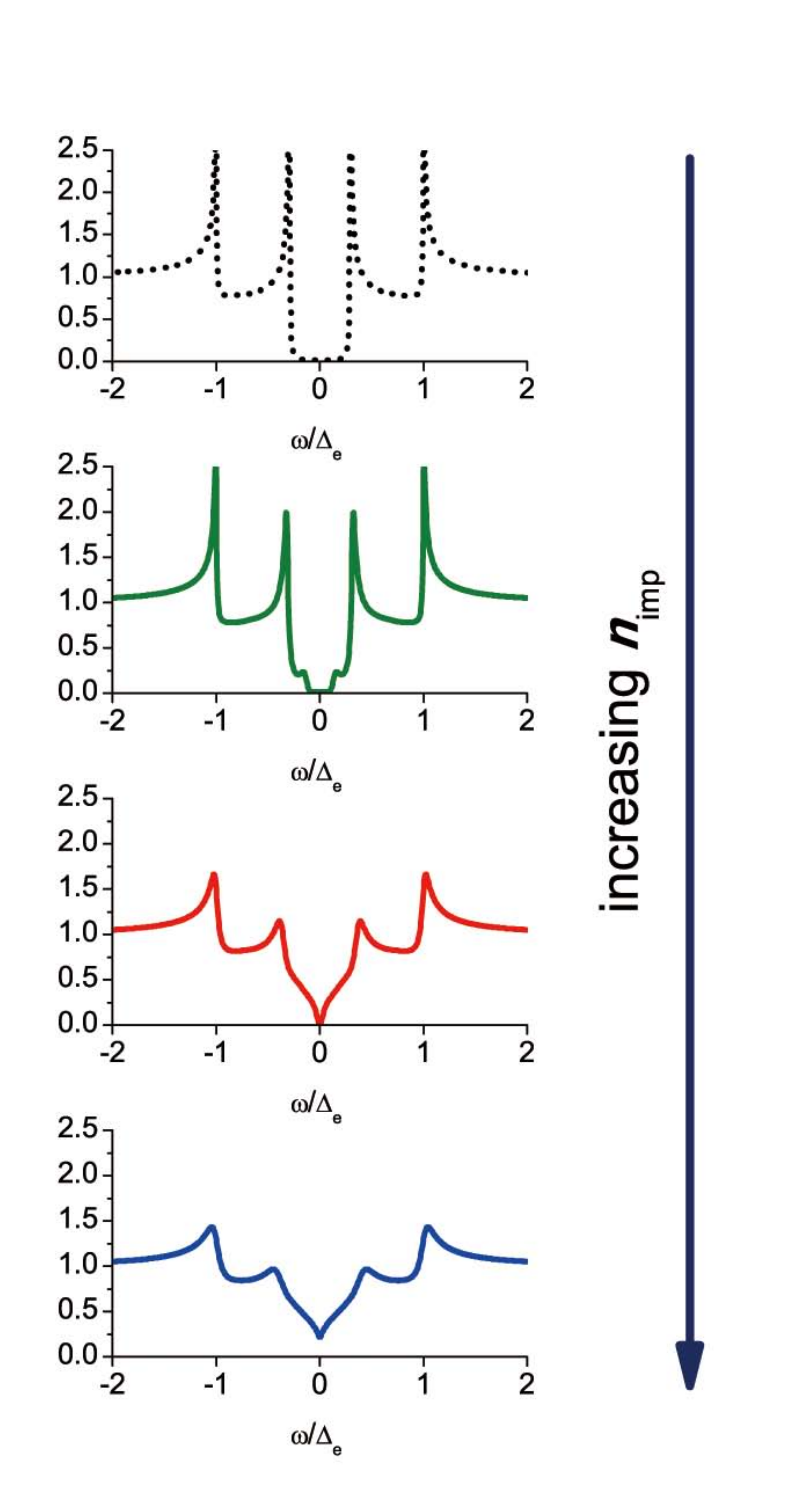}
\includegraphics[width=90mm]{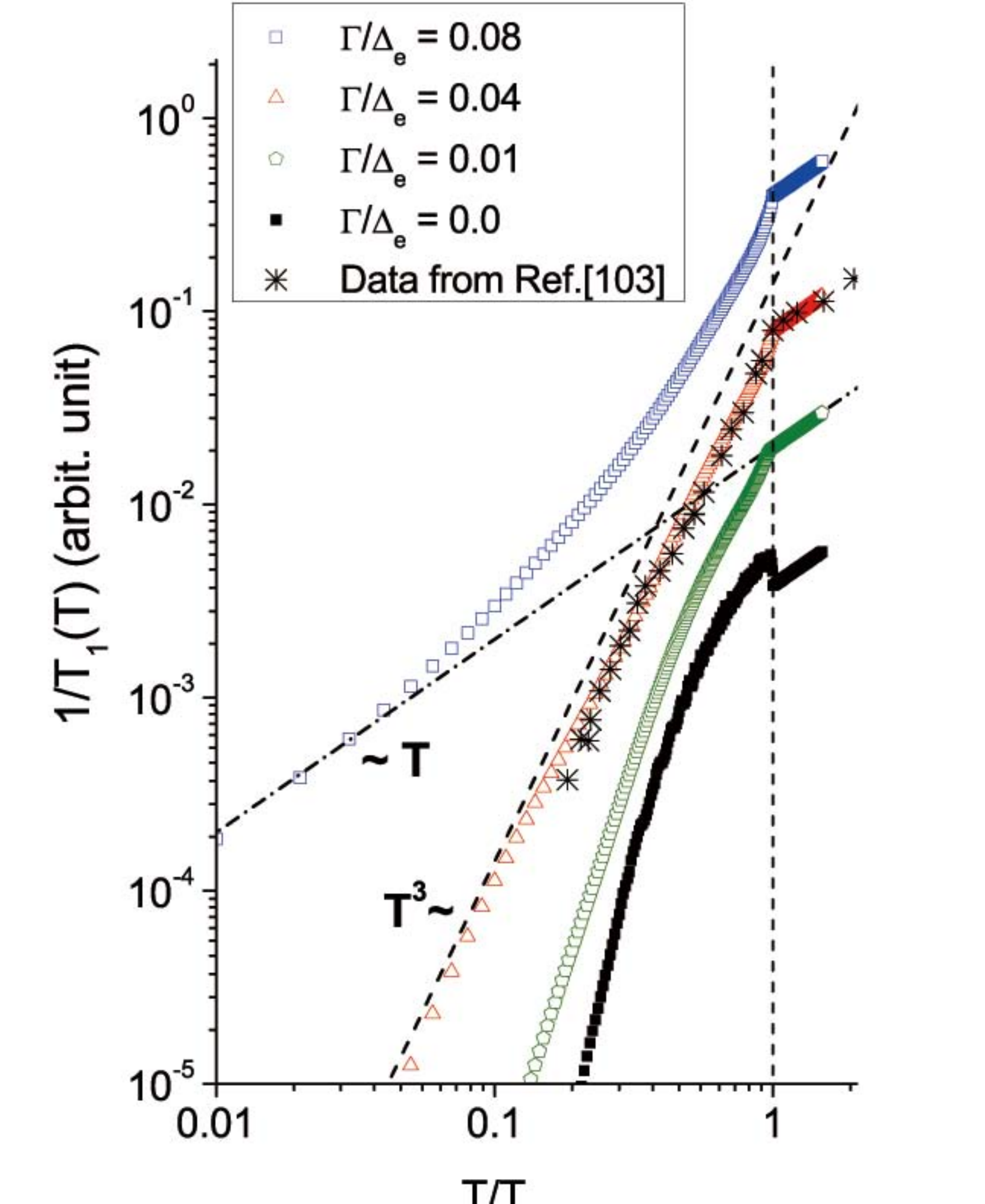}
\caption{(Color online)  (Right) Theoretical calculations of $1/T_1$(T) for $s^{\pm}$-wave SC state, with different impurity concentrations, $\Gamma / \Delta_e= 0.0, 0.01, 0.04, 0.08$, with $2 \Delta_h / T_c=3.0$ (with $\Delta_e / \Delta_h$ = $N_h /N_e =2.5$, hence $2 \Delta_e / T_c=7.5)$. Experimental data is from Ref.\cite{kawasaki2008two}. The curves are offset for clarity.  (Left) The corresponding evolution of DOS $N(\omega)$ with the corresponding impurity concentrations.  From \cite{bang2009imp}.
\label{T1_theo_imp}}
\end{figure}

\subsubsection{With impurities}
In order to include the impurity scattering effect in the spin lattice relaxation rate $1/T_1$, we use the same formula of $1/T_1$ of Eq.(\ref{1overT1}) but with renormalizing $\omega$ and $\Delta_{h,e}$ by  impurity selfenergies as $\tilde{\omega}$ and $\tilde{\Delta_{h,e}}$, which are calculated with Eqs.(13)--(16), respectively, using $\mathcal{T}$-matrix theory. We considered only non-magnetic impurities. The main effects of impurity scattering in the $s^{\pm}$-wave SC state is to create the in-gap states inside the gap energy as shown in Fig.(6) and Fig.(8), which directly affect $1/T_1(T)$ according to Eq.(\ref{1overT1}).

Figure \ref{T1_theo_imp} shows the calculation results of $1/T_1(T)$ with unitary impurities of concentrations:
$\Gamma/\Delta_e=0.0, 0.01, 0.04$, and $0.08$, respectively. Lefthand panel shows the systematic evolution of the DOS $N_{tot}(\omega)$ in the two band $s^{\pm}$-wave model due to the impurity bound states formed inside the gaps.
First, in the clean limit $\Gamma/\Delta_e=0.0$, the theoretical calculation of $1/T_1(T)$ (black squares) shows the $s$-wave features: Hebel-Slichter peak (although much reduced due to the sign-changing OPs) and the exponentially rapid drop at low temperatures. With a small increase of impurity density $\Gamma/\Delta_e=0.01$, the "U"-shape gap in the DOS $N(\omega)$ is reduced but not yet completely closed. The corresponding result of $1/T_1(T)$ (green pentagons) still shows some feature of $s$-wave superconductor such as a rapid drop of $1/T_1(T)$ at low temperatures, but Hebel-Slichter peak is completely wiped out.
At the critical impurity concentration $\Gamma/\Delta_e=0.04$, $N_{tot}(\omega)$ shows the "V"-shape DOS as in a clean $d$-wave superconductor. Accordingly, at this impurity concentration, the theoretical calculation of $1/T_1(T)$ displays the $T^3$ behavior (red triangles) over the entire temperature region. This $T^3$ behavior at low temperatures has the same origin as in the $d$-wave superconductor, i.e., the linearly rising DOS. However, the $T^3$ power behavior near $T_c$ down to roughly $T_c /3$ is in fact not the intrinsic property of the low energy DOS but controlled by the gap-to-$T_c$ ratio $R=2\Delta/T_c$. In Fig.\ref{T1_theo_imp}, the value $R= 2 \Delta_h / T_c=3.0$ (hence, $2 \Delta_e / T_c=7.5)$ were chosen to create the $T^3$ behavior over entire temperatures below $T_c$. By choosing larger or smaller values of $R$, the temperature dependence of  $1/T_1(T)$ near $T_c$ can be made steeper or slower. However, the low temperature part of  $1/T_1(T)$ is solely determined by the intrinsic property of the low energy DOS. For example, with higher impurity concentration $\Gamma/\Delta_e=0.08$ in the same model, $N_{tot}(\omega)$ shows "V"-shape DOS on top of a constant DOS $N_0$ (the bottom figure in the lefthand panel of Fig.\ref{T1_theo_imp}). Because of this finite DOS $N_0$,  $1/T_1(T)$ shows the $T$-linear behavior (blue squares) at low temperatures and the $T^3$ dependence near $T_c$ is due to the chosen value of $R$ used in the all calculations in Fig.\ref{T1_theo_imp}. Interestingly, the feature of the "V"-shape DOS doesn't show up its presence in $1/T_1(T)$ in this case because the constant DOS $N_0$ governs the low temperature behavior of $1/T_1(T)$.

It is clear that the puzzling $T^3$ behavior\cite{grafe200875,nakai2008evolution,kawasaki2008two,terasaki2008spin,mukuda2009doping} of $1/T_1$ in the Fe-based SC 1111-compounds can be understood by the $s^{\pm}$-pairing model with unitary impurities and it has the same origin as in the d-wave superconductor, i.e., the linearly rising DOS; however in the former case the $V$-shape DOS was dynamically created, but in the latter case it was formed by kinematic origin. We also emphasize that in order to capture this systematic evolution of $1/T_1$ with sample quality, it is absolutely necessary to include the non-trivial impurity scattering effects in the $s^{\pm}$-wave state.
Also notice that this wide range of variation in  $1/T_1$ can occur with the small variation of impurity concentration $0 < \Gamma/\Delta_e < 0.08$. For example, the reduction of
$T_c$, $\delta T_c / T_c ^0$, with this amount of impurity variation, which is proportional to $(\Gamma/\Delta_e)/[c^2+1]$ ($c=0$, for unitary impurity)\cite{bang2009imp}, is less than $10 \%$ reduction of $T_c ^0$ at most.

\subsubsection{$T^{5-6}$-power in $1/T_1(T)$.}

\begin{figure}
\hspace{0cm}
\includegraphics[width=150mm]{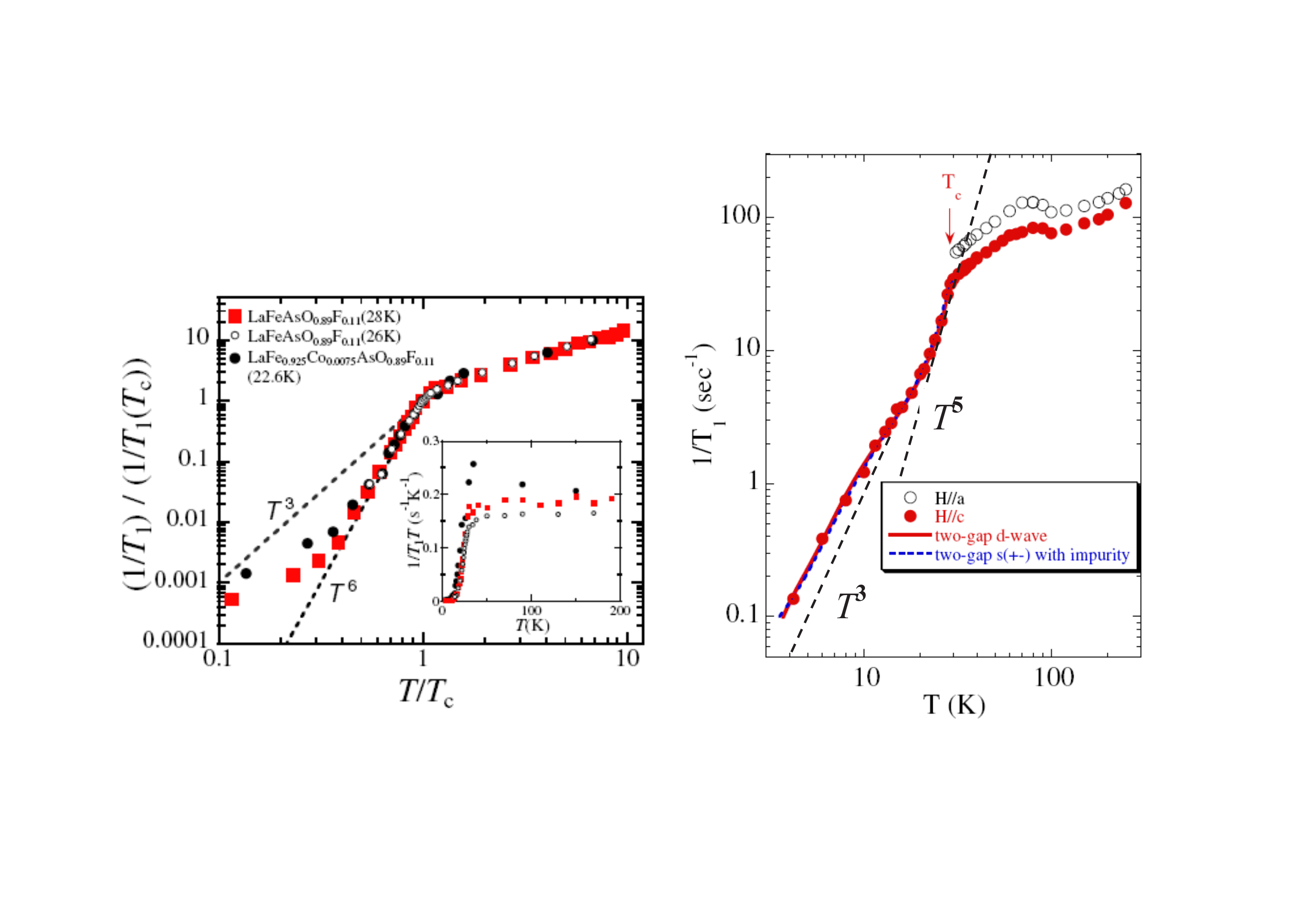}
\vspace{-2cm}
\caption{(Color online) (Left) $^{75}(1/T_1)$ data of LaFe$_{1-y}$Co$_y$AsO$_{1-x}$F$_x$ $(x=0.11)$. From \cite{sato2009studies}. (Right) $^{75}(1/T_1)$ data of Ba$_{0.72}$K$_{0.28}$Fe$_2$As$_2$.  Dashed lines are $\sim T^3$ and $\sim T^5$. From \cite{matano2009anisotropic}.
\label{T1_exp_power-to-5}}
\end{figure}

After the $T^3$-behavior in $1/T_1$ was explained with the impurity states in the $s^{\pm}$-wave state, several NMR experiments reported that the power law of $1/T_1(T)$ is not always $\sim T^3$, but can be much steeper as $\sim T^{5-6}$\cite{nakai2010p,matano2009anisotropic,sato2009studies,yashima2009strong,hammerath2010unusual} (see Fig.\ref{T1_exp_power-to-5}) and sometimes shows a step-like structure (Right panel in Fig.\ref{T1_exp_power-to-5}) in the middle of in between $T_c$ and $T=0$.
And some authors claimed that this is the evidence that the $s^{\pm}$-wave state is not the right pairing symmetry for the FeSCs.
However, notice that this steeper power law $T^{5-6}$ of $1/T_1$ observed in some of Fe-based SC compounds (e.g. La-1111\cite{matano2009anisotropic,sato2009studies,hammerath2010unusual}, (BaK)Fe$_2$As$_2$ \cite{yashima2009strong}, and BaFe$_2$(As$_{0.67}$P$_{0.33}$)$_2$ \cite{nakai2010p})
always occurs near $T_c$. As we explained above, this near-$T_c$ property has nothing to do with a pairing gap symmetry nor with the low energy DOS, but only reflects the gap-to-$T_c$ ratio $R$, which is a strong coupling effect in general.
Even in a clean $d$-wave superconductor, the genuine $T^3$ power law of $1/T_1$  is obeyed only at low temperatures for $T < T_c /2$, where the DOS $N(\omega) \sim \omega$ governs the thermodynamic properties, and the temperature slope of $1/T_1$ near $T_c$ can be made arbitrarily as steep as $\sim T^{5-6}$ by choosing a larger value of $R=2 \Delta_0/ T_c =8$, for example see Ref.\cite{bang2006nuclear}. Of course, then whether the value $R=8$ is physically plausible or not is another question; compared to the BCS value $R=3.5$, the value $R=8$ implies that this superconductor is a strong coupling superconductor and this value is quite possible with many strongly correlated SC materials.

\begin{figure}
\hspace{2cm}
\includegraphics[width=120mm]{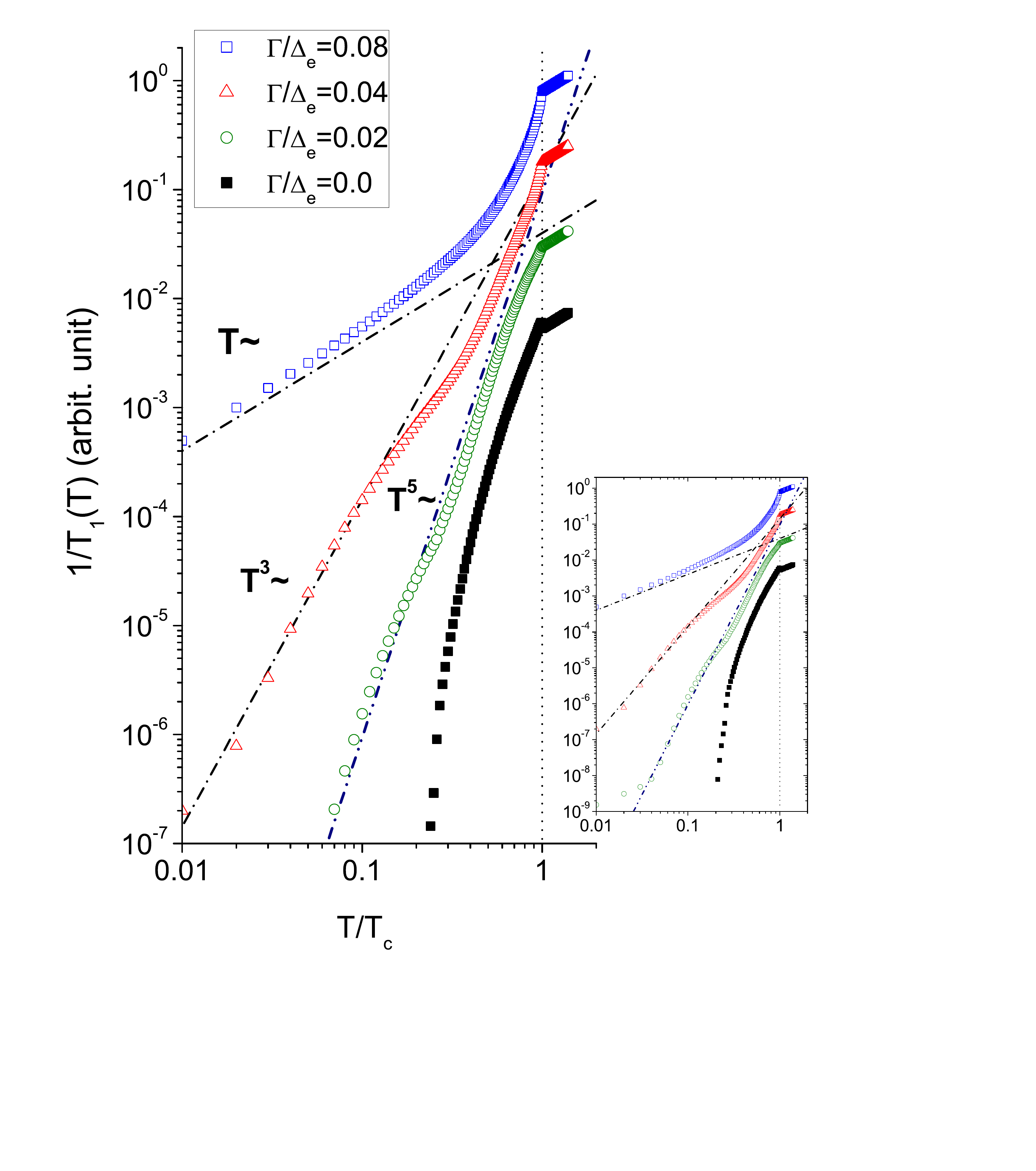}
\vspace{-2.5cm}
\caption{(Color online) The same calculations with the same parameters as in Fig.\ref{T1_theo_imp}, but with the larger gap-to-$T_c$ ratio $2 \Delta_h / T_c=5.0 (2 \Delta_e / T_c=12.5)$. The slope of $1/T_1$(T) near $T_c$ becomes steeper as $\sim T^5$, but the low temperature behaviors are the same as in Fig.\ref{T1_theo_imp}. The inset is a wide view.
\label{T1_theo_p5}}
\end{figure}

Similarly, the slope of $1/T_1(T)$ near $T_c$ of the $s^{\pm}$-wave state can be arbitrarily made steeper by choosing a larger value of $R_{h,e}=2 \Delta_{h,e}(T=0)/T_c $.  For illustration, we repeated the same calculations but only with larger $R$ value as $2 \Delta_h / T_c=5.0$ (hence, $2 \Delta_e / T_c=12.5)$. The results, plotted in Fig.\ref{T1_theo_p5}, show the same behaviors as in Fig.\ref{T1_theo_imp} in low temperatures for $T < T_c /3$ but steeper power laws ($\sim T^5$)  near $T_c$. For example, the result with the critical impurity concentration $n_{imp}^{c}$ ($\Gamma/\Delta_e =0.04$, red triangle symbols) show the $\sim T^5$ behavior near $T_c$ but with decreasing temperature, it evolves, after a short crossover, to the perfect $T^3$ behavior. With higher impurity concentration $\Gamma/\Delta_e =0.08$, $1/T_1$ near $T_c$ again shows the $\sim T^5$ behavior, but it quickly goes though a smooth crossover region and eventually becomes $T$-linear at low temperatures because of the finite DOS $N_0$. The most interesting behavior is for $\Gamma/\Delta_e =0.02$, in this case, $1/T_1(T)$ shows $\sim T^5$ over the entire temperature range of calculation and even shows the step-like structure at $\sim 0.2 T_c$; these features are quite similar to the data of Ba$_{0.72}$K$_{0.28}$Fe$_2$As$_2$ shown in Fig.\ref{T1_exp_power-to-5} (Right panel). Perhaps, the choice of the gap-to-$T_c$ ratio of $2 \Delta_h / T_c=5.0 (2 \Delta_e / T_c=12.5)$ used in the calculations in Fig.\ref{T1_theo_p5} might be too large for real Fe-based SC compounds. But this was for the demonstration to show that the slope near $T_c$ can be arbitrarily controlled by choosing only a different $R$ value, otherwise with the exactly same model as in Fig.\ref{T1_theo_imp}. For real Fe-based SC compounds, if we choose
a different ratio $\Delta_e / \Delta_h$ = $N_h /N_e$ (choosing a larger value, for example,  $\Delta_e / \Delta_h =4$), the $T^{5-6}$ behavior near $T_c$ can be easily obtained with a much smaller value of $2 \Delta_e / T_c \approx 6-8$.
Later more NMR experiments have also been performed and some data even detected the presence of a small Hebel-Slichter peak as well as a very rapid drop in $1/T_1(T)$ \cite{oka2012antiferromagnetic}, signatures of an $s$-wave superconductor. These behaviors are in fact quite similar to the plots of $\Gamma \approx 0$ cases in Fig.\ref{T1_theo_imp} and Fig.\ref{T1_theo_p5}. They also found a second bent in $1/T_1(T)$ at lower temperature in between $T_c$ and $T=0$, indicating the presence of multiple gaps with very different sizes $|\Delta_{h}|$ and $|\Delta_{e}|$.

\subsection{Summary}

The intrinsic behavior of $1/T_1(T)$ of the $s^{\pm}$-wave model should be like an $s$-wave superconductor but with a strongly suppressed Hebel-Slichter peak because of the sign-changing OPs  $\Delta_{h}$ and $\Delta_{e}$. The frequently observed $T^3$ power law behavior in $1/T_1(T)$ for many Fe-based SC compounds\cite{grafe200875,nakai2008evolution,kawasaki2008two,terasaki2008spin,mukuda2009doping} can be naturally understood with the $s^{\pm}$-wave model if
the resonant impurity scattering effect is included, which renormalizes the "U"-shape DOS into a "V"-shape DOS as $N(\omega) \sim \omega$ at low frequencies. Later found steeper power $T^{5-6}$ behavior near $T_c$ in $1/T_1(T)$ with some of Fe-based SC compounds \cite{matano2009anisotropic,sato2009studies,hammerath2010unusual,yashima2009strong,nakai2010p} is not an intrinsic property related to the pairing symmetry or the gap function, but a property controlled by the gap-to-$T_c$ ratio, $R$, so that this behavior can be fit with a larger value of $R$ within the $s^{\pm}$-wave state model.
Therefore, we can say that all experimental data of NMR Knight shift $K(T)$ and $1/T_1(T)$ in the FeSCs are consistently explained within the $s^{\pm}$-wave model with impurity scattering included. All early puzzles and challenges posed by NMR experiments actually have turned into strong evidences to support the correctness of the $s^{\pm}$-pairing state for the FeSCs. It is important to notice that the unequal size of the gaps $\Delta_{h,e}$ -- hence the unequal sizes of DOSs $N_{h,e}$ -- is a genuine property of the $s^{\pm}$-pairing state and it is a crucial factor to understand and fit the experimental data. This unequal size of gaps in the $s^{\pm}$-pairing state will repeatedly play a crucial role in understanding other SC properties of FeSCs.

\section{Specific Heat: temperature dependence of $C_{el}(T)$ near $T=0$.}
Specific heat (SH) measures all low energy excitations $E(T)$ or, in other words, the entropy variation $\Delta S(T)$. It is a standard and first experimental probe to confirm the truly bulk SC transition by identifying the specific heat jump at $T_c$. The size of the jump $\Delta C$ is a gauge to measure the SC volume fraction of the sample as well as the strong coupling character: the larger the jump $\Delta C$ is the larger the SC volume fraction is and the stronger the strong coupling character is.
As to probing the pairing symmetry, it is also an old and still excellent probe, if all non-electronic contributions are reliably subtracted. There is always some uncertainty to subtract the phonon part at high temperatures. However, as temperatures goes down all bosonic contributions, including phonons, to the SH are rapidly suppressed and if we are interested in the low temperature part of $C(T)$ near $T=0$, this subtraction of the non-electronic part is not an issue, and the $C_{el}(T)$ part will probe the electronic DOS $N(\omega)$ near $\omega=0$ which should reflect the SC gap structure. One complication though is often the case in the iron based superconductors, namely if there is a magnetic contribution (typically important below about 1-1.5 K).  Special care must be taken in subtracting such a contribution, since it is by nature magnetic field dependent.  See Ref.\cite{kim2014specific}.

\subsection{Clean limit and its evolution with impurities}

The formula for the SH coefficient (also called Sommerfeld coefficient) $C/T = \gamma(T)$ is written as follows
\ba
C(T)/T & = &  \gamma(T)  \nonumber \\
&=& - \int_0 ^{\infty} d \omega \frac{\partial f_{FD} (\omega)}{\partial \omega}
\Bigg(\frac{\omega}{T}\Bigg)^2 \Big[ N_h(\omega,T) + N_e(\omega,T) \Big] \nonumber \\
&&- \int_0 ^{\infty} d \omega S(\omega/T) \frac{d}{dT}\Big[ N_h(\omega,T) + N_e(\omega,T) \Big].
\label{CT_eq}
\ea
\noindent where $S(\omega/T)= \Big[(1- f_{FD}(\omega))\ln(1- f_{FD}(\omega)) +  f_{FD}(\omega) \ln f_{FD}(\omega)\Big]$ is the Fermionic entropy of excitation energy $\omega$.
We see $\gamma(T) $ consists of two parts: (1) the first term dominates at low temperatures near $T=0$, and (2) the second term dominates near $T_c$. In particular, the DOS $N_a(\omega, \Delta_a(T))$ near $T_c$ in the second term is rapidly changing with temperature, causing
the specific heat jump $\Delta C/T$. In order to identify the pairing gap symmetry, the low temperature behavior of $C(T)/T$ is more useful (see discussion below) and the SH jump $\Delta C$ is not as relevant.

\begin{figure}
\hspace{2cm}
\includegraphics[width=150mm]{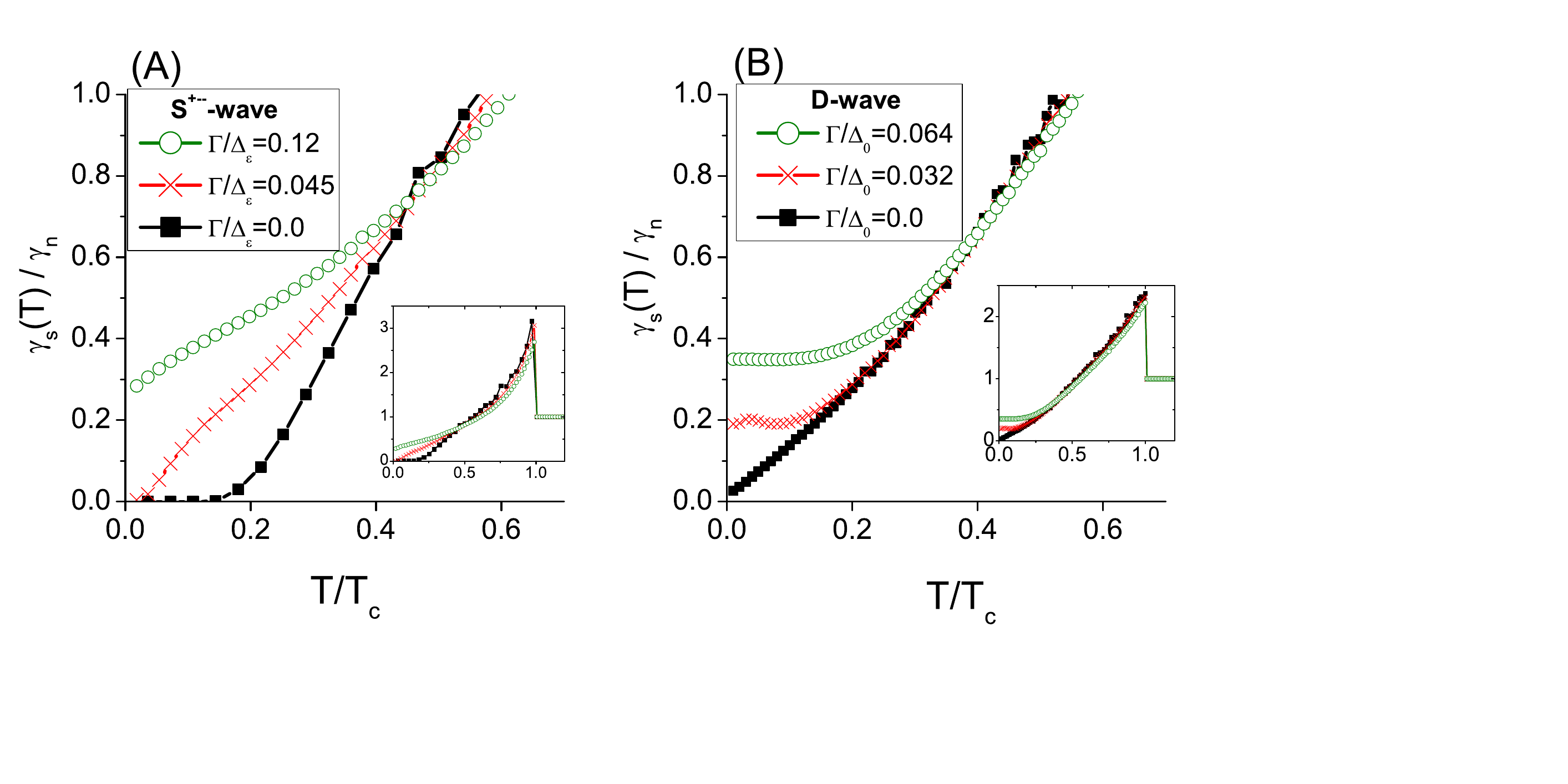}
\vspace{-2.0cm}
\caption{(Color online) Normalized specific heat coefficient $\gamma_s(T)/\gamma_n$ for $s^{\pm}$-wave and $d$-wave. (A) $s^{\pm}$-wave case for $\Gamma/\Delta_e=$ 0.12, 0.045, and 0.0, respectively, and $2 \Delta_{e}/T_c =7.5$ and $|\Delta_{e} / \Delta_{h}|=2.5$.
(B) $d$-wave case for $\Gamma/\Delta_{0}=$ 0.064, 0.032, and 0.0, respectively, and $2 \Delta_{0}/T_c =5$.
\label{SH_theor}}
\end{figure}

As to the low temperature behavior of $C(T)$, the first term in the above formula of $\gamma(T)$ is almost identical to the formula of Knight shift $K(T)$ (Eq.\ref{knight}) besides the difference of the weighting factor $\Big(\frac{\omega}{T}\Big)^2$. Therefore we expect  $\gamma(T)$ to behave similarly to the results of $K(T)$ in, e.g., Fig.\ref{knight_theor_imp}.  In fact, by a simple dimensional counting of the first term in Eq.\ref{CT_eq}, we can read $\gamma(T) \sim T^{\beta}$ if $N_a(\omega) \sim \omega^{\beta}$. Hence we can read the shape of the low energy DOS $N_a(\omega)$ from $\gamma(T)$.
In Figure \ref{SH_theor}, we show the theoretical calculations of the normalized specific heat coefficients $\gamma_s(T)/\gamma_n$ of the $s^{\pm}$-wave and $d$-wave cases with varying impurity scattering rates for comparison.
In the clean limit (black square symbols, $\Gamma_{imp}/\Delta=0$), shows the representative behaviors of each SC gap structure at low temperatures: an exponentially flat behavior for the $s^{\pm}$-wave and the $T$-linear behavior for the $d$-wave case. However, with the impurity scattering (we considered only non-magnetic impurities in the unitary limit),  the temperature dependencies of $\gamma_s(T)/\gamma_n$ for both SC cases become non-trivial.
For example, the $\gamma(T)$ of the $s^{\pm}$-pairing state for $\Gamma_{imp} > \Gamma_{imp}^{c}$ ($\Gamma_{imp}^{c}\approx0.045 \Delta_e$ in this particular example case) shows the $T$-linear behavior -- this is a common identifier for a nodal gap. On the other hand, the $\gamma(T)$ of the $d$-wave pairing state with impurities shows a flat $T$-dependence -- this is a common identifier for a $s$-wave full gap superconductor.

These results demonstrate that the typical temperature dependencies of $\gamma_s(T)/\gamma_n$ of the two representative SC states -- nodal and nodeless -- can be reversed with impurity scattering: the $s^{\pm}$-wave state shows $T$-linear behavior and the $d$-wave state shows flat-in-$T$ behavior, at low temperatures.
This reversing behavior with impurity happens in the exactly same manner with Knight shift $K(T)$ as explained in section 5.1.
Therefore, when the low temperature SH data $\gamma(T)$ is analyzed to identify the gap symmetry, it is important first to determine whether the SC samples are in the clean limit or not, and estimate how large the $\gamma_0 = \gamma(T=0)$ value is, since it should be remembered that the subtracted data $[\gamma(T) - \gamma(0)] \sim T^{\beta}$ do not follow the textbook behaviors of the clean $s^{\pm}$- and $d$-wave states.
The origin of this extreme sensitivity to the impurity scattering of two SC states is the sign-changing property of OPs in both cases.

\begin{figure}
\hspace{2cm}
\includegraphics[width=130mm]{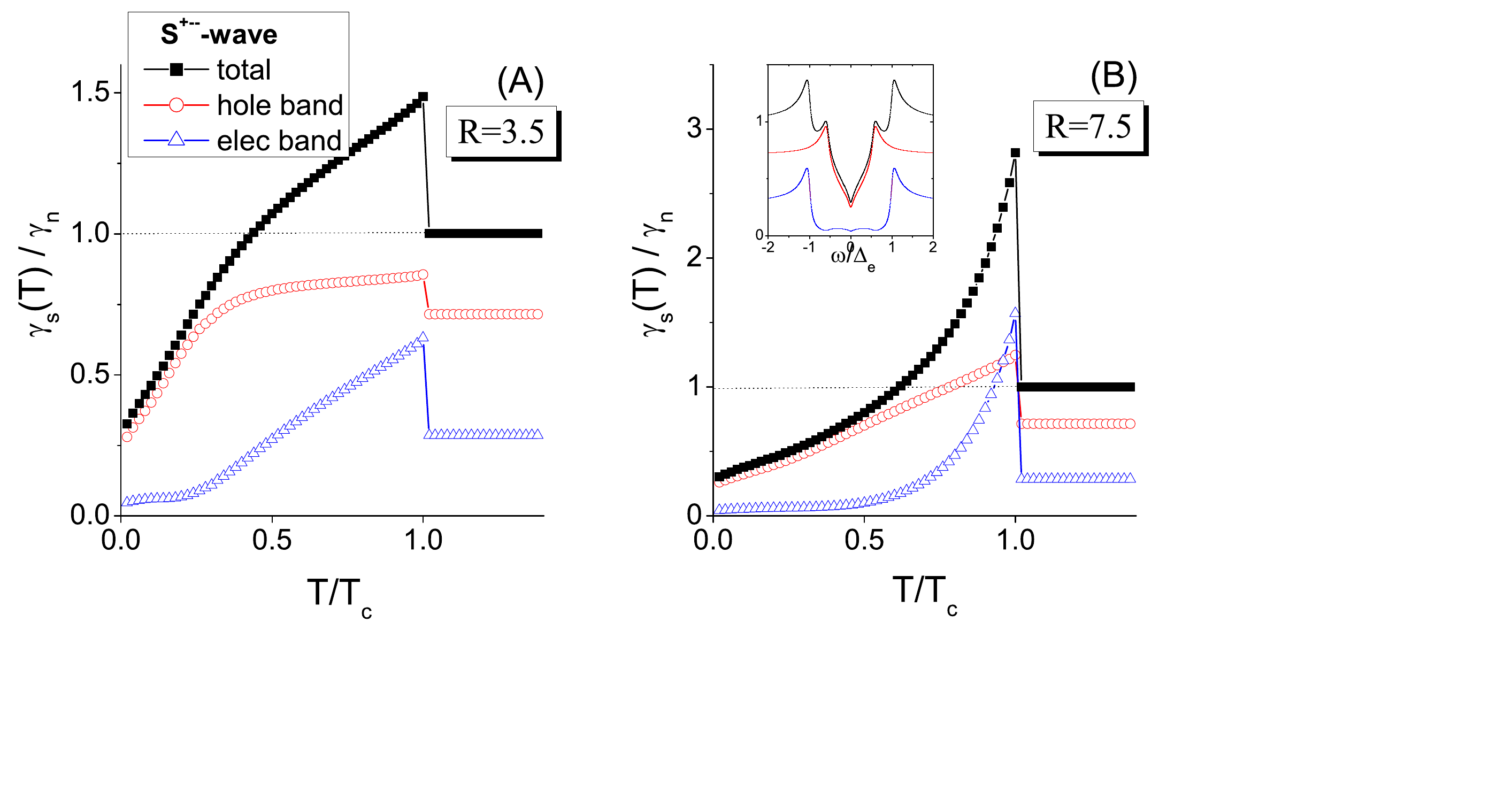}
\vspace{-1.5cm}
\caption{(Color online) Normalized specific heat coefficient $\gamma_s(T)/\gamma_n$ of $s^{\pm}$-wave showing the hole- and electron-band contributions. Common parameters for (A) and (B) are: $|\Delta_{e} / \Delta_{h}|=2.5$, $N_h /N_e =2.5$, and $\Gamma_{imp}/\Delta_e =0.1$.
(A) $R=2 \Delta_{e}/T_c =3.5$,  (B) $R=7.5$.  Inset of (B) is the common DOSs $N_{a=h,e} (\omega)$ and $N_{tot}(\omega)$ at $T=0$ both for (A) and (B).
\label{SH_theor_ratio}}
\end{figure}

In Figure \ref{SH_theor_ratio}, the overall temperature behavior of $\gamma(T)$ of the $s^{\pm}$-wave state are calculated. As in Knight shift $K(T)$ in section 5.1, the overall concavity of $\gamma(T)$ near $T_c$ is determined by the gap-to-$T_c$ ratio $R$: the larger $R$ is, the shape of $\gamma(T)$ is more concave up, and the smaller $R$ is, the shape of $\gamma(T)$ is more concave down. Fig. \ref{SH_theor_ratio} shows two example cases: (A) $R=2 \Delta_{e}/T_c =3.5$, and (B) $R=7.5$. Otherwise all other parameters are the same for both cases as $|\Delta_{e}|/|\Delta_{h}| = N_h / N_e = 2.5$, and the same impurity scattering rate $\Gamma_{imp}/\Delta_e =0.1$.
It can be seen that the small gap band (hole band (red circles) in this model calculations) is the one which is mostly modified with impurities to lead the $T$-linear behavior; this is also reflected in the $V$-shape DOS $N_h(\omega)$ in the inset of Fig.\ref{SH_theor_ratio}(B). The large gap band (electron band, blue triangles) appears to maintain the full gap-like behavior by showing a flat temperature dependence at low temperature region, but it is not exactly true because it has the finite value of constant $\gamma_s^{elec}(0)$ at low temperatures.
An interesting issue of the SH jump $\Delta C$ vs $T_c$ and the total condensation energy, which can be extracted from the SH data, too, will be discussed in section 11. Experimental Hints for Pairing Mechanism.

\subsection{Summary}
The low temperature SH, $C(T)/T \sim \gamma(T)$, is an excellent probe for the low energy DOS $N(\omega)$, so that the clean $s^{\pm}$-state should display an exponentially flat behavior as $\gamma(T) \sim \e^{- \Delta_s /T}$. However, with impurities, the low energy part of DOS $N(\omega)$ of the $s^{\pm}$-state drastically changes as shown in section 3. Increasing the concentration, the fully opened "U"-shape DOS in clean limit evolves to a "V"-shape DOS as in a clean $d$-wave state, and then a "constant"+"V"-shape DOS (see Fig. 6). Accordingly, the measured $T$-linear $\gamma(T)$ is not necessarily an evidence for a nodal gap, but it could be more a $s^{\pm}$-state with impurities. The important lessen of this section is that when the low temperature $\gamma_{exp}(T)$ is analyzed, it is primarily important to get a reliable estimation of the residual Sommerfeld coefficient $\gamma_0 = \gamma_{exp}(T=0)$. Without knowing the value of $\gamma_0$, just analyzing the temperature dependence of $\delta \gamma_{exp}(T)$ is totally misleading. Most of the SH experiments with FeSCs up to now appears to be consistent with the $s^{\pm}$-state, if the $\gamma_0$ value is properly taken into account.

\section{Volovik effect: specific heat $C(H)$ and thermal conductivity $\kappa(H)$}
In the previous section, we discussed that the temperature dependence of the SH $C_{el}(T)$ is a powerful probe for identifying the gap symmetry, if the non-electronic part contributions -- such as from phonons, spin fluctuations, etc -- are reliably subtracted. The same is true with the thermal conductivity $\kappa(T)$, which is another valuable probe for the entropy change (low energy thermal excitations) of the system. Therefore, for these experimental probes, it is always an issue how to extract only the electronic part, and one simple way of achieving it is to go to the lowest possible temperature $T \rightarrow 0$. At very low temperature far below $T_c$, the system is deep inside the SC phase and automatically $C(T)$ and $\kappa(T)$ contain only electronic contributions without any subtractions.

Then applying an uniform magnetic field $H$, the system enters the vortex state (also called as the mixed state) with a lattice of vortices. Most of unconventional superconductors are extreme type II, hence the Meissner phase exists only at very low field limit, so that we can ignore this region.
Therefore, measuring  $C(H, T \rightarrow 0)$ and $\kappa(H, T \rightarrow 0)$ with changing the field strength $H$ ($< H_{c2}$) can tell us how the low energy DOS $N(\omega, H)$ changes in the vortex state with magnetic fields $H$. The functional dependence of $N(\omega, H)$ in a vortex state sensitively depends on the SC gap structures, hence reveals information about the gap symmetry.
Typical structures of the DOS $N(\omega)$  are shown in Fig.\ref{DOS_power} for the representative pairing states. Now, we need to study how these DOSs $N(\omega)$ change with magnetic field $H$ in the vortex state to $N(\omega, H)$, which is called "Doppler effect" or "Volovik effect".

\subsection{Volovik effect in the $d$-wave state}
\begin{figure}
\hspace{3cm}
\includegraphics[width=110mm]{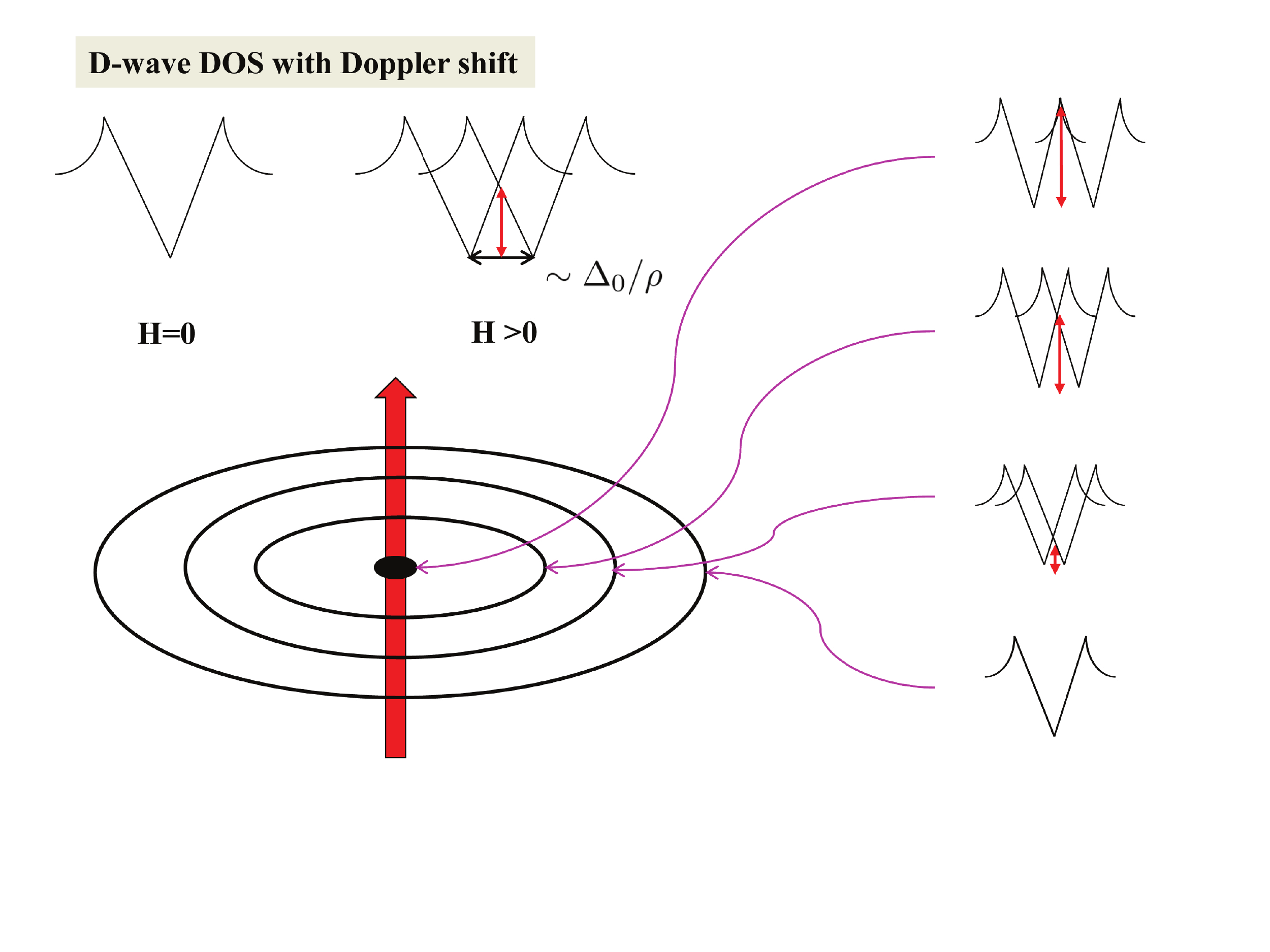}
\vspace{-1cm}
\caption{(Color online) Illustration of the local DOS $N_d (\omega,H,r)$ of the $d$-wave superconductor. The size of Doppler shift energy  is inversely proportional to the distance $r$ from the core as $\Delta E_{Doppler} \sim \Delta_0 / r$, and the DOS $N_d (\omega=0,H,r)$ at zero energy is increasing proportional to $\Delta E_{Doppler}$ because the $d$-wave DOS $N_d(\omega,H=0) \sim \omega$.
\label{d-wave_volovik_cartoon}}
\end{figure}

The field dependent DOS $N(\omega, H)$ was first studied with the $d$-wave cuprate superconductors by Volovik \cite{volovik1993} and soon was taken up by many researchers to investigate the SH and thermal conductivity of the cuprate superconductors\cite{kubert1998,vekhter1999}.
In a uniform (without fields) SC phase, Cooper pairs are formed by a pair of $(k \uparrow, -k \downarrow)$ states and their energies in normal states are degenerate as $\epsilon(k)=\epsilon(-k)$. When the SC condensation occurs as $\Delta(k)=-\sum_{k'} V_{k,k'} <c_{k \uparrow} c_{-k \downarrow}>$, the quasiparticles in SC phase are defined by the eigenenergies of the following BCS Hamiltonian matrix,

\begin{equation}
H(k) =
\left( \begin{array}{cc}
\epsilon (k)  & \Delta(k) \\
\Delta(k) & - \epsilon (-k) \end{array} \right)
\end{equation}

\noindent
whose energies are $\pm E(k)=\pm\sqrt{\epsilon^2(k)+\Delta^2(k)}$. In the vortex state with magnetic field $H$, the system is not uniform but has an array of the vortices and each vortex is carrying a circulating supercurrent $\vec{j}_s (r)=\rho_s \vec{v}_s(r)$. Now imagine a Cooper pair of $(k \uparrow, -k \downarrow)$ at position $"r"$, the distance from the vortex core. Their normal state energies are not anymore degenerate as $\epsilon(k)=\epsilon(-k)$, but are shifted opposite direction by riding on the supercurrent $\vec{v}_s(r)$ as $[\epsilon(\vec{k}+m\vec{v}_s(r));  \epsilon(-\vec{k}+m\vec{v}_s(r))]$. Since we are interested in near the Fermi level, in the limit $k_F \gg m v_s(r)$, the normal state energies of the $(k \uparrow, -k \downarrow)$ pair become $[\tilde{\epsilon}(k); \tilde{\epsilon}(-k)] \approx [\epsilon(k)+\vec{v}_s(r)\cdot \vec{k};  \epsilon(-k)-\vec{v}_s(r)\cdot \vec{k}]$. This is nothing but a Doppler effect and the quasiparticles in this vortex state are defined by the eigenenergies of the following modified BCS Hamiltonian matrix
\\
\begin{equation}
H_{mixed}(k,r) =
\left( \begin{array}{cc}
\epsilon (k)+\vec{v}_s(r)\cdot \vec{k}  & \Delta(k) \\
\Delta(k) & - \epsilon (-k)+\vec{v}_s(r)\cdot \vec{k} \end{array} \right)
\end{equation}
\noindent
The eigenenergies of $H_{mixed}(k,r)$ are $E_{1,2}(k)=-\vec{v}_s(r)\cdot \vec{k} \pm\sqrt{\epsilon^2(k)+\Delta^2(k)}$, which are not symmetric around the Fermi level but are shifted to one side. Most importantly, $E_{1,2}(k)$ are not always gapped but can hit the zero energy excitation. This is the result of the pair-breaking due to the mismatch of energies, $\tilde{\epsilon}(k) \neq \tilde{\epsilon}(-k)$, of the $(k \uparrow, -k \downarrow)$  pair at normal state. The single particle Green's function of $H_{mixed}(k,r)$ can be written as
\\
\be
G ({\bf k, r,\omega})=\frac{[\omega +  {\bf v_s (r)} \cdot
{\bf k}] \tau_0 + \epsilon (k) \tau_3 + \Delta({\bf k}) \tau_1}{[ \omega +
{\bf v_s (r)} \cdot {\bf k}]^2 - \epsilon ^2 (k) - \Delta ^2({\bf k}) }\\
\ee
\noindent
where $\tau_i$ are Pauli matrices. From the above Green's function
we obtain the local DOS $N(\omega,H,r)= - \frac{1}{\pi} {\rm Tr Im} \sum_k G ({\bf k,
r,\omega})$. The Doppler shifting energy is given as ${\bf v_s
(r)} \cdot {\bf k}=\frac{k}{m} \frac{1}{r} \cos{\theta} = b
\frac{\Delta_{0}}{\rho} \cos{\theta}$ with normalized distance
$\rho= r/ \xi$ ($\xi=$ coherence length) and "$b$" a constant of
order unity. Notice that the local DOS $N(\omega,H,r)$ is function of the distance $"r"$ from the vortex core as illustrated in Fig.\ref{d-wave_volovik_cartoon}.

The above discussion is for a single vortex. With increasing field $H$, the number of vortices is increasing as $\sim H$, or conversely the size of each vortex is decreasing as $\sim 1/H$. The typical size of the radius of a single vortex is called magnetic length $R_H = \alpha \sqrt{\frac{\Phi_0}{\pi H}}$ ($\Phi_0$ a flux quantum, $H$ magnetic field, and "$\alpha$" geometric factor of order unity) and the above Green's function is defined only for $1 \leq \rho \leq R_H /\xi$. When $\rho <1$, the Doppler shifting energy $\Delta_{Doppler}$ becomes larger than the maximum gap size $\Delta_0$, therefore the SC gap should collapse for $\rho <1$, which defines the vortex core. The thermodynamic averaged DOS is obtained by the magnetic unit cell averaged DOS as follows.
\be
\bar{N}_d (\omega,H)=<N_d (\omega,H,r)>_{cell} =\int_{\xi} ^{R_H} dr^2 N_d (\omega,H,r) / \pi R_H ^2
\ee
\noindent
Noticing that $N_d (\omega=0,H,r) \sim 1/r$ from Fig.\ref{d-wave_volovik_cartoon} for the $d$-wave superconductor, a simple dimensional counting of the above integral tells us that $\bar{N}_d (\omega=0,H) \sim 1/R_H \sim \sqrt{H}$, which is the famous Volovik result for the $d$-wave superconductor\cite{volovik1993}.

All the discussions here about the Doppler shift effect in the vortex state (or Volovik effect) is a semiclassical description and a phenomenological form of the field dependent gap size $\Delta_0(T)=\Delta_0 \sqrt{1-\frac{H}{H_{c2}}}$ is used.
This approximation is excellent for weak field $H < H_{c2}$, but certainly would break down when $H \rightarrow H_{c2}$ where quantum effect becomes more important. Nevertheless, we found empirically that this semiclassical approximation works well up to $H \approx 0.9 H_{c2}$. For a full quantum theory -- presumably should work up to $H_{c2}$ -- we refer to Ref.\cite{vekhter1999,mishra2009,brandt1967,pesch1975}.

Once $\bar{N}_d (\omega,H)$ is calculated, thermodynamic quantity like specific heat $C(T,H)$ can be calculated as
\be C_d(T,H) = \int_0 ^{\infty} d \omega \Big(\frac{\omega}{T}\Big)^2 ~
\frac{\bar{N}_d (\omega, H)}{{\rm cosh}^2 (\frac{\omega}{2 T})}. \ee
Similarly, thermal conductivity is calculated with
\cite{ambegaokar1964}
\ba \kappa_d (T,H,r) &\propto & N ^0 v_F ^2 \int_0 ^{\infty} d
\omega \Big(\frac{\omega}{T}\Big)^2 \frac{K_d (\omega,T,H,r)}{{\rm cosh}^2 (\frac{\omega}{2 T})}, \\
K_d (\omega,T,H,r) &=& \Bigg<\frac{1} {Im \sqrt{\tilde{ z}^2 - \tilde{\Delta}^2(k)}}
\times \Big(1 + \frac{|\tilde{ z}|^2 - |\tilde{\Delta}(k)| ^2}{|\tilde{ z}^2 - \tilde{\Delta}^2(k)|} \Big) \Bigg>_k,
\ea
\noindent where $\tilde{z} =\tilde{\omega} +{\bf v}_s ({\bf r)}\cdot {\bf k_F} $ and $<...>_k$ means the Fermi surface average.
And then longitudinal and transversal thermal conductivities are calculated as
\ba
\kappa _{\parallel}(T,H) &=& \int_{cell} d^2 r \kappa(T,H,r)  / \pi R_H ^2 ,\\
\kappa^{-1}_{\perp}(T,H) &=& \int_{cell} d^2 r \kappa^{-1}(T,H,r)  / \pi R_H ^2 ,
\ea, respectively.

\begin{figure}
\hspace{2cm}
\includegraphics[width=110mm]{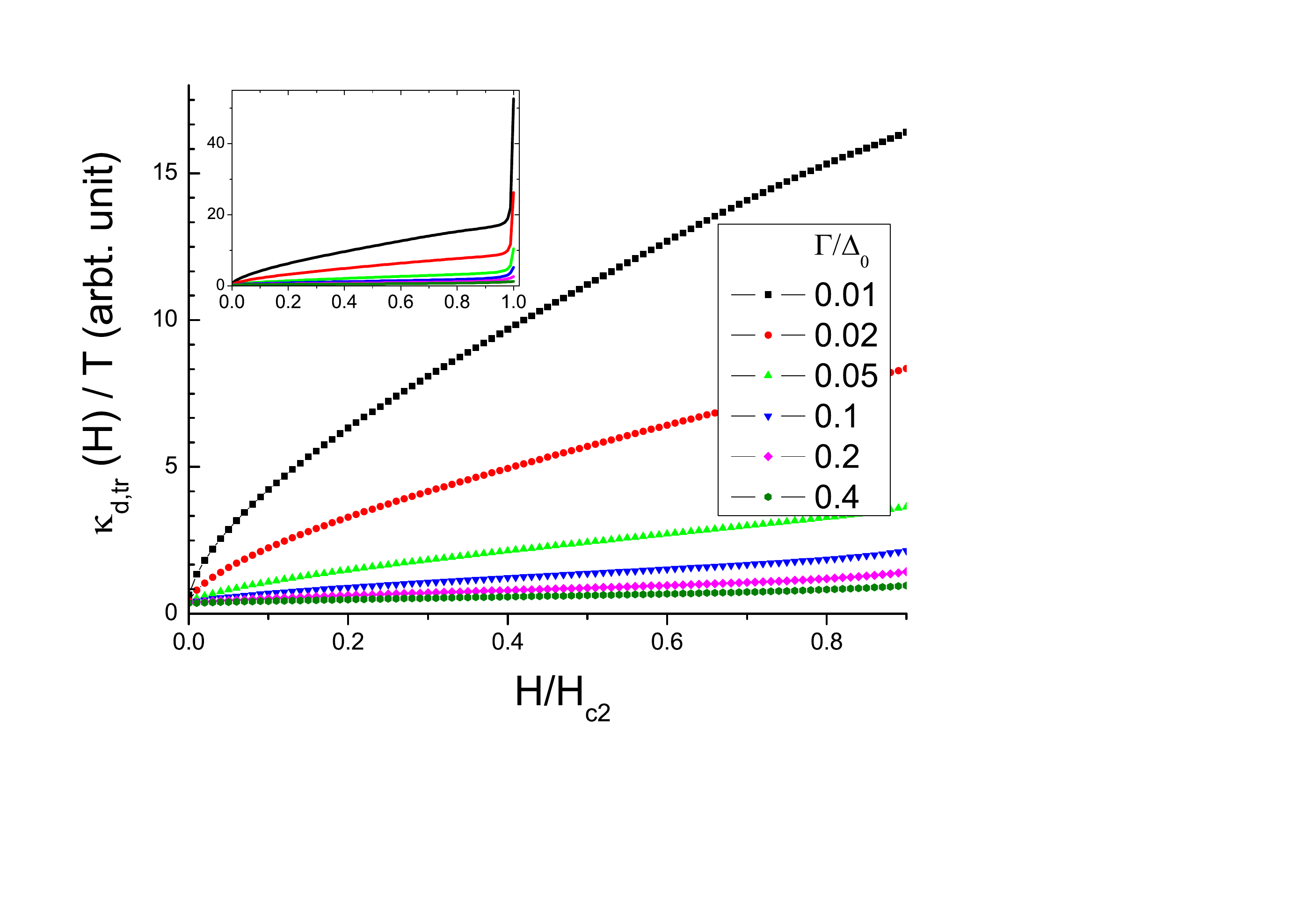}
\vspace{-1cm}
\caption{(Color online) Transversal thermal conductivity $\kappa_{d \perp} (H)/T$ vs the normalized fields $H/H_{c2}$ of the $d$-wave SC state,
calculated at $T=0.02 \Delta_0$ for various impurity concentrations $\Gamma/\Delta_0= 0.01, 0.02, 0.05, 0.1, 0.2,$ and
0.4. (unitary impurity). The inset shows the full range of fields up to $H/H_{c2}=1$.
All $\kappa_{d \perp} (H \rightarrow 0)/T$ approaches to a universal value. From \cite{bang2012there}.
\label{d-wave_kappa_theor}}
\end{figure}

\begin{figure}
\hspace{2cm}
\includegraphics[width=110mm]{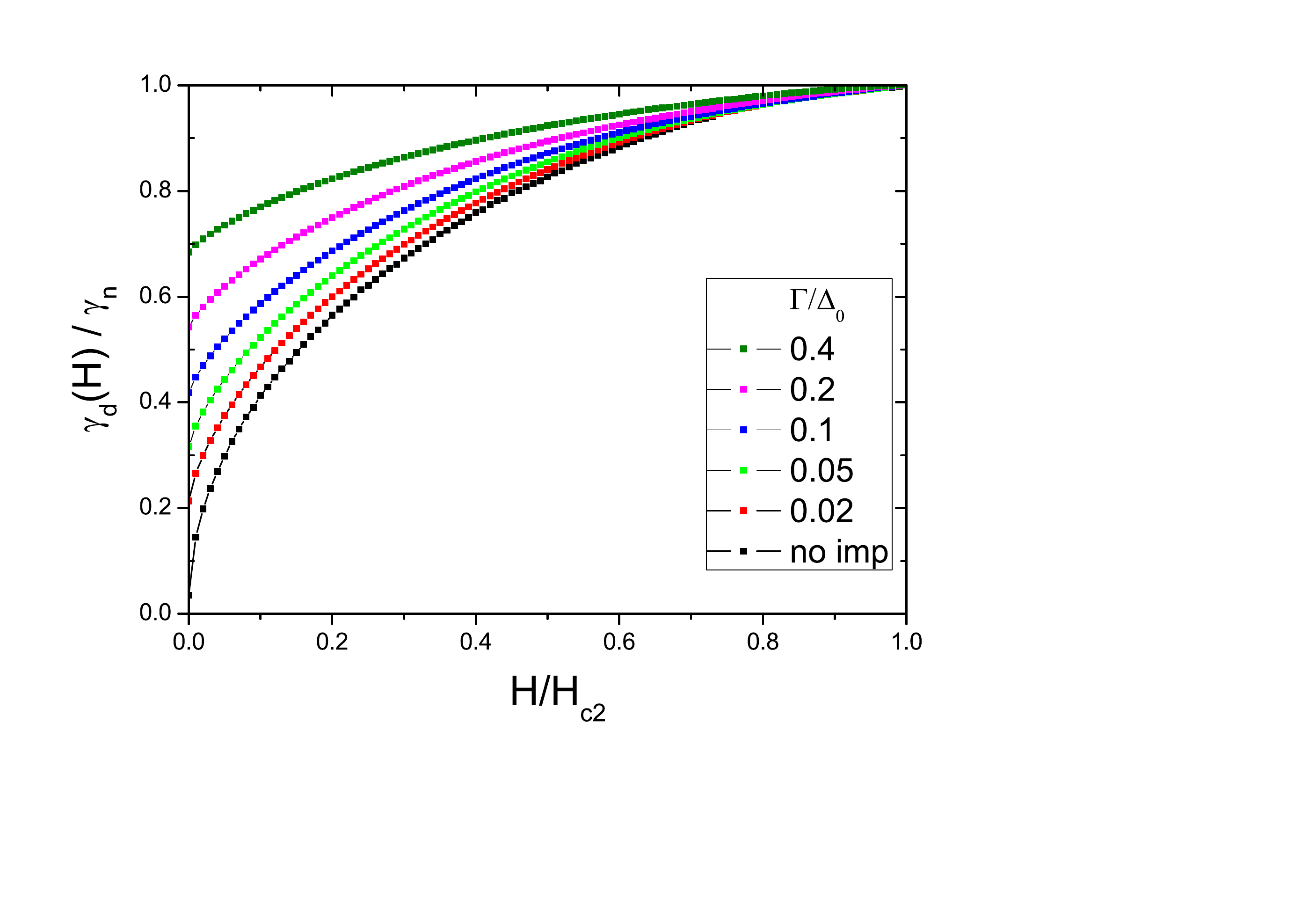}
\vspace{-1cm}
\caption{(Color online) Normalized specific heat coefficient $\lim_{T \rightarrow 0} C(T,H)/T=\gamma_d (H)$ vs fields $H$
calculated at $T=0.02 \Delta_0$ of the $d$-wave SC state for
various impurity concentrations $\Gamma/\Delta_0= 0, 0.02, 0.05, 0.1, 0.2,$ and 0.4 (unitary impurity).
$\gamma_d (H=0)$ has no universal value. From \cite{bang2012there}.
\label{d-wave_SH_theor}}
\end{figure}

Figure \ref{d-wave_kappa_theor} shows the numerical calculations of the transverse thermal conductivity $\kappa_{d \perp} (H)/T$ vs the normalized fields $H/H_{c2}$ of the $d$-wave SC state for different impurity scattering rate $\Gamma/\Delta_0 = 0.01, 0.02, 0.05, 0.1,$ and 0.2. It indeed confirms the theoretical prediction of $\kappa_{d} (H)/T \sim \sqrt{H}$ and the universal value (a constant value independent of the impurity scattering rate $\Gamma$) of $\kappa_{d} (H \rightarrow 0)/T \approx \frac{v_F}{v_1}$.
Figure \ref{d-wave_SH_theor} shows the results of the specific heat coefficient $C(H)/T =\gamma(H)$ vs $H/H_{c2}$ of the same model as in Fig.\ref{d-wave_kappa_theor}. It also shows the $\sqrt{H}$-dependence. However, the values $C(H \rightarrow 0)/T$ are not universal but increase with the scattering rate $\Gamma$ as $\gamma(H \rightarrow 0) \sim \sqrt{\Delta_0 \Gamma}$.

All these so-called Volovik effects have been well studied and confirmed with the high-$T_c$ cuprates as well as several heavy fermion superconductors\cite{kubert1998,vekhter1999,junod2000direct}. The origin of these phenomena is due to the thermodynamically averaged DOS $\bar{N}(H) \sim \sqrt{H}$ in the vortex state of nodal gap superconductors. On the other hand, a simple reasoning of the Doppler effect on an $s$-wave superconductor trivially tells us that $\bar{N}(H < H_{c2}) \approx 0$, hence measurements of $\gamma(H)$ and $\kappa(H)/T$ with $s$-wave superconductors should yield exponentially flat or activated behavior with field as $\sim e^{- \Delta_0 /H}$ at low fields. As a result, it became a standard practice to take
the observation of the $\sqrt{H}$-dependence in $\gamma(H)$ and $\kappa(H)/T$ as an evidence for a nodal gap superconductor.
Therefore the frequent observation of strong field dependencies (in fact, approximately close to $\sim \sqrt{H}$) of $\gamma(H)$ and $\kappa(H)/T$ in many FeSCs has been taken as the evidence that these FeSCs are nodal superconductors, and cast a doubt on the $\pm s$-wave pairing scenario. However, it was soon proven that a multiple band $s$-wave superconductor with different gap sizes $\Delta_{s1}$ and $\Delta_{s2}$ can also yield a strong field dependence on $\bar{N}(H)$ in the vortex states as $\bar{N}(H) \sim H$ but not as $\sim \sqrt{H}$\cite{bang2010volovik,wang2011volovik}.

\subsection{Volovik effect in the $s^{\pm}$-wave state}
\begin{figure}
\hspace{3cm}
\includegraphics[width=110mm]{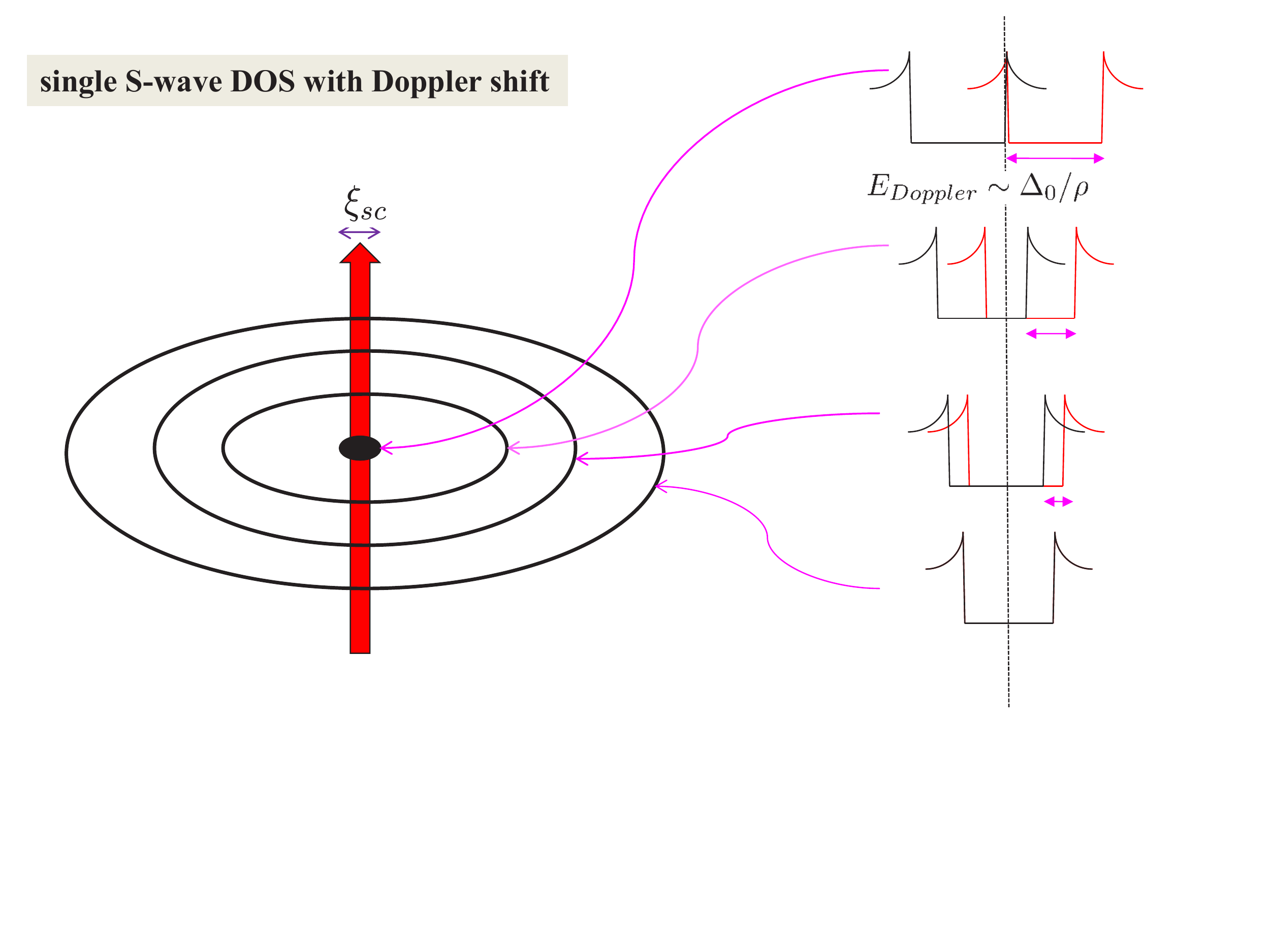}
\vspace{-2cm}
\caption{(Color online) Illustration of the local DOS $N_s (\omega,H,r)$ of the $s$-wave superconductor. The size of Doppler shifting energy  is inversely proportional to the distance $r$ from the core as $\Delta E_{Doppler} \sim \Delta_0 / r$, but the condition $\Delta E_{Doppler} <  \Delta_0$ always holds outside the core, hence the DOS $N_s (\omega=0,H,r)$ remains to be zero outside the core. \label{s-wave_Volovik_cartoon}}
\end{figure}

At the semiclassical level, which we found works well in the $d$-wave case, it is easy to understand the field dependence (Volovik effect) in the $s^{\pm}$-wave state. Fig.\ref{s-wave_Volovik_cartoon} illustrates that the Doppler shifting occurs even in a single band $s$-wave superconductor as strongly as in the $d$-wave superconductor, but its consequence to low energy responses is null because the uniform gap $\Delta_0$ is always larger than the Doppler shifting energy $\Delta E_{Doppler}$ everywhere outside the vortex core region: in other words, the region where $\Delta E_{Doppler} > \Delta_0$ is by definition the vortex core where the superconductivity breaks down.

\begin{figure}
\vspace{0cm}
\hspace{3cm}
\includegraphics[width=110mm]{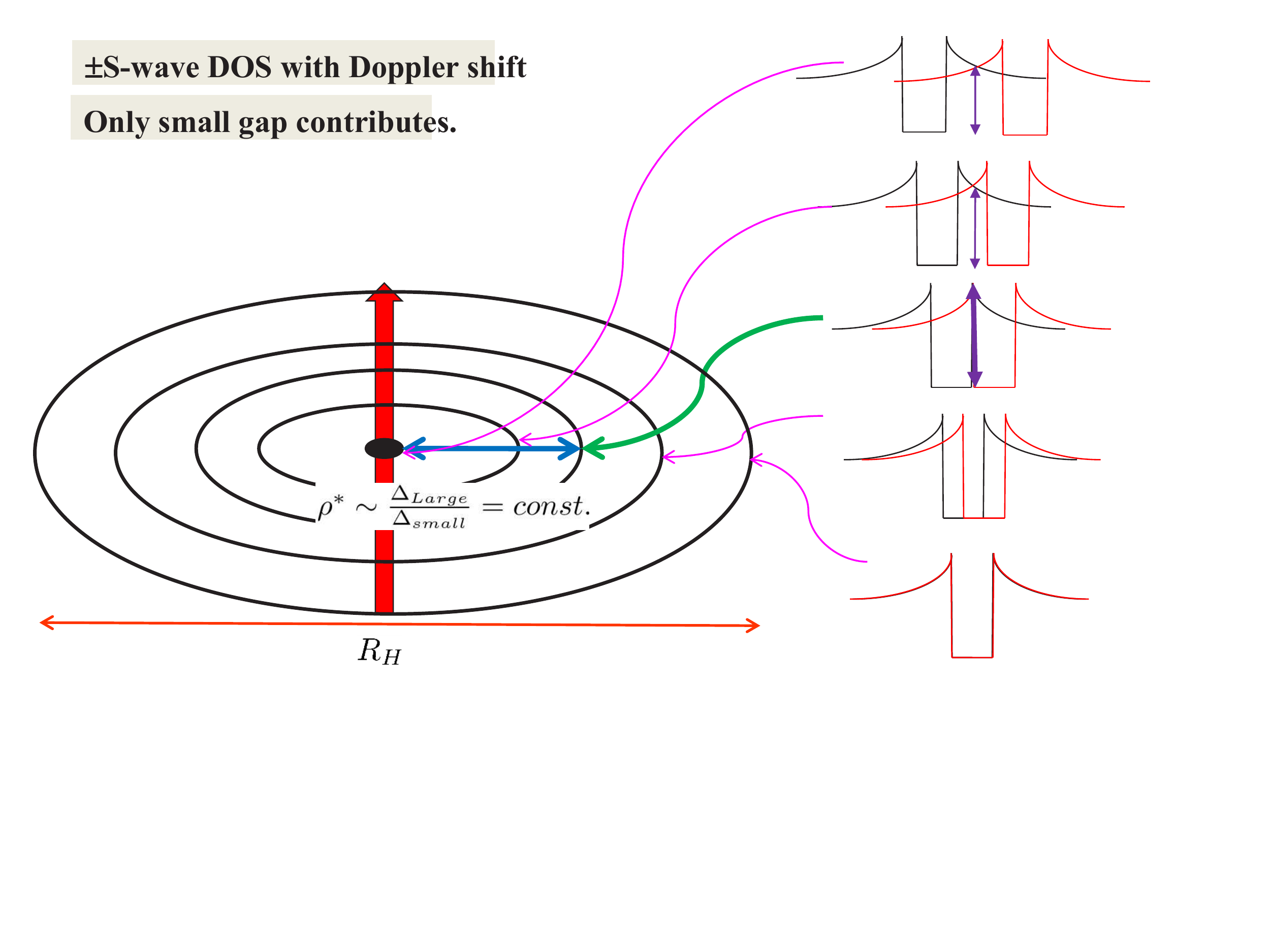}\\
\vspace{-2.5cm}
\caption{(Color online) Illustration of the local DOS $N_{s1} (\omega,H,r)$ of the small gap band ($s1$) of the $s^{\pm}$-wave superconductor. Within the finite range ($r < r^{\ast}$), the small gap band develops the zero energy excitations as $N_{s1} (\omega=0,H,r) \sim N_{s1}^{normal}$.  \label{pmS-wave_volovik_cartoon}}
\end{figure}

Now the question is what happens with a two gap $s$-wave superconductor as in the $s^{\pm}$-wave state ? If two $s$-wave OPs $\Delta_{L}$ and $\Delta_{S}$ (a larger and a small gaps) are independent, everything is the same as a single band $s$-wave superconductor. But if two SC OPs are coupled by an interband pairing interaction $V_{inter}$ as in the $s^{\pm}$-wave state, there exist a finite region ($ r < r^{\ast}$, with $r^{\ast}=b \xi \frac{\Delta_L}{\Delta_S}$; $\xi=$ coherence length, $b=$ a constant of $\sim O(1)$) outside the vortex core, where $\Delta E_{Doppler} > \Delta_{S}$ but still $\Delta E_{Doppler} < \Delta_{L}$ with $\Delta_{S} < \Delta_{L}$ as depicted in Fig.\ref{pmS-wave_volovik_cartoon}.

In this region ($ r < r^{\ast}$), the quasiparticle excitations $N_{S}(\omega,H,r)$ of the small gap band allows the zero energy excitations with $\Delta_{S}$ remained uncollapsed, because the SC OP $\Delta_{S}$ is sustained by the larger gap $\Delta_{L}$, through the interband pairing $V_{inter}$, which still survives because $\Delta E_{Doppler} < \Delta_{L}$ in this region. The local DOS of the larger gap band,  $N_{L} (\omega,H,r)$, behaves the same as the single band $s$-wave case in Fig.\ref{s-wave_Volovik_cartoon}. With this observation, the magnetic unit cell averaged DOS $\bar{N}_a (\omega,H)=<N_a
(\omega,H,r)>_{cell} =\int_{\xi} ^{R_H} dr^2 N_a (\omega,H,r) /\pi R_H ^2$ is readily obtained at $\omega=0$ as follows.
\ba \bar{N}_L (\omega=0,H) &=& \frac{0}{\pi R_H ^2} =0 \\
\bar{N}_S (\omega=0,H) &=& N_S ^{normal} \frac{[ (b
\frac{\Delta_L}{\Delta_S})^2 - 1] \xi^2} {R_H ^2} \propto H \ea
The above Eq.(34)-(35) holds as far as $\Delta_S < \Delta_L$ and shows that
Volovik effect immediately creates a finite DOS in the isotropic $\pm$s-wave state and there is no threshold value of magnetic
field $H^{*}$ to create the zero energy excitations. Its generic field dependence is linear in $H$ and its slope is proportional to
$\approx (\frac{\Delta_L}{\Delta_S})^2$. It was found that impurity scattering will smooth this
generic linear-in-$H$ field dependence and make it more sublinear and closer to $\propto \sqrt{H}$. Therefore the impurity effect is
important to understand experiments.

\begin{figure}
\hspace{3cm}
\includegraphics[width=60mm]{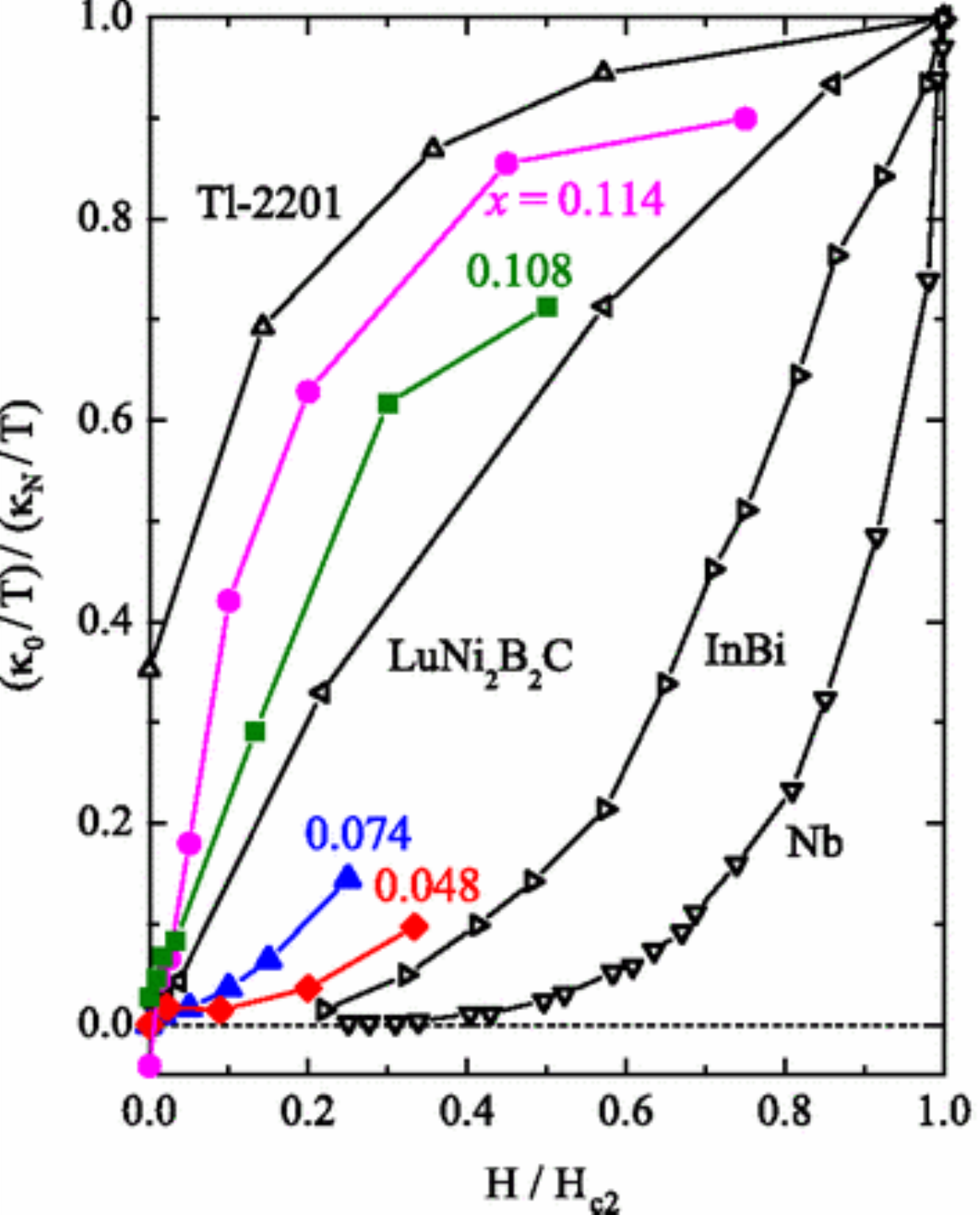}\\
\vspace{-0.5cm}
\caption{(Color online) The electronic thermal conductivity data of Ba(Fe$_{1-x}$Co$_x$)$_2$As$_2$ (full symbols) $[\kappa(H)/T] / [\kappa_{N}/T]$
as a function of normalized magnetic field $H/H_{c2}$. From \cite{tanatar2010thermal}\label{tanatar_kappa}}
\end{figure}

Now having calculated the local DOS $N_{a}(\omega,H,r), a=S,L$, it is straightforward to calculate the specific heat coefficient $\gamma(H) = \lim_{T \rightarrow 0}C(H, T)/T$ and the thermal conductivity $\lim_{T \rightarrow 0}\kappa(H,T)/T$ of the $\pm$s-wave state\cite{bang2010volovik}. What was found the most interesting was that the slope of the field dependence, which is linear-in-$H$ in clean limit, of $\lim_{T \rightarrow 0}\kappa(H,T)/T$ continuously increases from a very flat (when $|\Delta_S / \Delta_L |\approx 1$) to a very steep one (when $|\Delta_S / \Delta_L | \ll 1$). Therefore, the overall behavior of the $\lim_{T \rightarrow 0}\kappa(H,T)/T$ vs. $H$ looks like evolving from a standard $s$-wave superconductor to a nodal superconductor only by changing the relative size of the two gaps $|\Delta_S |$ and $|\Delta_L |$. This behavior was exactly captured in experiment for Ba(Fe$_{1-x}$Co$_x$)$_2$As$_2$ with a systematic change of Co doping $"x"$\cite{tanatar2010thermal} as shown in Fig.\ref{tanatar_kappa}.

\begin{figure}
\vspace{-0.5cm}
\hspace{2cm}
\includegraphics[width=110mm]{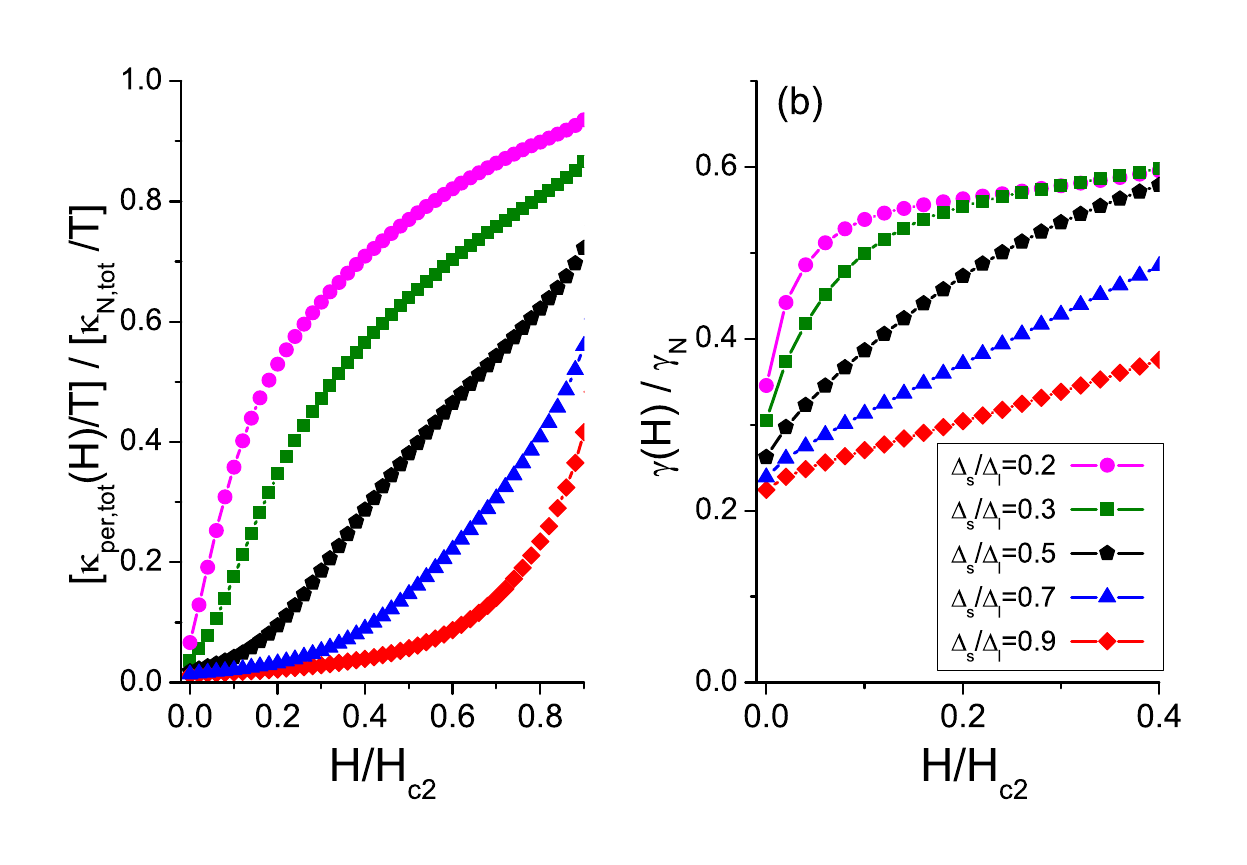}\\
\vspace{-1cm}
\caption{(Color online) (a) Normalized total transverse thermal conductivity coefficient
$\lim_{T \rightarrow 0}[\kappa_{\perp,tot}(H)/T] / [\kappa_{N,tot}/T]$ for
different gap size ratios, $|\Delta_S ^0 / \Delta_L ^0|= 0.2, 0.3, 0.5, 0.7$, and $0.9$, respectively.
(b) The Volovik part of specific heat coefficient $\gamma_{volovik}(H)/ \gamma_{tot,N}$
for different gap size ratios as in (a). All calculations include the
same concentration of impurities $\Gamma/ \Delta_L=0.05$ with unitary scattering limit ($c=0$)
both for intra- and interband scattering. from\cite{bang2010volovik}\label{kappa_theor}}
\end{figure}

Figure \ref{kappa_theor}(A) shows theoretical calculations of the transverse (current in the $ab$-plane and $H$ field along $c$-axis) thermal conductivity $\lim_{T \rightarrow 0}\kappa(H,T)/T$ vs $H$ for varying $|\Delta_S / \Delta_L | =0.2, 0.3, 0.5, 0.7,$ and 0.9, respectively. The normal DOSs $N_a (a=S,L)$ for each band were assumed equal and the unitary impurity ($c=0$) with the concentration $\Gamma/\Delta_L=0.05$ was included.
The overall behavior of this theoretical result is very similar to the experimental measurements of $\kappa(H)/T$ for Ba(Fe$_{1-x}$Co$_x$)$_2$As$_2$\cite{tanatar2010thermal,dong2010thermal} shown in Fig.\ref{tanatar_kappa}. In particular, the systematic increase of the slope of $\kappa(H)/T$ vs $H$ with increasing Co doping ($"x"$) and the evolution of the overall shape of the field dependence from a flat (concave up) for a smaller $"x"$ to a steep (concave down, hence looks almost $\sim \sqrt{H}$) for a larger $"x"$ is exactly captured by this simple two band $s^{\pm}$-wave pairing model.

Another important feature is that the values of $\kappa_{\perp}(H)/T$  (also $\kappa_{\parallel}(H)/T$) in the zero field limit are negligibly small for all cases despite substantial impurity scattering induced DOS accumulated at $\omega=0$ as seen in the data of $\gamma(H=0)$ in \ref{kappa_theor}(B);
this is even true with the $|\Delta_S ^0 /\Delta_L ^0|=0.2$ case which shows the behavior $\sim \sqrt{H}$ as in the nodal d-wave case.
In fact, these extremely small values of the thermal conductivity coefficient $\kappa_{\perp}(H)/T$ in the zero field limit were
argued as an evidence of an isotropic s-wave gap nature \cite{tanatar2010thermal,dong2010thermal}. However, it is a very puzzling feature when we note that the several experiments \cite{mu2009low,gang2010sizable} observed substantial values of the specific heat
coefficients $\gamma(H \rightarrow 0)$ with the same compounds with similar dopings.

This seemingly conflicting feature can be understood by considering the difference of the coherence factors between $\gamma(H)$ and  $\kappa(H)/T$.
Although the same DOSs $\bar{N}_{L,S}$ contribute to both specific heat and thermal conductivity, the kernel of thermal conductivity (see Eq.(31)), being an energy current-energy current correlation function, contains a destructive coherence factor ("$-$" sign in the numerator of the last term in Eq.(31)), but the specific heat does not have such a destructive coherence factor. Therefore we can expect a substantial difference between $\gamma(H)$ and  $\kappa(H)/T$ particularly at  low frequencies and low field.
In the case of d-wave pairing, the same destructive coherence factor for thermal conductivity becomes very weak in the low energy limit
because the nodal gap $\Delta_d (\theta)$ linearly disappears, so that $\gamma(H \rightarrow 0)$ and $\kappa(H \rightarrow 0)/T$
behave rather similarly.
The theoretical result of $\gamma(H \rightarrow 0)/\gamma_{tot,N}$ in Fig.\ref{kappa_theor}(B) indeed show substantial values in the zero field limit $\gamma(0,0)/\gamma_{tot,N}$ as $H/H_{c2} \rightarrow 0$, while the zero field limit of  $\kappa(H)/T$ in Fig.\ref{kappa_theor}(A) is approximately zero.

The second important difference between $\gamma(H)$ and $\kappa(H)/T$ is that $\gamma$ is the thermodynamic quantity and therefore it contains contributions both from the extended states outside vortices (which are calculated using Eq.(34)-(35) into Eq.(29)) and also from the normal states inside vortex cores\cite{hussey2002low}.
These core localized states has no contribution to the transverse thermal conductivity $\kappa_{\perp}(H)/T$ due to the geometry ($J \perp H$), which is displayed in Fig.\ref{kappa_theor}(A). In principle, theoretically, the longitudinal thermal conductivity $\kappa_{\parallel}(H)/T$  ($J \parallel H$) should have a similar contribution from the core states but there is experimental difficulty and uncertainty to measure the ideal longitudinal thermal conductivity, therefore we will not consider this core correction to the longitudinal thermal conductivity $\kappa_{\parallel}(H)/T$.
However, the SH coefficient $\gamma(H)$ always has the normal state contributions from the vortex cores and we need to correct $\gamma_{tot}(H)$ as follows.
\be
\gamma_{tot}(H) = (1-H/H_{c2})\gamma_{volovik}(H)+(H/H_{c2})\gamma_n.
\ee

\begin{figure}
\vspace{0cm}
\hspace{2cm}
\includegraphics[width=120mm]{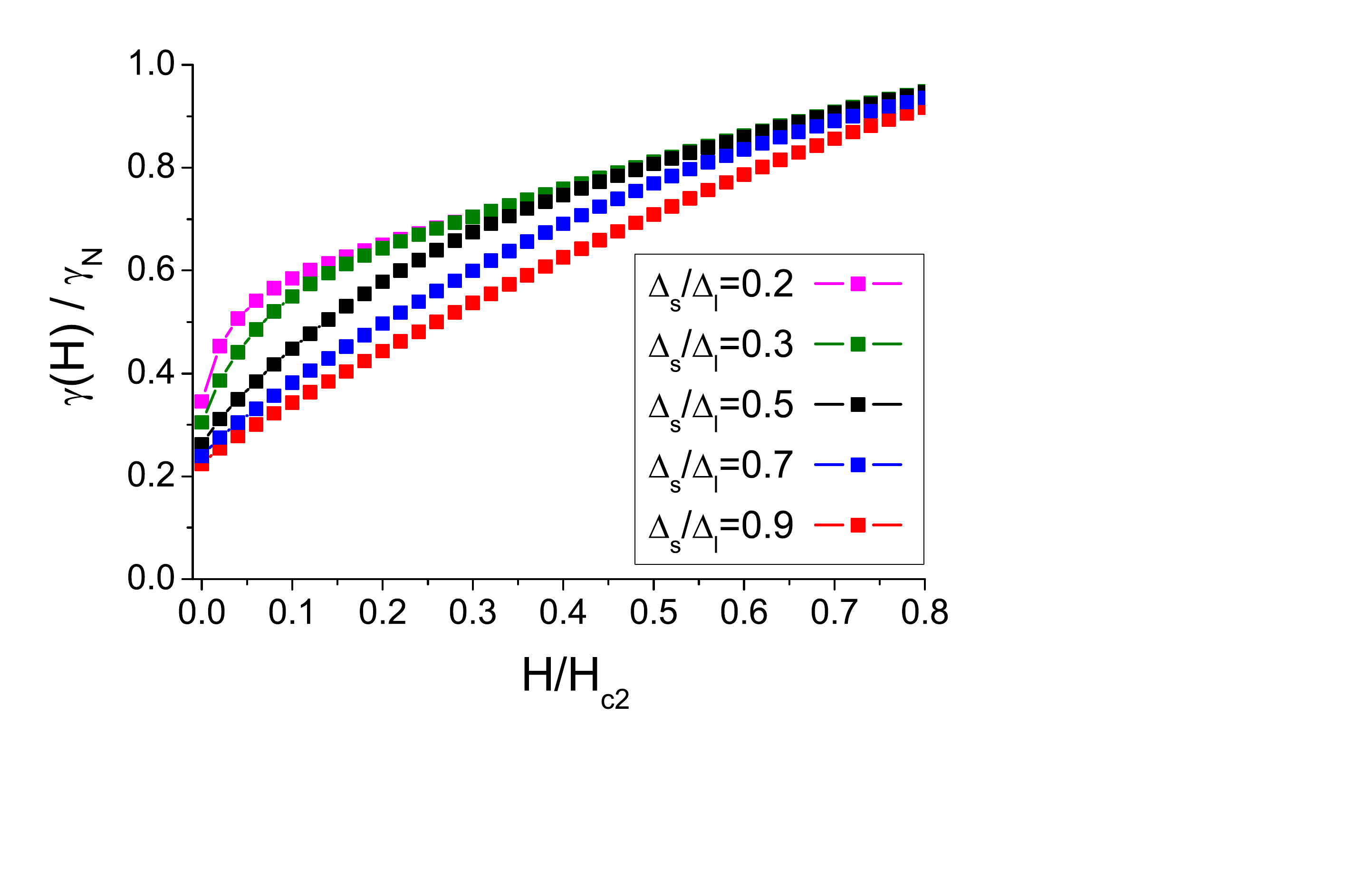}\\
\vspace{-2cm}
\caption{(Color online) Replot of the data $\gamma_{tot}(H)/ \gamma_{tot,N}$ in Fig.\ref{kappa_theor}(B) but now corrected
by adding the vortex core contribution $(H/H_{c2}) \gamma_{N}$. \label{gamma_corrected}}
\end{figure}

\begin{figure}
\vspace{0cm}
\hspace{3cm}
\includegraphics[width=80mm]{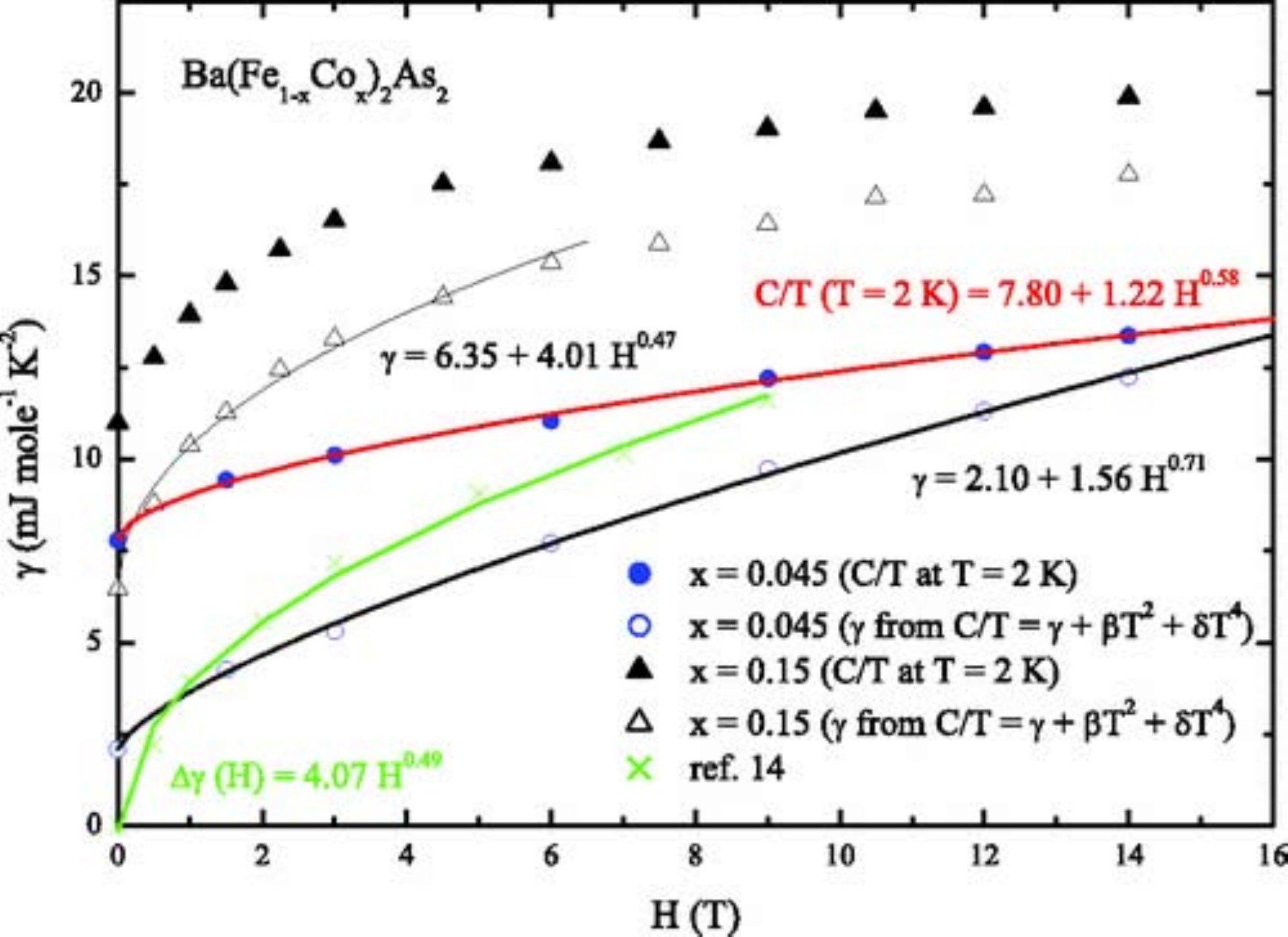}\\
\vspace{-0.5cm}
\caption{(Color online) $C/T$  at 2 K vs field as an indication of $\gamma$ vs  $H$  for annealed single crystals of Ba(Fe$_{0.955}$Co$_{0.045}$)$_2$As$_2$  and Ba(Fe$_{0.85}$Co$_{0.15}$)$_2$As$_2$. From \cite{kim2012specific}. \label{jskim_SH}}
\end{figure}

\begin{figure}
\vspace{0cm}
\hspace{3cm}
\includegraphics[width=80mm]{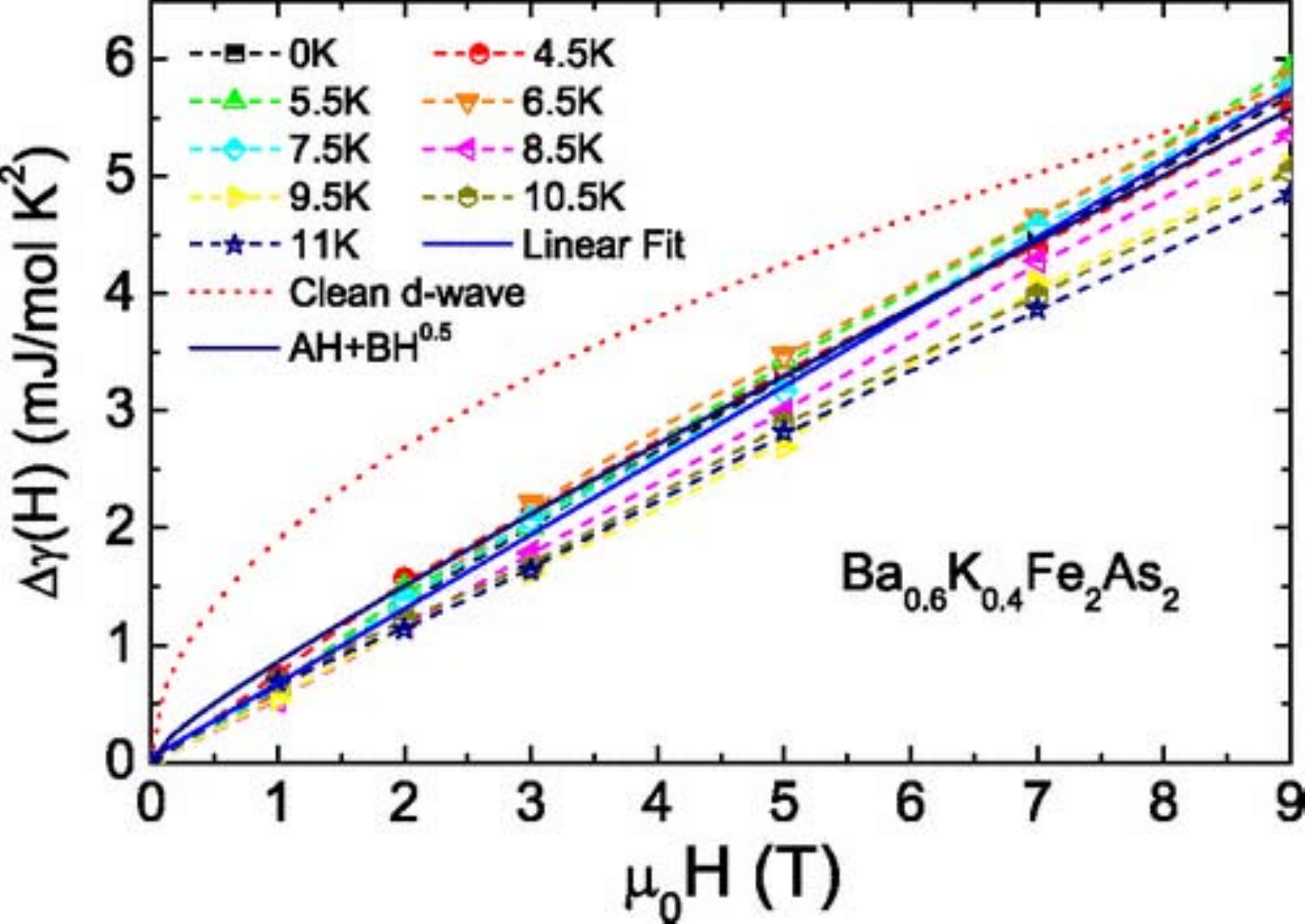}
\vspace{0cm}
\caption{(Color online) $\Delta \gamma(H)=[C(T,H)-C(T,0)]/T$ vs $H$ for Ba$_{0.6}$K$_{0.4}$Fe$_2$As$_2$.
Almost $H$-linear behavior of $\Delta \gamma(H)$ means that Ba$_{0.6}$K$_{0.4}$Fe$_2$As$_2$ is a full gap superconductor with
$|\Delta_L| \approx |\Delta_S|$. From \cite{mu2009low}. \label{mu_SH}}
\end{figure}

In Fig.\ref{gamma_corrected}, this corrected $\gamma_{tot}$ with core contribution from the result $\gamma_{volovik}$ in Fig.\ref{kappa_theor}(B) is plotted. It shows that for relatively similar gap size cases, such as $|\Delta_S ^0 / \Delta_L ^0 |=$ 0.7 and 0.9, $\gamma(H)$ is very linear in $H$ for a substantial region of fields (up to $\approx H_{c2}/2$) despite a finite $\gamma(H \rightarrow 0)$.
Decreasing the gap size ratio to $|\Delta_S ^0 / \Delta_L ^0 |=$ 0.5, 0.3 and 0.2, the field
dependence of $\gamma(H)$ becomes gradually more concave down.
This behavior is in excellent agreement with the measurements of Ba$_{0.6}$K$_{0.4}$Fe$_2$As$_2$ \cite{mu2009low} ($\approx H$), (Fe$_{0.92}$Co$_{0.08}$)$_2$As$_2$ \cite{gang2010sizable} (sublinear in $H$), (Fe$_{0.955}$Co$_{0.045}$)$_2$As$_2$ \cite{kim2012specific} ($\approx H$),
(Fe$_{0.85}$Co$_{0.15}$)$_2$As$_2$ \cite{kim2012specific} ($\sim \sqrt{H}$), and LaO$_{0.9}$F$_{0.1-\delta}$FeAs \cite{mu2008condensed} ($\sim \sqrt{H}$). Some representative experimental data for (Fe$_{1-x}$Co$_{x}$)$_2$As$_2$ and Ba$_{0.6}$K$_{0.4}$Fe$_2$As$_2$ are shown in Fig.\ref{jskim_SH} and Fig.\ref{mu_SH}.

\vspace{4cm}

\subsection{Summary}
The theoretical discovery of non-trivial Doppler effect (Volovik effect) in the $s^{\pm}$-wave state with the magnetic field $H$ was an unexpected surprise because the Volovik effect was considered as a unique feature of a nodal gap superconductor as in the $d$-wave SC state\cite{volovik1993}. Therefore early experiments of a strong field dependence observed with  $\gamma(H \rightarrow 0)$ and $\kappa(H \rightarrow 0)/T$ in the FeSC were interpreted as strong evidences for a nodal gap SC state in these compounds, while other experimental probes, in particular, the ARPES experiments were clearly indicating for an isotropic $s$-wave full gap superconductor. However, this conflict and puzzle were nicely resolved by the Volovik effect in the $\pm$s-wave SC state\cite{bang2010volovik,wang2011volovik}. In particular, the systematic evolution of the field dependence of $\gamma(H \rightarrow 0)$ and $\kappa(H \rightarrow 0)/T$ as a function of the gap size ratio $R_{gap}=|\Delta_S ^0 / \Delta_L ^0 |$ and its
excellent agreement with experiment of Ba(Fe$_{1-x}$Co$_x$)$_2$As$_2$\cite{tanatar2010thermal,dong2010thermal} not only resolve the experimental puzzle but also it strengthened the validity of the $\pm$s-wave pairing scenario for the FeSC.
Of course, now it is quite certain that a few Fe-based SC compounds indeed have nodes. However, strong field dependencies in $\gamma(H)$ and $\kappa(H)/T$ are not to be understood as a "Hallmark" evidence for a nodal gap superconductor anymore as we have explained in this section. In particular, because the $\pm$s-wave pairing state shows many unexpected nodal gap like behaviors in SC properties although it is nominally a full $s$ gap superconductor, we need more than one piece of experimental evidence -- also needs cross checking for a self-consistency among different data -- in order to confirm a nodal gap superconductor. This issue will be discussed more in the next section.

\section{Penetration Depth}

\subsection{Evolution of $\delta \lambda(T)$ of the $s^{\pm}$-wave state with impurities}
\begin{figure}
\noindent
\hspace{3cm}
\includegraphics[width=70mm]{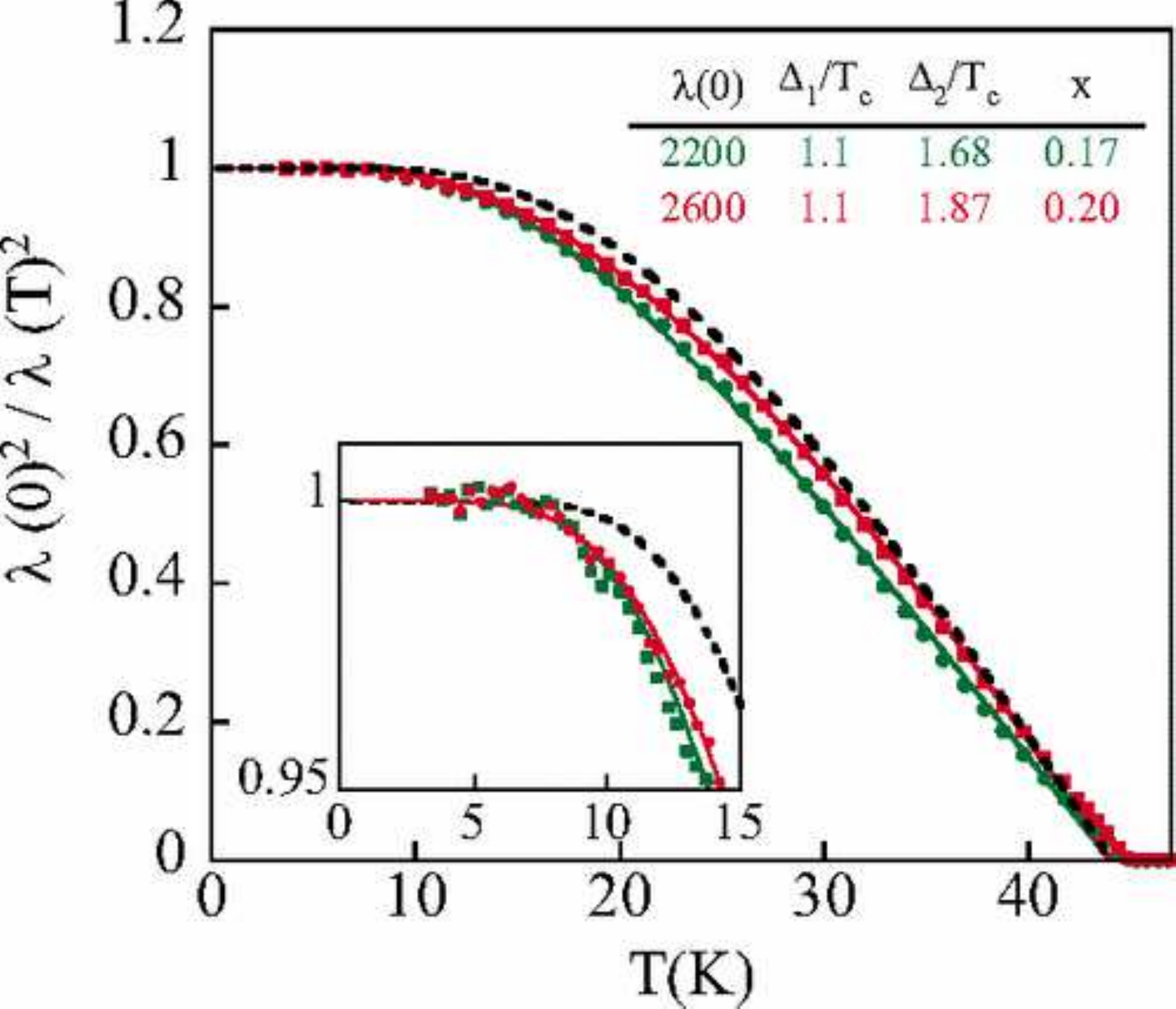}
\vspace{-0.0cm}
\caption{(Color online) Experimental data of the normalized superfluidity density $[\lambda^2(0)/\lambda^2(T)]$ vs $T$ of SmFeAsO$_{0.8}$F$_{0.2}$ (color symbols) and theoretical fittings (color solid lines) with the two-gap $s$-wave superconductor models, and the inset table shows the fit parameters. The black dashed line is the single gap $s$-wave model fitting with $\Delta_0/T_c =1.76$. The inset shows the low temperature data on an expanded scale, making clear the deviation of the black dashed line (single s-wave model) fit from the observed data. From \cite{malone2009magnetic} } \label{malone}
\end{figure}

Perfect diamagnetism (Meissner effect) is the hallmark of superconductivity, therefore magnetic field should decay exponentially inside the superconductor. The typical decay length of the magnetic field is called the penetration depth $\lambda$ and it is a function of temperature. Through the London equation,  it is also related to the definition of superfluidity density $\rho_{s}(T)$ as follows.

\ba
{\bf J} &=& {\bf J_p} + {\bf J_d} = -\frac{1}{4 \pi \lambda^2(T)} {\bf A} =  -\frac{\rho_s (T) e^2}{m} {\bf A}\\
&=&  {\bf K_p}(T) \cdot {\bf A} -\frac{\rho_{tot} e^2}{m} {\bf A},
\ea
where we use the units $c=1$ and $\hbar = 1$. The paramagnetic current kernel ${\bf K_p}(T)$ is the current-current correlation function $<j_a  j_b >_T$ with  ($a,b=x,y,z$) at finite temperature $T$.
At $T=0$, $\lambda(T=0)$ reaches its minimum value and the corresponding $\rho_s (T=0)$ is total electron density $\rho_{tot}$ because ${\bf K_p}(T=0)=<j_a  j_b >=0$.
Therefore, we can interpret the quantity $\frac{1}{\lambda^2(T)}$ as a measure of the reduction of the superfluidity density by quasiparticle thermal excitation as follows,  $\frac{1}{\lambda^2(T)} \sim \rho_s (T) = \rho_{tot} - \rho_{n}(T)$. And we can interpret measurement of the penetration depth $\lambda(T)$ at low temperatures as probing the temperature dependence of the thermally excited quasiparticle density $\rho_{n}(T)$, which is governed by the shape of DOS in SC state $N(\omega)$ (see Fig.3).
As a result, we can predict that the low temperature variation of $\rho_{n}(T)$ is exponentially small for $s$-wave superconductor but it increases as $T$-linear with a nodal gap superconductor.
This expectation is indeed correct, therefore the gap symmetry can be identified by measuring $\lambda(T)$ at low temperatures: $\lambda(T)$ as $T \rightarrow 0$ is flat for $s$-wave superconductor and increases as linear-in-$T$ fashion for a line nodal gap superconductors. Figure \ref{malone} shows the typical data of $1/\lambda^2(T) \sim \rho_s (T)$ in SmFeAsO$_{0.8}$F$_{0.2}$ compound\cite{malone2009magnetic}. It shows the full gap $s$-wave behavior, $\lambda(T)$ flat at low temperatures, consistent with the $s^{\pm}$-wave pairing scenario.

\begin{figure}
\noindent
\hspace{3cm}
\includegraphics[width=100mm]{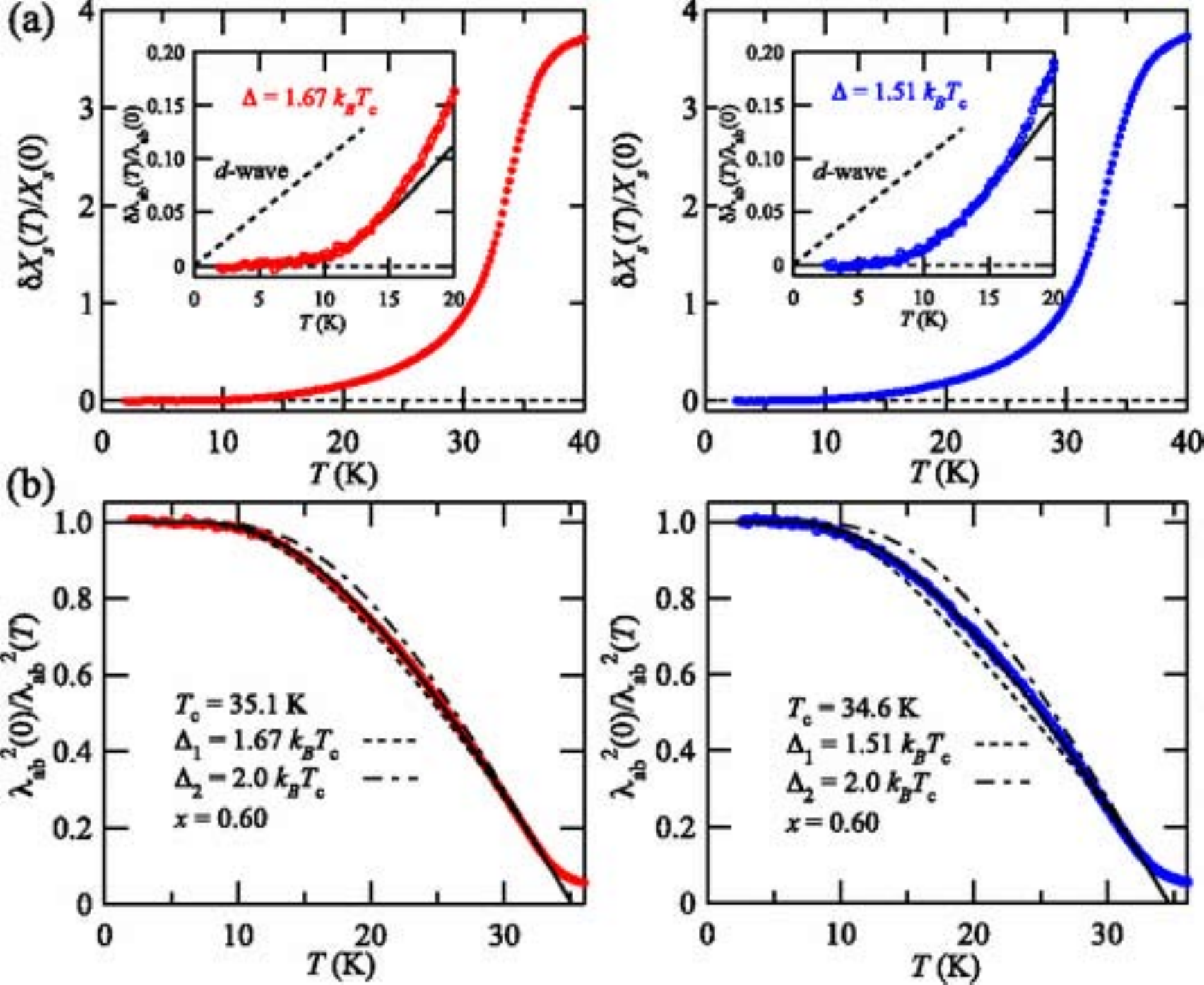}
\caption{(Color online) (a) Experimental data of  $\delta \lambda(T)/\delta \lambda(0)$ of PrFeAsO$_{1-y}$.
(b) Experimental data of  superfluidity density $\lambda^2(0)/\lambda^2(T)$. Solid lines are the fit with the two gap model, and
the dashed and dashed-dotted lines are the single gap model fittings. From \cite{hashimoto2009microwave1111}} \label{pene_exp2}
\end{figure}

With impurity scattering, however, the DOS $N(\omega)$ are modified in its own unique way for different gap states (see Fig.5 and Fig.6). From the above discussion that $\rho_{n}(T) \sim <j_a j_a>_T \sim  <N(\omega)>_T$, we may expect the temperature behavior of $\rho_s (T)$ approximately to a constant minus of Knight shift (see Fig.13) or specific heat (see Fig.21). However, this is not the case for the penetration depth $\lambda(T)$\cite{bang2009superfluid}. The reason is that the relation $<j_a j_a>_T \sim  <N(\omega)>_T$ is only true for the non-interacting quasiparticles. Ref.\cite{bang2009superfluid} showed that the DOS probed by thermodynamic quantities like specific heat and Knight shift and the DOS probed by transport quantities like currents are not exactly the same in the interacting case.
To see this, we write the general formula for the theoretical expression of the superfluidity density in clean limit as\cite{abrikosov2012methods}
\ba
{\bf J}(q) &=& - \frac{\rho_{tot}e^2}{m} K(q,T) {\bf A}, \\
K(q,T) &=& 1+ \pi T\sum_{\omega_n} \int^{\infty}_{-\infty} d \xi \frac{(i \omega_n+\xi_{+})(i \omega_n+\xi_{-}) + \Delta^2}{(\omega_n^2 +\xi_{+}^2+\Delta^2) (\omega_n^2 +\xi_{-}^2+\Delta^2 )}.\label{K_eq1}
\ea
The kernel Eq.[\ref{K_eq1}] is the formula for a $s$-wave superconductor and can be further simplified as\cite{abrikosov2012methods}
\be
K(q=0,T)=\pi T\sum_{\omega_n} \frac{\Delta^2}{(\omega_n^2+\Delta^2)^{3/2}}=2 \int^{\infty}_{0} d \omega f_{FD}(\omega) Re \frac{\Delta^2}{(\omega_n^2+\Delta^2)^{3/2}}
\label{K_eq2}
\ee
The final expression of Eq.[\ref{K_eq2}] indeed can be shown as $1- <N(\omega)>_T$, namely as $\sim (\rho_{tot} - \rho_n(T))$.
We can generalize the above formula to the $s^{\pm}$-wave superconductor with impurity scattering as follows.

\begin{equation}
K(T)=  \sum_{a=h,e} N_{a} \pi T \sum_n v_{a \|} ^2 {\rm Re}
\frac{\tilde{\Delta}_{a}^2}{(\tilde{\omega}_n ^2 +
\tilde{\Delta}_{a}^2)^{3/2}}. \label{K_eq3}
\end{equation}
where $\tilde{\omega}_n$ and $\tilde{\Delta}_{a}$ are the quantities renormalized with the impurity selfenergies $\Sigma_{imp}^{0,1}$ as described in section 3, and $N_{a}$ are the normal state DOS of two bands $h$ and $e$, respectively. This quantity $K(T)$ is directly proportional to the superfluid
density $\rho_s (T)$ and 1/$\lambda_{L} ^2 (T)$ in the London limit.

\begin{figure}
\hspace{1cm}
\includegraphics[width=70mm]{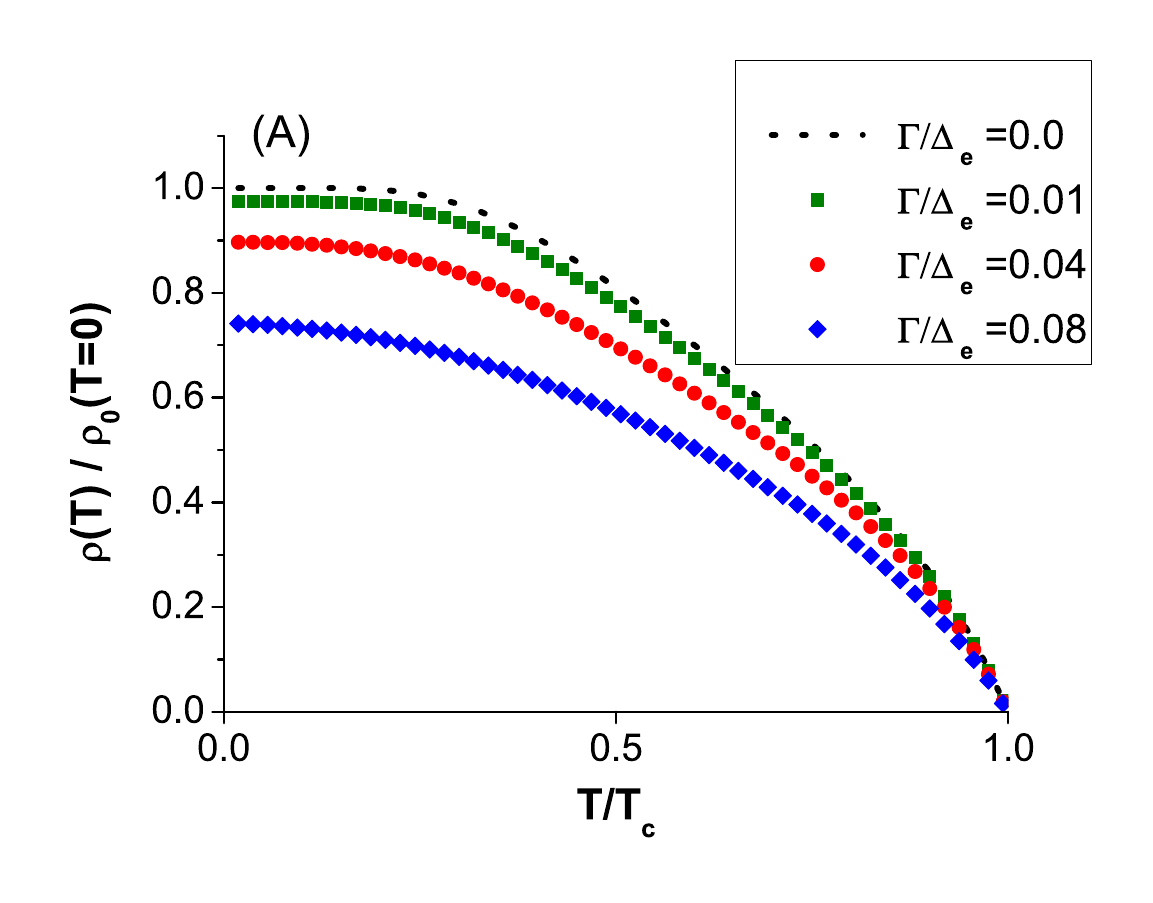}
\includegraphics[width=70mm]{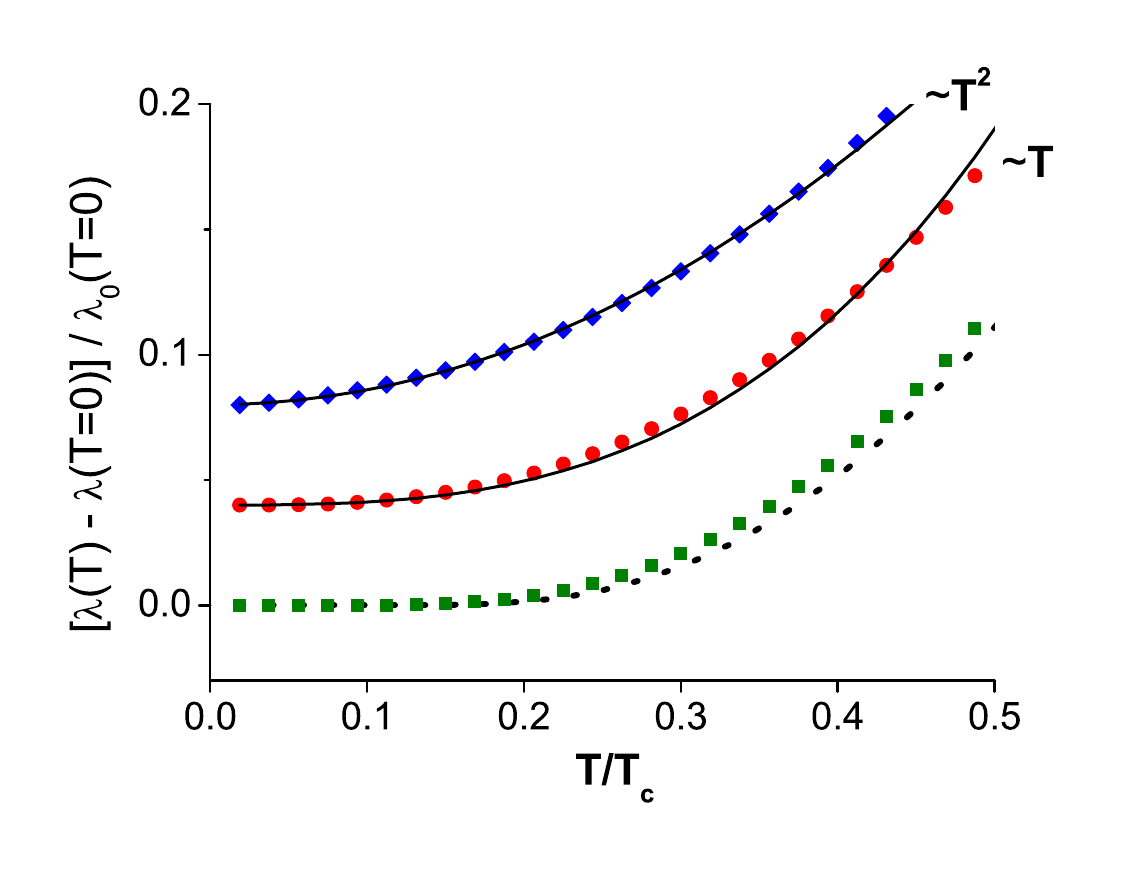}
\caption{(Color online) (a) Superfluid density $\rho_{s}(T)$ for different impurity concentrations, $\Gamma / \Delta_e= 0.0, 0.01,
0.04, 0.08$, normalized by $\rho_s ^0(T=0)$ of the pure state. (b) Corresponding penetration depth $\Delta \lambda(T)=
\lambda(T)-\lambda(T=0)$ normalized by $\lambda_0(T=0)$ of the pure state. The data for $\Gamma / \Delta_e= 0.04, 0.08$ are offset
for clarity. The power law lines (sold black lines) of $T^2$ and $T^3$ are shown for comparison. From \cite{bang2009superfluid}.
\label{pene_theor}}
\end{figure}

Figure \ref{pene_theor} shows the theoretical results of $\rho_s (T)$ and $\lambda_{L}(T)$ for the typical $s^{\pm}$-wave superconductor with varying impurity scattering rates $\Gamma_{imp}/\Delta_e = 0.0, 0.01, 0.04,$ and 0.08, respectively. With these impurity scattering rates, the total DOS $N(\omega)$ systematically changes from a full gap $s$-wave type $\rightarrow$ "V"-shape DOS $\rightarrow$ a dirty limit DOS as shown in Fig.8(B).
The corresponding $\rho_s (T)$ and $\lambda_{L}(T)$ at low temperatures continuously evolves in a sequence of the forms:
exponentially flat $\rightarrow  \propto T^3 \rightarrow \propto T^2$ with increase of impurity concentration.
What is surprising is that the case with the critical impurity scattering rate $\Gamma_{imp}/\Delta_e =0.04$ which has the "V"-shape DOS just as in a clean $d$-wave superconductor displays $\rho_s (T)$, $\lambda_{L}(T) \sim T^3$ (red circles in Fig.\ref{pene_theor}), instead of $\sim T$ as expected in the $d$-wave superconductor. This result tell us that although the DOS $N(\omega)$ looks the same, the dynamically shaped DOS (e.g. by self energy correction) and kinematically shaped DOS (e.g. by the Bogoliubov quasiparticles in $d$-wave superconductor) respond differently for the transport properties.

\begin{figure}
\noindent
\hspace{4cm}
\includegraphics[width=80mm]{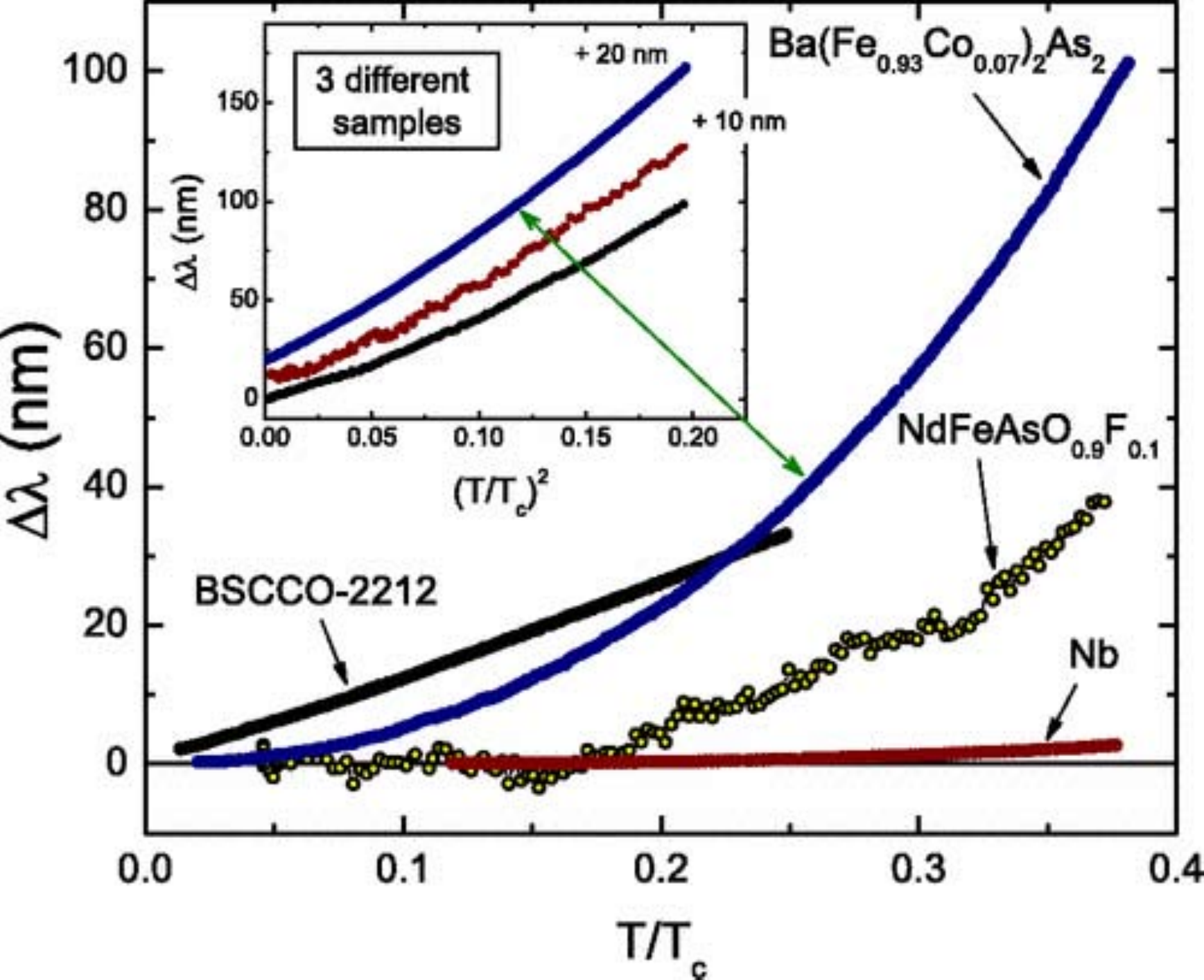}
\caption{(Color online) (a) Experimental data of  $\delta \lambda(T)$ for three different samples of Ba(Fe$_{0.93}$Co$_{0.07}$)$_2$As$_2$. Note the inset plot of $\delta \lambda(T)$ vs $T^2$, in which slight positive curvatures indicate $n > 2$ for $\delta \lambda(T) \sim T^n$, and indeed the best fittings produced the exponent $n \sim 2.15 - 2.42$.  From \cite{gordon2009unconventional}}
\label{gordon_exp}
\end{figure}

These results consistently explain the various temperature dependencies of the experimental data of $M$-1111 ($M$=Pr,Sm,Nd) \cite{malone2009magnetic,hashimoto2009microwave1111,prozorov2009anisotropic} (flat), (Ba,K)Fe$_2$As$_2$ \cite{hashimoto2009microwave122} (flat), $R$FeAsO$_{0.9}$F$_{0.1}$ ($R$=La,Nd) \cite{martin2009nonexponential} ($\propto T^{2}$), and Ba(Fe,Co)$_2$As$_2$
\cite{gordon2009london,gordon2009unconventional,bobowski2010precision} ($\propto T^{2-2.5}$). More recent experiments with intentional electron irradiation on (Ba,K)Fe$_2$As$_2$ also confirmed the predicted evolution of power law $\lambda_{L}(T) \sim T^{\alpha}$ with increasing dirtiness\cite{cho2014effects}.
Vorontsov et al. \cite{vorontsov2009superfluid} have performed the theoretical studies on the same problem, and obtained a similar result $\rho_s(T)\propto T^2$ for high concentration of impurities but obtained a different result $\rho_s(T)\propto T^{1.6}$ for the critical impurity concentration. This difference arises from the different methods of studying the impurity scattering effects -- weak coupling theory \cite{vorontsov2009superfluid} and strong coupling theory \cite{bang2009superfluid} -- when calculating the expression $K(T)$ above.

\subsection{Possible nodal gap evidence: $\delta \lambda(T) \sim T$}
As discussed above, most of the Fe-based SC compounds display the temperature dependence of penetration depth $\delta \lambda(T)$ as either exponentially flat or high power $n >2$ in $\delta \lambda(T) \sim T^n$, consistent with the $s^{\pm}$-wave gap state.
However, there exist a few Fe-based SC compounds which show quasi-linear-in-$T$ behavior down to very low temperatures in $\delta \lambda(T)$, which is a kind of hallmark evidence for a nodal gap superconductor\cite{hardy1993precision}. These are LaFePO \cite{fletcher2009evidence,hicks2009evidence}, BaFe$_2$(As$_{0.67}$P$_{0.33}$)$_2$ \cite{hashimoto2010line}, and KFe$_2$As$_2$ \cite{hashimoto2010evidence}.

\begin{figure}
\noindent
\hspace{3cm}
\includegraphics[width=80mm]{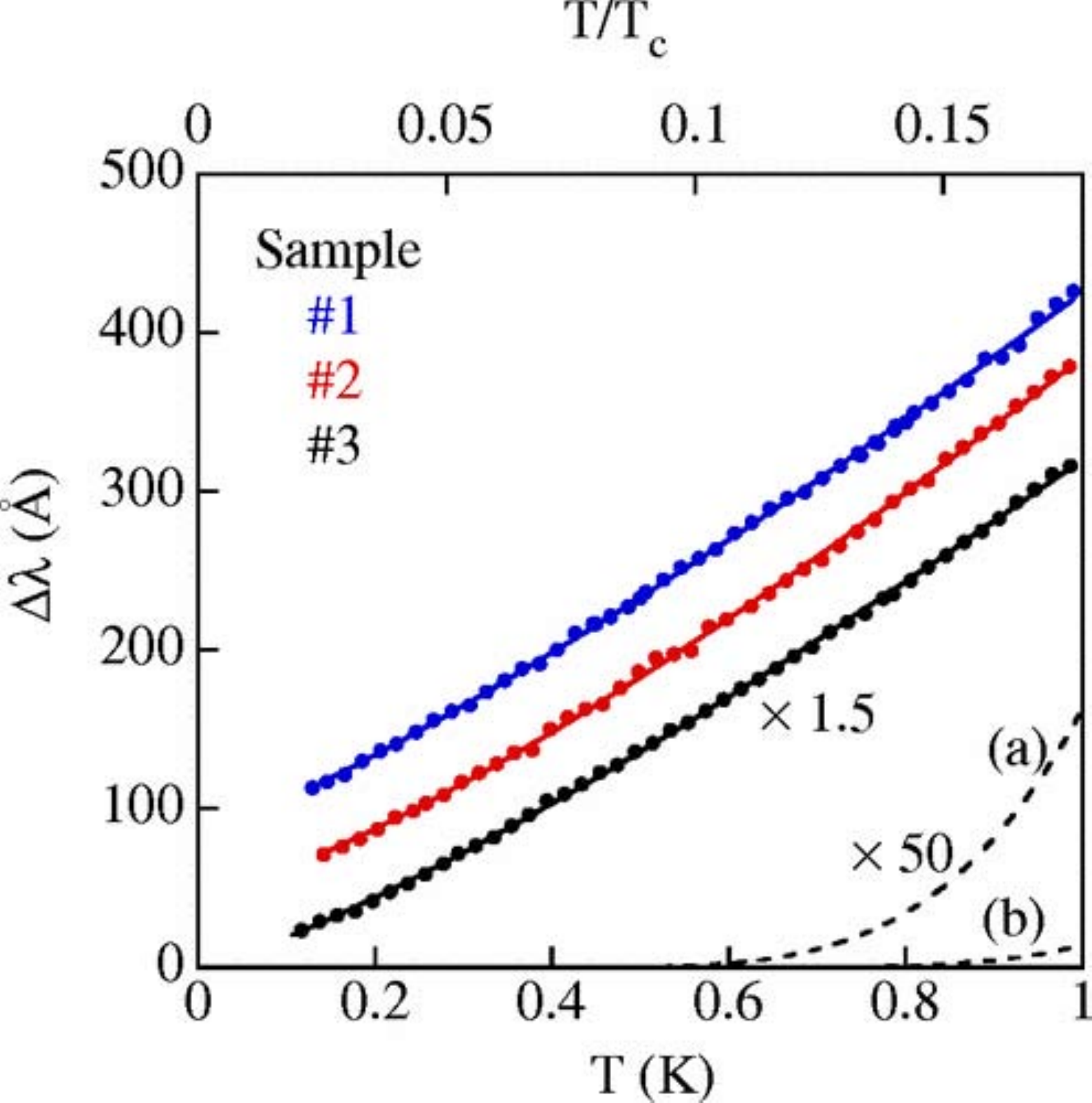}
\caption{(Color online) Experimental data of  $\delta \lambda(T)$ in three single crystals of LaFePO. Solid lines are power-law fits giving an exponent of $1.2 \pm 0.1$.  From \cite{fletcher2009evidence}} \label{LFPO_exp}
\end{figure}

As the data of $\delta \lambda(T)$ for LaFePO shows in Figure\ref{LFPO_exp}, these three compounds display the temperature dependence of $\delta \lambda(T)$ quite close to linear in $T$. Although there are always some uncertainties to determine the power-law exponent due to the uncertainty of the absolute value $\lambda(T=0)$, all three compounds produced the exponent $n \sim 1.1-1.2$. These values of power-law exponent are not compatible with the full-gap $s^{\pm}$-wave pairing state unless one of the gap value $\Delta_s$ is extremely small and carrying a substantial DOS\cite{bang2012there}. Therefore, it is reasonable to suspect that theses three Fe-based SC compounds indeed possess line nodes in their gap functions. However, even if there exist line nodes in these Fe-based SC compounds, it can still be understood as a smooth evolution of the full-gap $s^{\pm}$-wave state to the $s+g$-wave gap as depicted in Fig.1(C), and does not imply a qualitatively new pairing mechanism.

However, in order to confirm the existence of the line nodes, independent experimental evidences, other than the penetration depth, need to be  tested. Careful cross-checking analysis with the penetration depth $\lambda(T)$ and the thermal conductivity $\kappa(T,H)$ was carried out in Ref.\cite{bang2012there}. The main check point was that the $T$-linear behavior of  $\lambda(T)$ down to $T/T_c < 0.05$ \cite{fletcher2009evidence,hicks2009evidence,hashimoto2010line,hashimoto2010evidence} implies a nodal gap but in extremely clean limit;  it is well known that a tiny amount of impurity would immediately change $\lambda(T)$ from $T$-linear to $T^2$-behavior for a nodal gap superconductor\cite{hirschfeld1993effect}. Quantitative estimate of the impurity scattering rate $\Gamma_{imp}/\Delta_0$ compatible with the measured $\lambda(T)$ -- if it is assumed from a nodal gap  -- of these three compounds can be extracted from the data, and it was shown to be as clean as $\Gamma_{imp}/\Delta_0 < 0.02$ \cite{bang2012there}. Then the Ref.\cite{bang2012there} cross checked with the thermal conductivity data $\kappa(T=0,H)$ whether these three Fe-based SC compounds are indeed in such clean limit.

In a nodal gap superconductor, it was well known that the $\kappa(T=0,H \rightarrow 0)$ obtains a universal value regardless of the amount of impurity because of the cancelation between the impurity induced DOS $\rho_{imp} \sim \Gamma_{imp}$ at zero frequency and the relaxation time of the quasiparticle due to the same impurity scattering $\tau_{imp} \sim 1/\Gamma_{imp}$: then themal/electric conductivity obtains a universal value as $\sim \rho_{imp} \cdot \tau_{imp} \sim const.$\cite{lee1993localized,graf1996electronic}, independent of the impurity scattering rate. Hence the values of $\kappa(T=0,H \rightarrow 0)$ cannot tell us about the dirtiness of the superconducting samples. However, if normalized by the normal state value of $\kappa_n$, the value $\kappa_s(T=0,H \rightarrow 0)/\kappa_n \approx \Gamma_{imp}/\Delta_0$ becomes an excellent measure of the dirtiness of the nodal gap SC samples\cite{bang2012there}.

\begin{figure}
\noindent
\hspace{2cm}
\includegraphics[width=120mm]{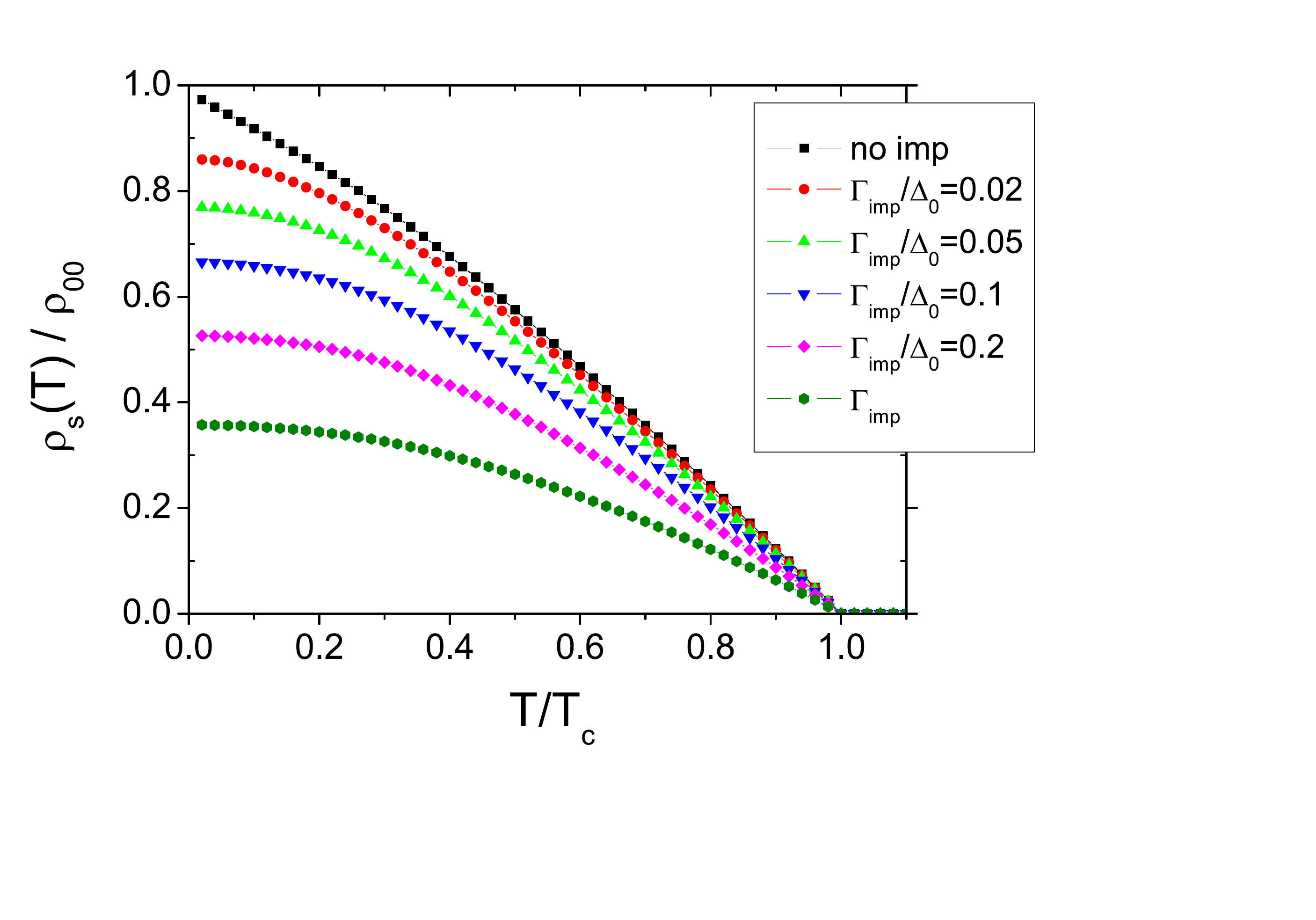}
\vspace{-2cm}
\caption{(Color online)  Normalized superfluid density $\rho_S(T)$ vs $T/T_c$ of the $d$-wave SC state for various impurity
concentrations $\Gamma_{imp}/\Delta_0= 0, 0.02, 0.05, 0.1, 0.2,$ and 0.4 (unitary impurity). $2 \Delta_0 /T_c =4$ is used.
From \cite{bang2012there}.} \label{d-wave_pene}
\end{figure}

Using this criterion, Ref.\cite{bang2012there} concluded that only KFe$_2$As$_2$ is compatible with a clean nodal gap superconductor, but LaFePO and BaFe$_2$(As$_{0.67}$P$_{0.33}$)$_2$ compounds are in fact in extremely dirty limit having $\Gamma_{imp}/\Delta_0 \approx 0.4$ from the normalized thermal conductivity data $\kappa_s(T=0,H \rightarrow 0)/\kappa_n \approx \Gamma_{imp}/\Delta_0$, so that the observed $T$-linear $\lambda(T)$ with these compounds cannot be understood  with a nodal gap superconductor with this much impurity scattering rate $\Gamma_{imp}/\Delta_0 \approx 0.4$; for a nodal gap scenario to be compatible with the experimental $T$-linear $\lambda(T)$, the sample's dirtiness should be less than $\Gamma_{imp}/\Delta_0 \approx 0.02$ (see Fig.\ref{d-wave_pene}).
Indeed for KFe$_2$As$_2$, laser ARPES experiment independently confirmed that there exists eight nodal points ($A_{1g}$-nodes) around the middle hole pocket \cite{okazaki2012octet}.
As for the issue of the possible nodal gap in LaFePO and BaFe$_2$(As$_{0.67}$P$_{0.33}$)$_2$, not to bias the readers, we would like to remark that the above description and conclusions are only one viewpoint and there are many active researchers who have confidence to interpret their data as strong evidences for the nodal gap SC states in LaFePO\cite{fletcher2009evidence,hicks2009evidence}, BaFe$_2$(As$_{0.67}$P$_{0.33}$)$_2$\cite{yamashita2011nodal,hashimoto2012sharp}, and also in Ba(Fe$_{1-x}$Co$_x$)$_2$As$_2$\cite{reid2010nodes}. Therefore the possible nodal gap issue with some of FeSCs is still not completely settled.

\subsection{Summary}
The $s^{\pm}$-wave pairing model is consistent with the temperature dependence of the penetration depth $\lambda(T)$ of most of the Fe-based SC compounds.  Having the sign-changing OPs, this full-gap superconductor quickly develops in-gap state with impurities (magnetic and non-magnetic), which then causes the systematic evolution of the temperature dependence of  $\lambda(T)$ in a sequence of the forms: exponentially flat $\rightarrow  \propto T^3 \rightarrow \propto T^2$ with increasing impurity concentration. This theoretical prediction is in excellent accord with the various temperature dependencies of the experimental data of $M$-1111 ($M$=Pr, Nd, Sm) \cite{malone2009magnetic,hashimoto2009microwave1111,martin2009nonexponential} (flat), (Ba,K)Fe$_2$As$_2$ \cite{hashimoto2009microwave122} (flat), and Ba(Fe,Co)$_2$As$_2$ \cite{gordon2009london,gordon2009unconventional,bobowski2010precision} ($\propto T^{2-2.5}$) and many others\cite{cho2014effects}.
Finally, there are a few Fe-based SC compounds showing $T$-linear $\lambda(T)$, hence appearing not compatible with the full-gap $s^{\pm}$-wave pairing model. In particular, the nodal gap possibility with LaFePO and BaFe$_2$(As$_{0.67}$P$_{0.33}$)$_2$ compounds is strongly supported with accumulated experiments, hence this issue is not yet settled. However, even if these compounds are confirmed to be a nodal gap superconductor, that does not necessarily imply that a qualitatively different pairing mechanism other than the $s^{\pm}$-wave pairing model is realized in these compounds.

\section{$T_c$ suppression with impurities in the $s^{\pm}$-wave state}

\begin{figure}
\noindent
\hspace{2cm}
\includegraphics[width=65mm]{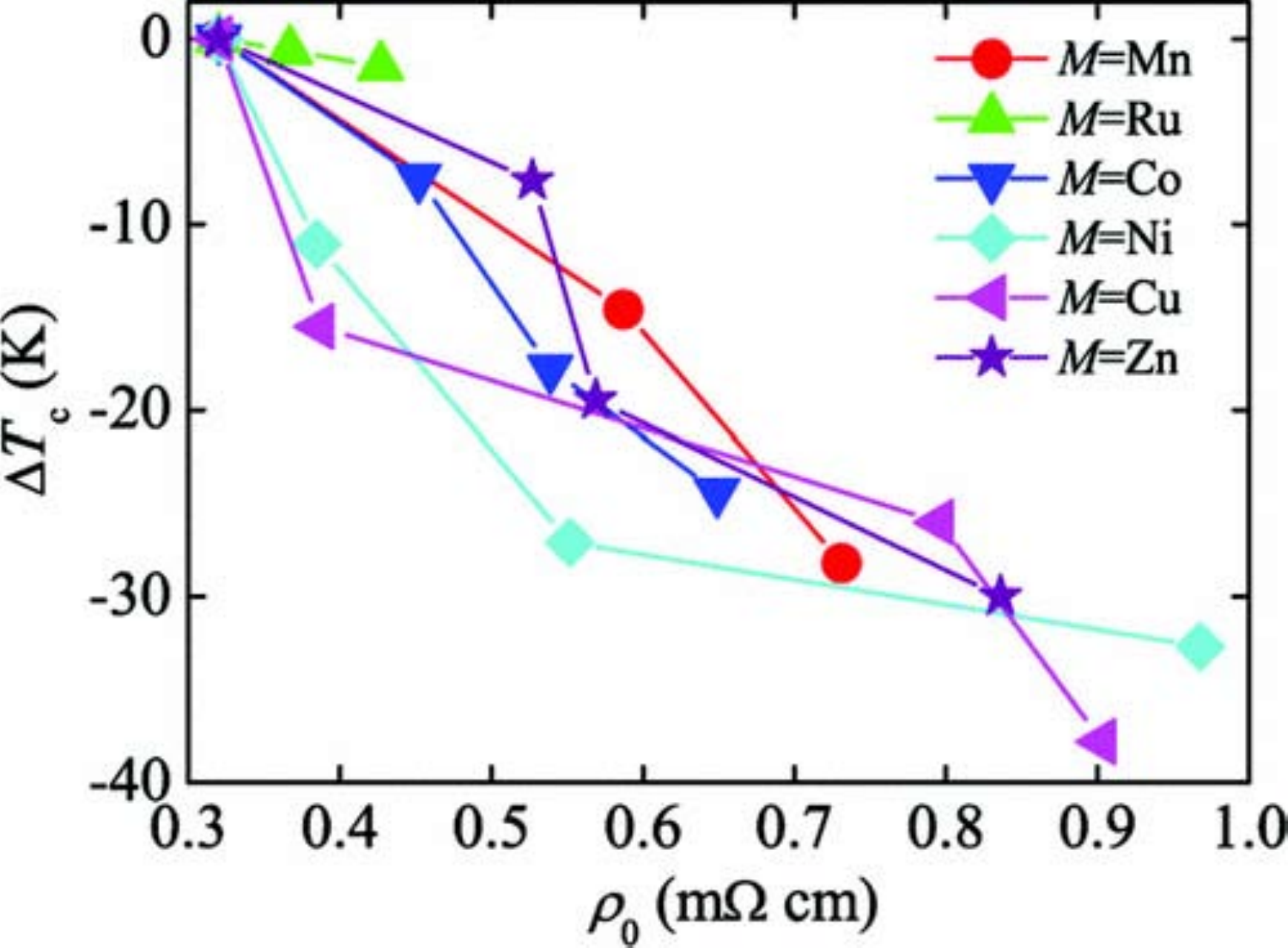}
\includegraphics[width=65mm]{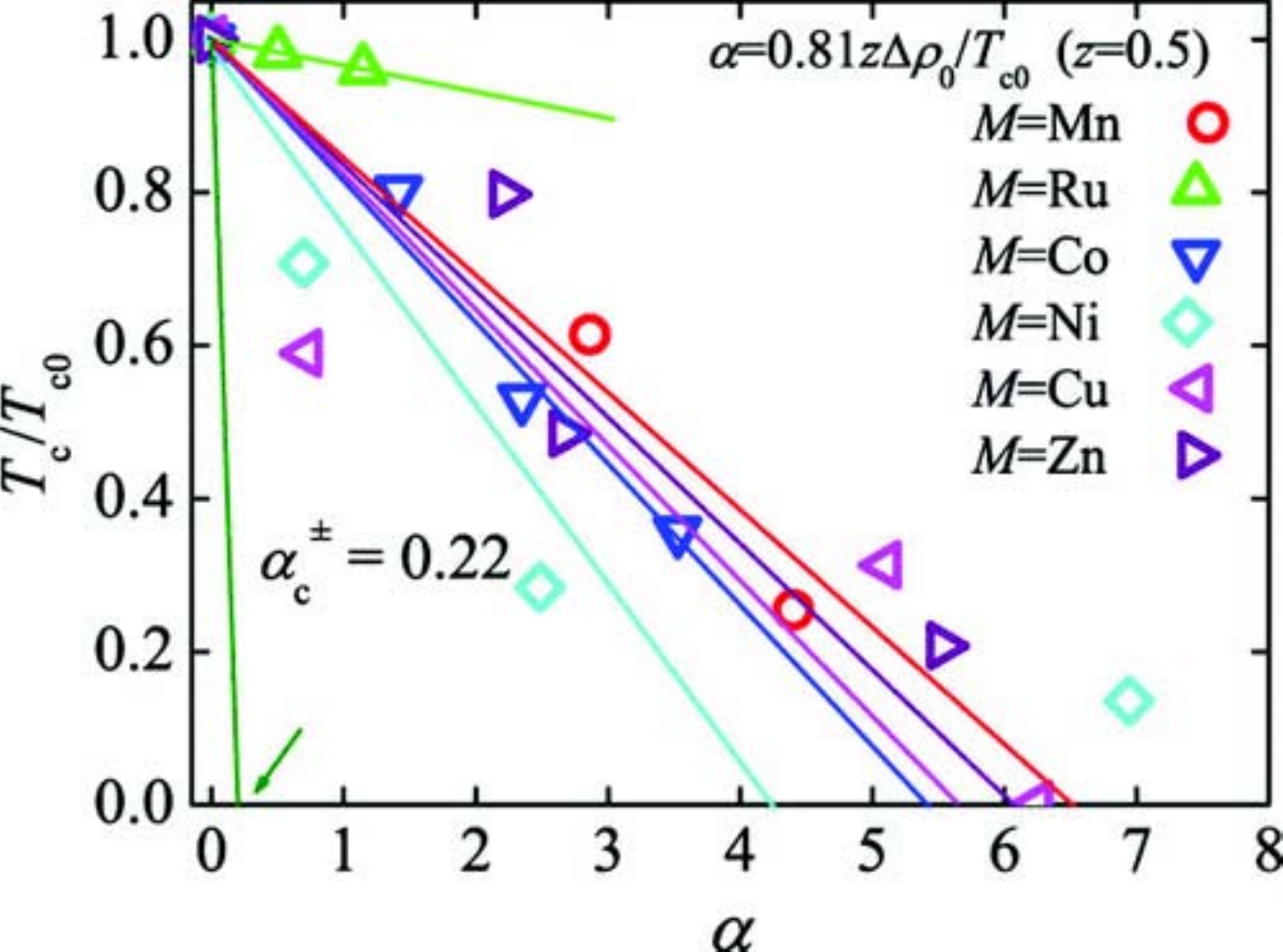}
\caption{(Color online) (Left) $T_c$-reduction, $\Delta T_c$, as a function of residual resistivity ($\rho_0$) for the superconductors Ba$_{0.5}$K$_{0.5}$Fe$_{2-2x}M_{2x}$As$_{2}$ ($M$ = Mn, Ru, Co, Ni, Cu, and Zn); (Right) The same data plotted as normalized $T_c/T_c^0$ vs dimensionless pair-breaking parameter $\alpha=0.88 z \Delta \rho_0 /T_c^0$.
From \cite{li2012superconductivity}.} \label{Tc_Li_exp}
\end{figure}

The $s^{\pm}$-wave SC state, having the sign-changing OPs $\Delta_h$ and $\Delta_e$, is expected to have a similar $T_c$-suppression rate with pointlike defects as in the $d$-wave case\cite{bang2009imp,onari2009violation}. In early period, this theoretical prediction of the fast $T_c$-suppression appeared inconsistent with the experimental observations\cite{li2012superconductivity,sato2009studies,matsuishi2009effect,han2009superconductivity}, where these experiments introduced various transition metal elements (Mn, Co, Ni, Cu, Zn, Ru) substituting the Fe sites in various Fe-1111 and Fe-122 compounds. Direct doping on the Fe-sites with transition metals is expected to introduce strong random  potentials onto the Fe-As plane, and expected to suppress $T_c$ fast.
However, the above mentioned experiments show very slow decay of $T_c$  (e.g. see Fig.\ref{Tc_Li_exp}), often an order of magnitude slower than the theoretical prediction with the pointlike strong impurity potentials on he $s^{\pm}$-wave pairing state\cite{bang2009imp,onari2009violation}.

To resolve this discrepancy, two options were attempted. The first one was pursued by Kontani and coworkers\cite{kontani2010orbital,onari2012self} who claimed that this is the evidence that the $s^{\pm}$-wave model is not compatible with FeSCs and proposed the $s^{++}$-wave model as the pairing state of FeSCs, instead, which immediately predicts a slow or no $T_c$-suppression due to the Anderson's theroem\cite{anderson1959}. The second option is to try to understand the slow $T_c$-suppression experiments within the $s^{\pm}$-wave model.
The first option, the $s^{++}$-wave state, is a quick solution to explain the slow $T_c$-suppression. However this option created many more new problems which needed separate resolutions with specific mechanisms and respective fine tunings: (1) first of all, as to the $T_c$-suppression, this model is not compatible with the almost equal $T_c$-suppression rates with magnetic (Mn) and non-magnetic impurities\cite{li2012superconductivity}; (2) this model has to invent all the specific theories to explain other SC properties of FeSCs such as NMR, penetration depth, neutron resonance, etc., which were naturally explained with the $s^{\pm}$-wave model. In this review, we will focus on the second option how the seemingly slow $T_c$-suppression observed in experiments can be understood with the $s^{\pm}$-wave model. This approach is mainly pursued by Hirschfeld and coworkers\cite{wang2013using,prozorov2014effect} and we follow the main results of their recent paper\cite{prozorov2014effect}.

\subsection{Pointlike impurities: $U^{imp}_{intra-band}= U^{imp}_{inter-band}$}

\begin{figure}[t]
\noindent
\hspace{3cm}
\includegraphics[width=60mm]{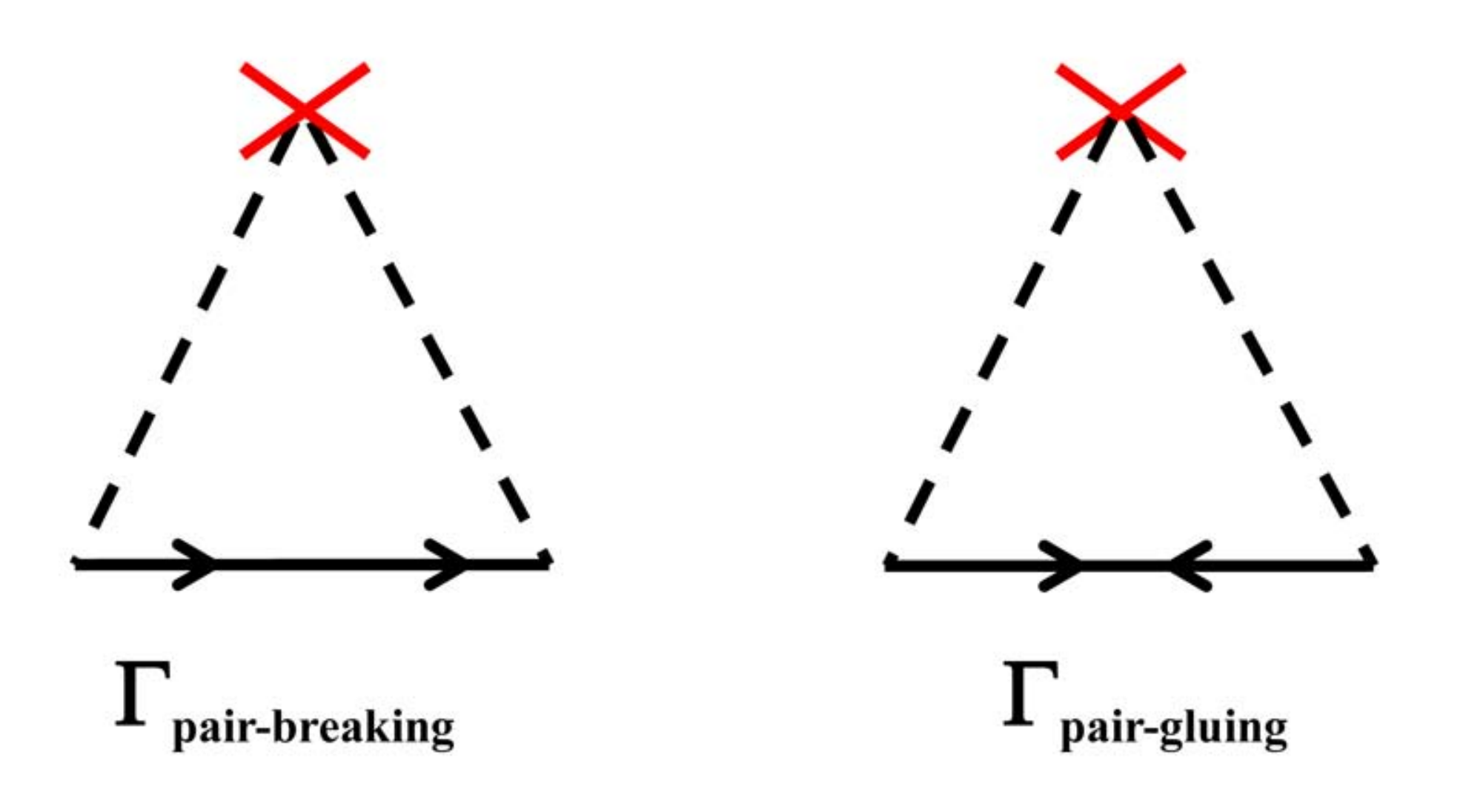}
\caption{(Color online) Schematic diagrams of impurity scattering processes in SC state in leading order (Born approximation): (Left) pair-breaking (normal) scattering, and
(Right) pair-gluing (anomalous) scattering.} \label{imp_scatt}
\end{figure}

To study the $T_c$-suppression in SC state by impurities, the key concept is to distinguish two different processes: pair-breaking ($\Gamma_{pb}$) and pair-gluing ($\Gamma_{pg}$)  scatterings\cite{abrikosov1960contribution}; their leading order processes are shown in Fig.\ref{imp_scatt}. The physical meaning of these processes is that the impurity scattering is not always acting as a pair-breaker ($\Gamma_{pb}$) but it also act as a pair-gluing interaction ($\Gamma_{pg}$) depending on the gap function $\Delta(k)$. For the two band $s^{\pm}$-wave state, the impurity potentials can be conveniently parameterized with two potentials, $U^{imp}_{intra-band}$ and $ U^{imp}_{inter-band}$, and the two scattering rates are calculated as
\ba
\Gamma_{pb} &=& \Sigma^{0}_h(\omega_n) + \Sigma^{0} _e(\omega_n) \\
\Gamma_{pg} &=& \Sigma^1 _{h} (\omega_n) + \Sigma^1 _{e} (\omega_n).
\ea
where $\Sigma^{0,1} _{h,e}$ are defined in Eq.(13)-(16). Using the Born approximation as depicted in Fig.\ref{imp_scatt}, a simplification occurs for the calculations of $\Sigma^{0,1} _{h,e}$ and the two scattering rates are given, in the limit of $T \rightarrow T_c$  ($\Delta_{h,e} \rightarrow 0$), as follows.

\ba
\Gamma_{pb} &=& \Gamma_{imp}~ sgn(\omega_n) = \Gamma_{imp}  \frac{\omega}{|\omega|}, \label{pb}\\
\Gamma_{pg} &=& \Gamma_{imp} \frac{[\tilde{N}_h <\Delta_h> + \tilde{N}_e <\Delta_e>]}{|\omega|}. \label{pr}
\ea
where $\tilde{N}_a = N_a/N_{tot}$, the normalized DOS for band $a$, and $\Gamma_{imp}=n_{imp} \pi N_{tot} U^2_{imp}$.
Once $\Gamma_{pb}$ and $\Gamma_{pg}$ are calculated, the final $T_c$ suppression is written as \cite{abrikosov1960contribution}
\be
\ln{\frac{T_{c0}}{T_c}} = \psi(\frac{1}{2}+\frac{\rho}{2}) - \psi(\frac{1}{2})
\ee
where $\rho=\Gamma^{eff}_{pb}/\pi T_c$ with $\Gamma^{eff}_{pb} = \Gamma_{pb}- \Gamma_{pg}$ and for small scattering limit ($\Gamma^{eff}_{pb} < T_c$), we have
\be
T_c = T_{c0} - \frac{\pi}{4}\Gamma^{eff}_{pb}.
\ee

\noindent
For a $s$-wave superconductor, $\Gamma^{eff}_{pb}=0$ because $\Gamma_{pb}=\Gamma_{pg}$, hence the $T_c$-suppression becomes zero, consistent with Anderson's theorem\cite{anderson1959theory}. For the $s^{\pm}$-wave state, because of the sign-changing OPs, the $\Gamma_{pg}$ in Eq.(46) becomes almost zero because $[\tilde{N}_h <\Delta_h(k)>_{FS} + \tilde{N}_e <\Delta_e(k)>_{FS}] \sim 0$ (almost but not exactly zero)\cite{bang2008possible}, hence $\Gamma^{eff}_{pb} \approx \Gamma_{pb}$ and the $T_c$-suppression becomes maximum as in the $d$-wave case; the maximum $T_c$-suppression in the $d$-wave occurs because of the exactly same mechanism as $N_0<\Delta_{d-wave}(k)>_{FS} =0$.

\begin{figure}
\noindent
\hspace{1cm}
\includegraphics[width=80mm]{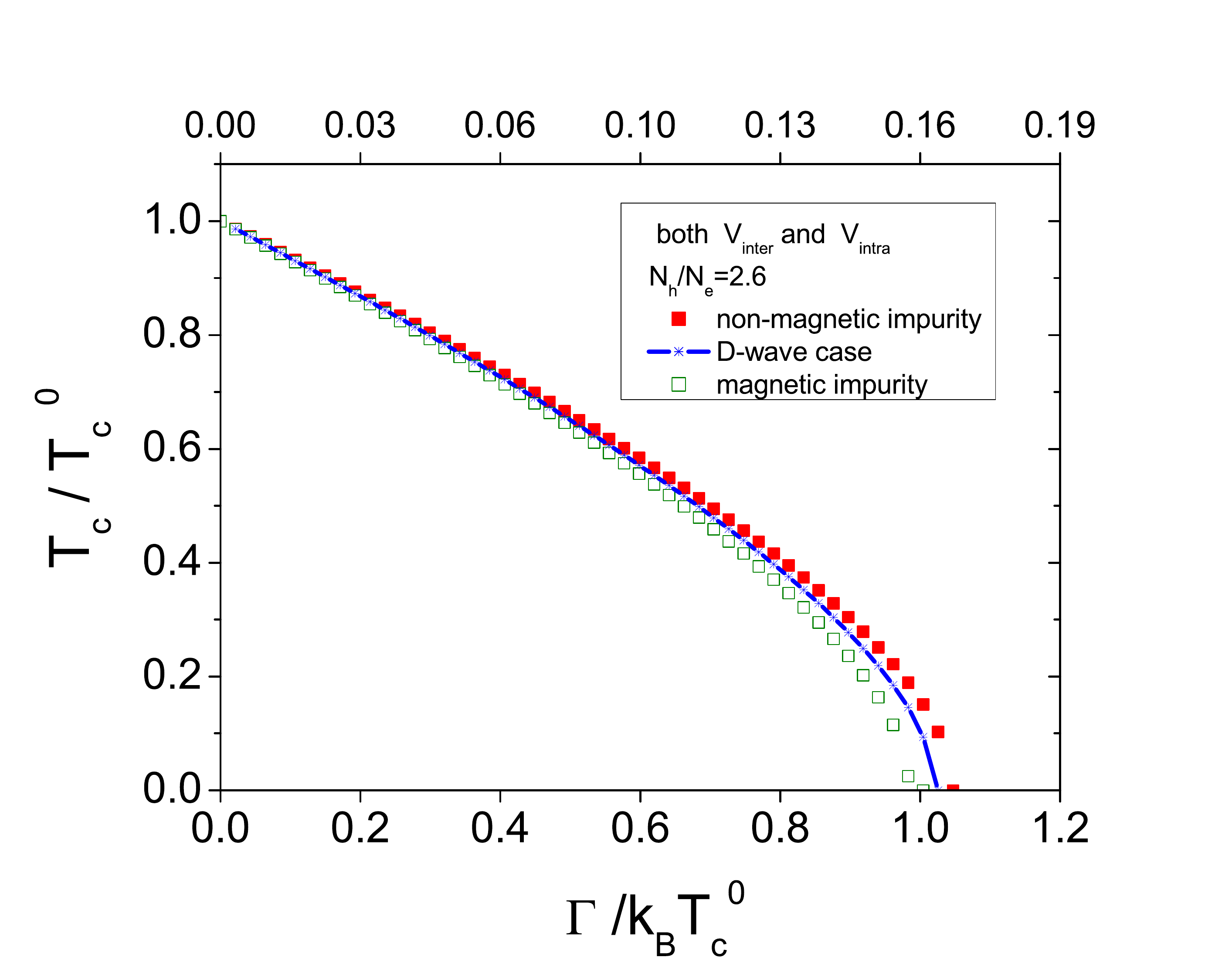}
\includegraphics[width=80mm]{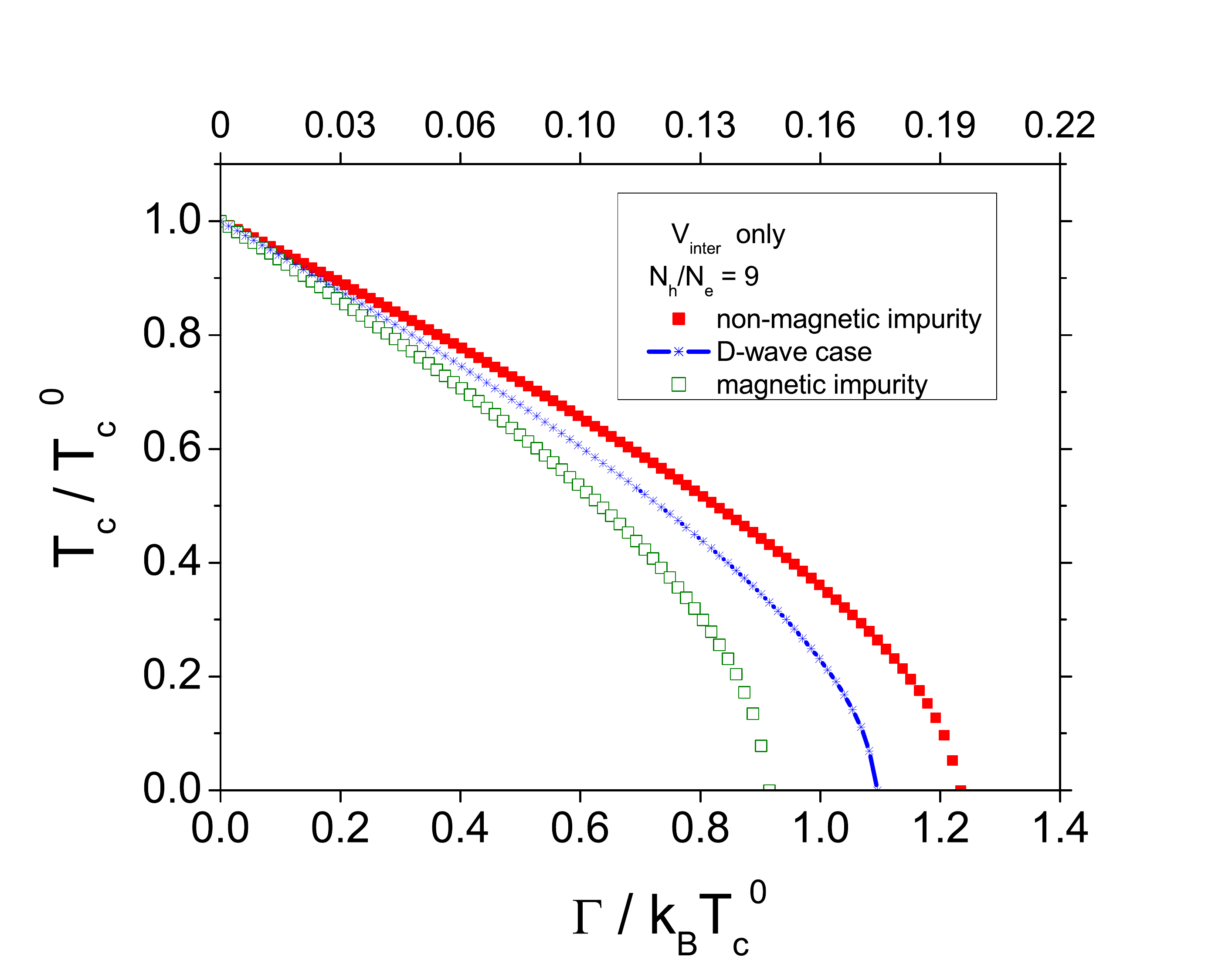}
\caption{(Color online) (Left) Normalized critical temperature $T_c /T_c^0$ vs normalized impurity scattering rate $\Gamma_{imp}/k_B T_c ^0$ for the $s^{\pm}$-wave and $d$-wave superconductors with pointlike unitary ($c=0$) scatterers. The calculations are with both inter- and intra-band pairing interactions, and with $N_h /N_e \approx 2.6$. (Right) The same calculations with $N_h /N_e \approx 9$ and the inter-band pairing interaction only. From \cite{bang2009imp}} \label{Tc_imp}
\end{figure}

Figure \ref{Tc_imp} shows the numerical results of $T_c /T_c^0$ vs $\Gamma_{imp}/k_B T_c ^0$ of the $s^{\pm}$-wave model. Indeed, the $T_c$-suppression rates for the $s^{\pm}$-wave and $d$-wave states are almost equal as expected. The difference between the magnetic and non-magnetic impurities is also negligible, demonstrating the maximum pair-breaking effect of ordinary non-magnetic potential scatterers in the sign-changing OP superconductors. In the righthand panel, the same calculations were done with an extreme DOS ratio of $N_h /N_e \approx 9$, possibly realized with heavily overdoped FeSCs either by holes or electrons. The $T_c$-suppression rates change only by about 20$\%$.  To facilitate comparison with experimental data, the top $x$-axes of Fig.\ref{Tc_imp} are marked with commonly used dimensionless pair-breaking parameter $g_p$\cite{prozorov2014effect} $ = \alpha$\cite{li2012superconductivity} $= \Gamma_{imp}/[2 \pi k_B T_c^0]$. The critical impurity scattering rate is shown to be $g_p ^c =\alpha_c \sim 1/2 \pi  \approx 0.16$.   Li et al. \cite{li2012superconductivity} has calculated the 5 orbital (5 bands) $s^{\pm}$-wave model to find a value of $\alpha_c \approx 0.22$, which is a similar parameter as $g_p ^c$ and defined as $\alpha=0.88 z \Delta \rho_0 /T_c^0$.
As shown in Fig.\ref{Tc_Li_exp}, the comparison between theory and experiments shows that the $s^{\pm}$-wave model with {\it pointlike impurities} ($U^{imp}_{intra-band}= U^{imp}_{inter-band}$) has definitely much faster $T_c$-suppression rate than the experimental data of real FeSCs.
However, it should be noted that the chemical doping experiments as in \cite{li2012superconductivity,sato2009studies,matsuishi2009effect,han2009superconductivity}, where the impurities are introduced by doping with various transition metals (Mn, Co, Ni, Cu, Zn, Ru), have many unknown parameters and effects which will affect $T_c$ by other than impurity potential itself such as change of carrier density, change of pairing interactions, etc. Therefore, it is more ideal to compare the theoretical results with the irradiation experiments such as proton\cite{nakajima2010suppression,taen2013pair}, $\alpha$-particle\cite{tarantini2010suppression}, and electron irradiations\cite{van2013electron,prozorov2014effect}.

\begin{figure}
\noindent
\hspace{3cm}
\includegraphics[width=70mm]{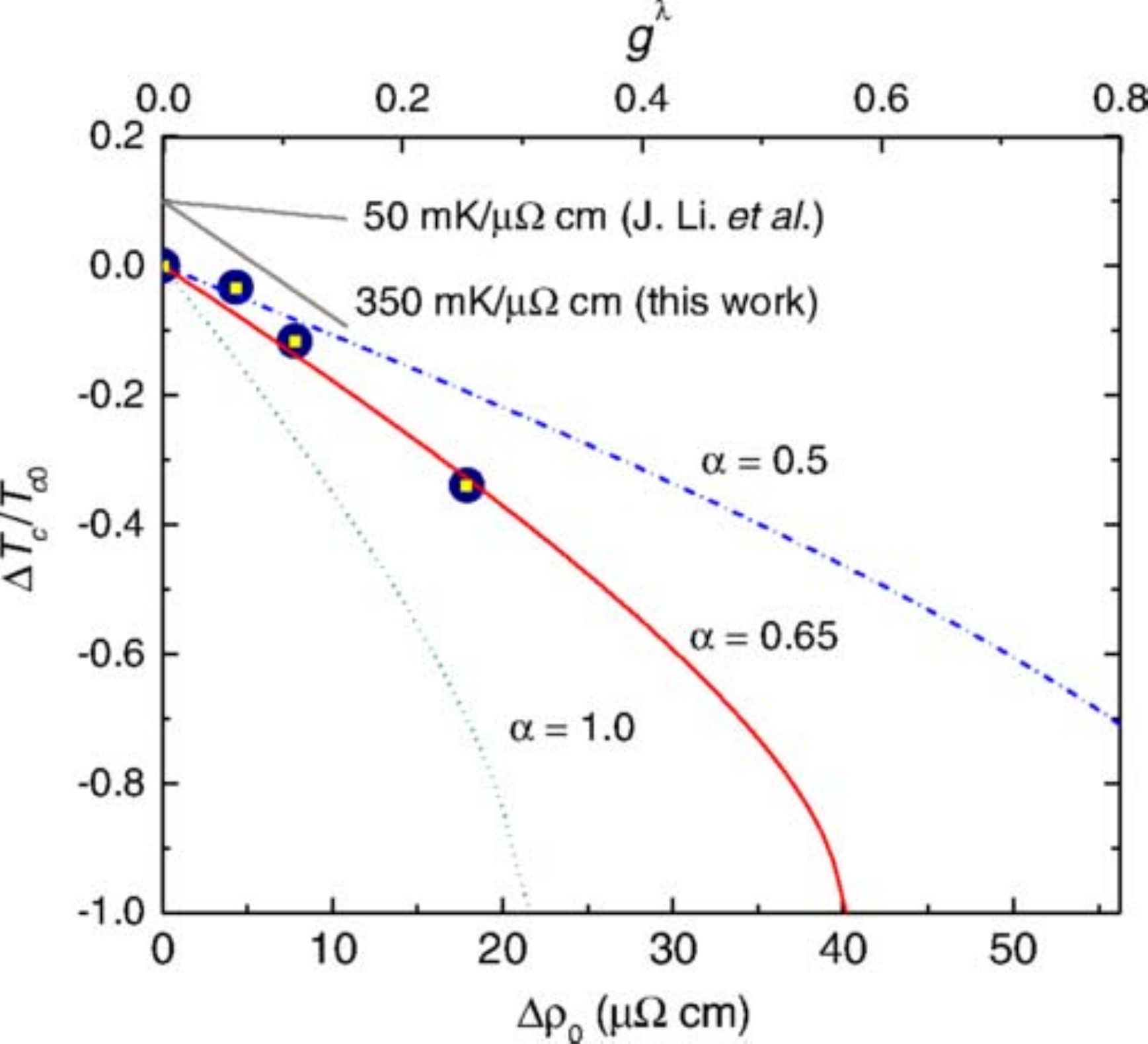}
\caption{(Color online) Data of $\Delta T_c/T_{c0}$ versus $\Delta \rho_0$. Symbols are experimental data from Ba(Fe$_{1-x}$Ru$_x$)$_2$As$_2$ ($x = 0.24$) with electron irradiation. Solid curve lines are theoretical results with different $\alpha= U^{imp}_{inter}/U^{imp}_{intra}=1.0, 0.65$, and 0.5. From \cite{prozorov2014effect}} \label{prozorov_2014}
\end{figure}

\subsection{Finite-size impurities: $U^{imp}_{intra-band} > U^{imp}_{inter-band}$}

In order to resolve the above discrepancy between theory and experiments, Hirschfeld and coworkers \cite{prozorov2014effect,wang2013using} have invoked finite ranged impurity potentials, i.e. $U^{imp}_{intra-band} > U^{imp}_{inter-band}$, which is more realistic for impurities having a finite size.
It is intuitively obvious that if $U^{imp}_{intra-band} > U^{imp}_{inter-band}$, the $T_c$-suppression of the $s^{\pm}$-wave model should become much slower because the pair-gluing impurity scattering rates $\Gamma_{pg}$ of Eq.(46) rapidly increase to finite values as
\ba
\Gamma_{pg,h} &=& \frac{[\Gamma_{intra} \tilde{N}_h <\Delta_h> + \Gamma_{inter} \tilde{N}_e <\Delta_e>]}{|\omega|} \nonumber \\
\Gamma_{pg,e} &=& \frac{[\Gamma_{intra} \tilde{N}_e <\Delta_e> + \Gamma_{inter} \tilde{N}_h <\Delta_h>]}{|\omega|}. \label{pg_2}
\ea
where $\Gamma_{intra, inter}= (U^{imp}_{intra, inter})^2\pi N_{tot} n_{imp}$ are the inter- and intra-band impurity scattering rates in Born approximation. Then the effective pair-breaking rate $\Gamma^{eff}_{pb} = \Gamma_{pb}- \Gamma_{pg}$ also rapidly decreases, hence the $T_c$-reduction is also reduced according to Eq.(47) or Eq.(48).

Prozorov et al.\cite{prozorov2014effect} have performed the systematic calculations of $T_c$ of the two band $s^{\pm}$-wave model with $U^{imp}_{intra-band} > U^{imp}_{inter-band}$, and at the same time, they have calculated the theoretical residual resistivity $\rho_0$ using the same parameters, so that they have produced the consistent theoretical data of $T_c /T_{c0}$ vs $\rho_0$ to be directly compared to experimental data.
Figure \ref{prozorov_2014} shows these results. The results show that the $s^{\pm}$-wave model with finite-ranged impurity potentials ($U^{imp}_{intra-band} > U^{imp}_{inter-band}$) can perfectly fit the experimental data from Ba(Fe$_{1-x}$Ru$_x$)$_2$As$_2$ with electron irradiations, with a moderate ratio of $\alpha= U^{imp}_{inter}/U^{imp}_{intra}=0.65$ (this $\alpha$ is a different parameter than the previously defined $\alpha=0.88 z \Delta \rho_0 /T_c^0$). And reducing the ratio $\alpha$, the $T_c$-suppression rate can be easily slowed down by a couple of factors. These authors also confirmed that the irradiated electrons become non-magnetic impurities, therefore their results of the $T_c$-suppression with electron irradiation have ruled out the possibility of the $s^{++}$-wave state in Ba(Fe$_{1-x}$Ru$_x$)$_2$As$_2$ and support the $s^{\pm}$-wave state.
In Fig. \ref{prozorov_2014}, we see that there is still some large discrepancy between the transition metal doping experiments of Li et al. \cite{li2012superconductivity} (which shows the average $T_c$-suppression rate $\approx 50 mK/\mu \Omega cm$ that is 7 times faster than the electron irradiation data of Fig.\ref{prozorov_2014}) and the theory. However, as we have mentioned, transition metal doping experiments contain unknown factors/effects other than pure impurity scattering.

\subsection{Summary}
The message of this section is simple. The $s^{\pm}$-wave pairing model is intrinsically sensitive to the impurity scattering (both magnetic and non-magnetic) because of the sign-changing OPs as in the $d$-wave pairing state. On the other hand, it is also true that most of experiments of the $T_c$-suppression with transition metal dopings and irradiations show much slower rate of the $T_c$-suppression compared to the theoretical prediction of the $T_c$-suppression in the $s^{\pm}$-wave state with pointlike impurities. Initially, this discrepancy between theory and experiments was taken as the evidence for the inadequacy of the $s^{\pm}$-wave pairing model for the FeSCs. However, more realistic consideration of the impurity potentials -- which should have a finite size (not a pointlike) -- leads to a finite ranged impurity potential $U^{imp}(q)$ in momentum space. It implies $U^{imp}_{intra-band} > U^{imp}_{inter-band}$ in the two band $s^{\pm}$-wave model, and the systematic theoretical calculations of $T_c$ with $U^{imp}_{inter-band}/U^{imp}_{intra-band} =\alpha (< 1$) can produce arbitrarily slow $T_c$-suppression rate with the $s^{\pm}$-wave state by choosing a smaller $\alpha$ value \cite{wang2013using,prozorov2014effect}. Although this problem of the $T_c$-suppression in FeSCs is still under debate among researchers with different opinions, we can say that the $s^{\pm}$-wave pairing model can be compatible with experiments if the introduced impurities in real materials are not pointlike but finite sized defects.

\section{Experimental hints for Pairing Mechanism}
\begin{figure}[h]
\noindent
\hspace{1cm}
\includegraphics[width=150mm]{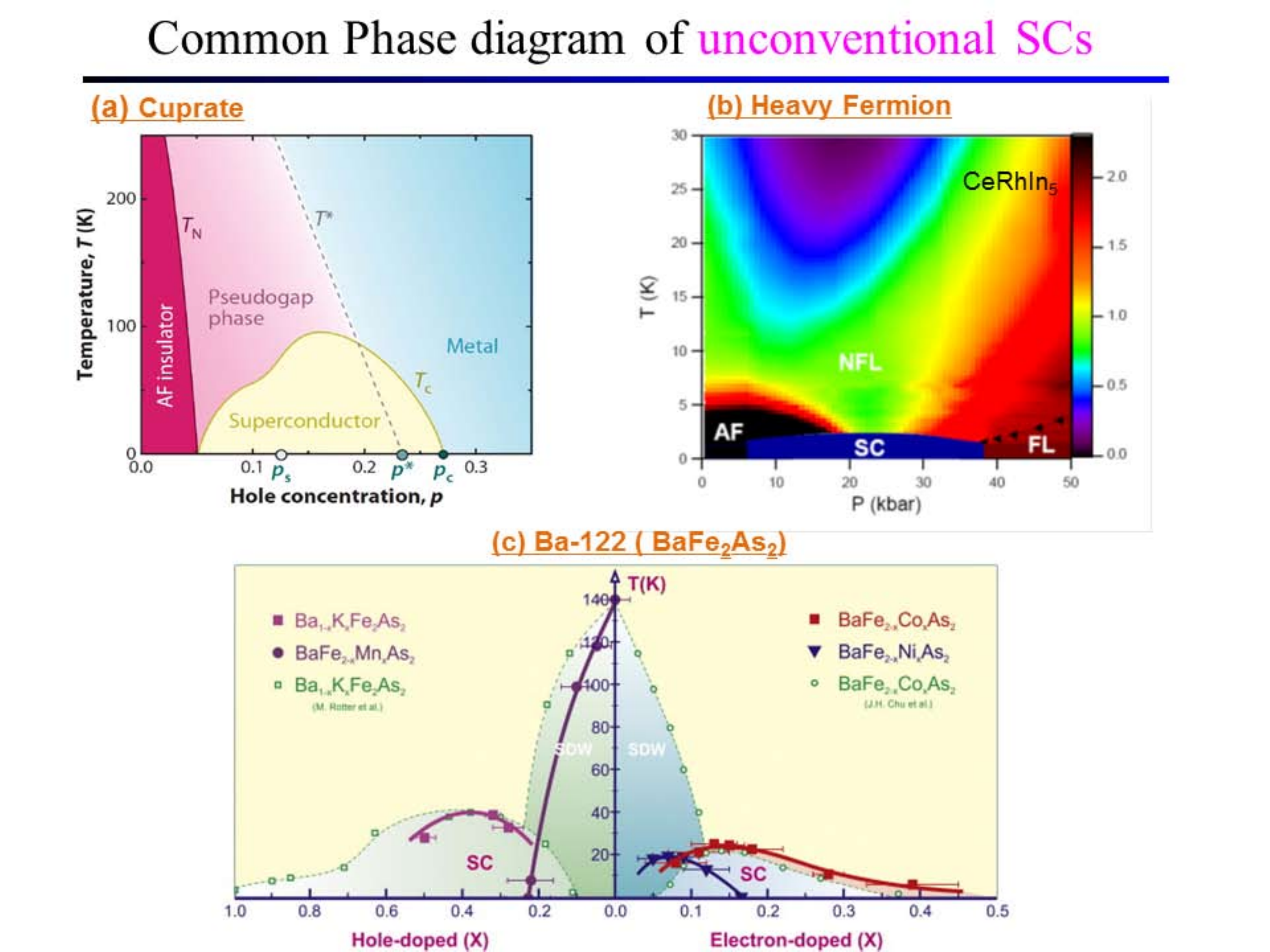}
\caption{(Color online) The phase diagrams of three most studied classes of unconventional superconductors. (a) Cuprate superconductor (YBCO). From \cite{doIron2007quantum}, (b) Heavy fermion superconductor (CeRhIn$_5$). From \cite{park2008isotropic}, and (c) Iron-based superconductor (Ba-122). From \cite{liu2010aliovalent}. The commonality is obviously that the $T_c$ vs tuning parameter (dopings and pressure) relation defines a dome shape phase having a maximum $T_c ^{max}$ around a hypothetical QCP at which the correlation effect is strongest.} \label{ph_diagram}
\end{figure}

Up to this section, we have investigated the various SC properties of the $s^{\pm}$-wave SC state and their compatibility with experimental data, and we didn't question much about the possible pairing mechanism of the $s^{\pm}$-wave SC state. In section 2, we have only briefly described a minimal two band BCS model, in a way to visualize the succinct features of the $s^{\pm}$-wave pairing state, but didn't imply that this is the ultimate pairing mechanism of the FeSCs. Even as a BCS theory, this two band model, being a minimal phenomenological model,  ignored all the details of bands (5 or 10 bands, depending on the choice of the unit cell, and orbital degrees of freedom as well as the details of the pairing interactions $V(k,k')$ and their coupling matrix elements $M_{\alpha, \beta}^{a,b}(k,k')$ ($\alpha, \beta=$ orbital indices; $a,b=$ band indices), etc. Furthermore this minimal BCS model has completely ignored any correlation effects, which should be reasonably strong in the Fe-based SC materials. Therefore, up to this section, although we have demonstrated the compatibility of the $s^{\pm}$-wave pairing state with the SC properties of almost all available experiments with the FeSCs, it doesn't provide much hint as to the nature of the pairing mechanism.

On the other hand, many researchers believe that the Fe-based SC materials are intermediately to strongly correlated systems. Roughly speaking, its correlation is weaker than the cuprates superconductors because the parent undoped compounds of FeSCs are still metallic SDW state while the cuprate parent compounds are Mott AFM insulators, and also weaker than the heavy fermion SC systems because the renormalization factor of the quasiparticle (q.p.) masses of the FeSCs -- in particular, of Fe-pnictides\cite{tamai2010strong} -- are a factor of 2-5 at most\cite{lu2008electronic,qazilbash2009electronic,coldea2008fermi,analytis2009fermi} while the typical renormalization factor of the heavy fermions are from several tens to several hundreds. Nevertheless, these three classes of unconventional SC compounds can display quite similar phase diagrams: namely, the common AFM correlation -- seen in all the cuprates, most of the FeSCs, and some of the heavy fermions -- is weakened by tuning the system parameters such as doping, fields, pressure, etc, and at the point of $T_N \rightarrow 0$ or at some distance from it, the q.p. are maximally renormalized (or completely broken down) defining "Quantum Critical Point" (QCP). Then all these three classes of SC materials have the maximum $T_c$ around the hypothetical QCP (although where the pseudogap (PG) temperature $T^{\ast} \rightarrow 0$ in the phase diagram is still under discussion) and display a dome shape of the SC phase in the $T_c$ vs tuning parameter phase diagram.

Hence many researchers suspected that the superconductivity in the FeSCs --regardless of the SC gap symmetry -- should be organized by some novel and unconventional pairing mechanism, or at least a non-BCS type pairing mechanism. And indeed there exist strong experimental indications supporting this idea, which are the anomalous scaling relations of (1) specific heat jump  $\Delta C$ vs. $T_c$, and (2) Condensation Energy (CE)  $\Delta E$ vs. $T_c$. Pioneered by Canfield and coworkers\cite{bud2009jump} and supported by the same and other researchers\cite{bud2013heat,bud2014heat,kim2011specific,kim2012specific,kim2014specific,hardy2010doping,gofryk2010doping,
gofryk2011effect,walmsley2013quasiparticle,grinenko2014superconducting} is the anomalous scaling law of  $\Delta C \sim T_c ^3$, obeyed by over 50 Fe-based SC samples, while the standard BCS theory predicts $\Delta C \sim T_c$. This observation is indeed quite non-BCS-like and appears in accord with the idea of a quantum critical (QC) fluctuations\cite{zaanen2009specific} driven superconductivity. And more recently, J. Xing et al.\cite{xing2014power}, and J. S. Kim et al. \cite{kim2015universal} have advanced this observation further to collect the data of the CE ($\Delta E$) vs $T_c$, and also found a scaling relation of $\Delta E \sim T_c^{3.5}$ for about 30 Fe-based SC samples, again strongly deviated from a BCS prediction $\Delta E \sim T_c^{2}$. These seemingly very non-BCS like scaling relations together indicate that the pairing mechanism of the FeSCs should be, at least, a non-BCS type and most probably should be intimately connected to the QC fluctuations.

However, in this section, we will discuss how these two seemingly non-BCS like scaling relations can be consistently explained with the minimal two band BCS model of the $s^{\pm}$-wave state described in section 2 \cite{bang2016anomalous,bang2016origin}. This is a surprising result, but on the other hand it just demonstrates the fact that the multiband BCS superconductor can have many novel and qualitatively new SC properties, not expected in a single band BCS superconductor.
And as to the pairing mechanism, the simultaneous explanation of two anomalous scaling relations with the BCS two band model of the $s^{\pm}$-wave state strengthened the speculation that the fundamental pairing mechanism of the FeSCs is basically a BCS theory -- in a very general sense, i.e., the itinerant fermionic carriers (quasiparticles) are glued into Cooper pairs by an exchange of non-phononic boson fluctuations\cite{note1} and this process is described by a general BCS-Eliashberg formalism.

\subsection{BNC scaling of specific heat jump  $\Delta C$ vs. $T_c$}

\begin{figure}
\noindent
\hspace{3cm}
\includegraphics[width=90mm]{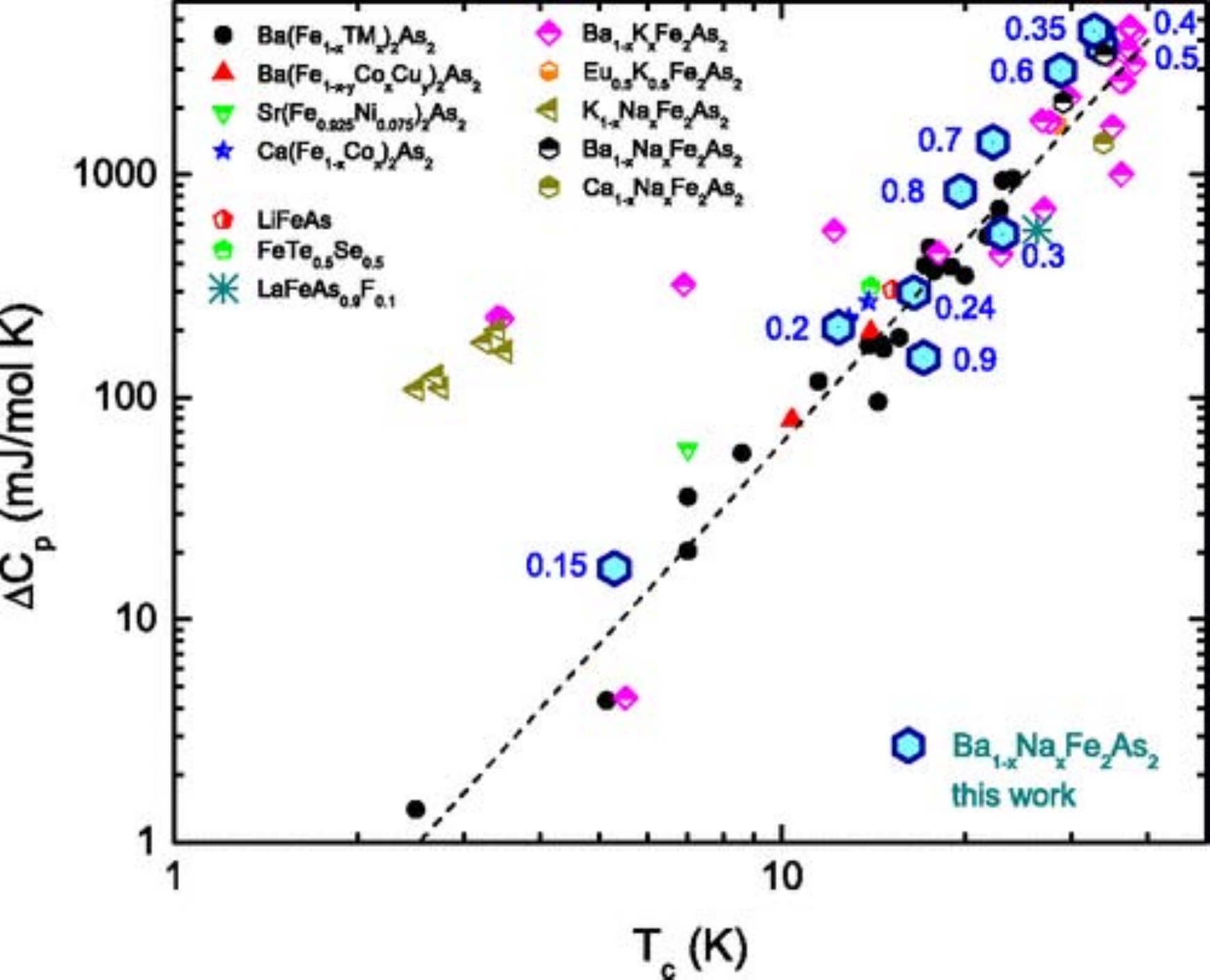}
\caption{(Color online) $\Delta C_p$ at the SC transition vs $T_c$ for the Ba$_{1-x}$Na$_x$Fe$_2$As$_2$ series, plotted together with literature
data for various FeAs-based superconducting materials. The plot from \cite{bud2013heat} was updated to include published data for K$_{1-x}$Na$_x$Fe$_2$As$_2$
($0 \leq x \leq 0.31$), Ca$_{1-x}$Na$_x$Fe$_2$As$_2$, Ba$_{1-x}$Na$_x$Fe$_2$As$_2$ ($x = 0.35, 0.4$), and LaFeAs$_{0.9}$F$_{0.1}$ \cite{zhao2011doping,pramanik2011multigap,aswartham2012hole,abdel2013evidence,grinenko2014superconducting}. The line corresponds to $\Delta C_p \sim T_c^3$. Numbers near the symbols are Na concentrations $x$. From \cite{bud2014heat}} \label{bnc_2014}
\end{figure}

The strong power law behavior of the specific heat jump $\Delta C$ {\it vs.} $T_c$ ($\Delta C \sim T_c ^{\alpha}, \alpha\approx 3$), first observed by Bud'ko, Ni, and Canfield (BNC)\cite{bud2009jump}, has been confirmed with several families of the Fe-based superconducting compounds with various dopings by several  research groups \cite{bud2013heat,bud2014heat,kim2011specific,kim2012specific,kim2014specific,hardy2010doping,gofryk2010doping,
gofryk2011effect,walmsley2013quasiparticle,grinenko2014superconducting} by now.

It is well known that the BCS theory\cite{BCS} of superconductivity predicts the universal ratio $\Delta C /T_c =1.43 \gamma$ ($\gamma = \frac{2\pi^2}{3} N(0)$ is the Sommerfeld coefficient of the normal state), hence the BCS scaling law should be $\Delta C \sim T_c$ -- from a na{\"i}ve point of view assuming that $\gamma$ and $T_c$ are not related\cite{kim2015universal}.
Therefore, the experimental observation by Bud'ko, Ni, and Canfield (BNC)\cite{bud2009jump}, $\Delta C \sim T_c^3 $ for a family of
doped Ba(Fe$_{1-x}$TM$_x$)$_2$As$_2$ compounds with $TM=$Co, Ni is a very intriguing behavior and stimulated active investigations both
experimentally and theoretically. After this original work\cite{bud2009jump}, this so-called BNC scaling relation was expanded
with an increasing list of the iron pnictide and iron chalcogenide SC compounds\cite{kim2011specific,kim2012specific,
hardy2010calorimetric,hardy2010doping,gofryk2010doping,gofryk2011effect,walmsley2013quasiparticle,chaparro2012doping,bud2014heat,xing2014power},
hence strengthens the speculation that some generic mechanism must exist behind this unusual scaling behavior.
However, more recent works showed that this BNC scaling is not a perfect relation and there exist a few compounds showing varying degree of deviations. For example, the observation of a strong deviation from the BNC scaling in the K-doped Ba$_{1-x}$K$_x$Fe$_2$As$_2$ for
$0.7 < x \leq 1$ \cite{bud2013heat} is contrasted to the Na-doped Ba$_{1-x}$Na$_x$Fe$_2$As$_2$ ($0.1 \leq x \leq
0.9$)\cite{bud2014heat} which displays an excellent BNC scaling. And the authors of recent measurement\cite{grinenko2014superconducting} with Na-doped K$_{1-x}$Na$_{x}$Fe$_2$As$_2$ claimed that $\Delta C \sim T_c^{2}$ fits better the data instead of $\sim T_c^{3}$ although data of this compound are limited to a very narrow range of $T_c$ variation.

Therefore, it is fair to say that, even including some deviating materials, all reported data of the Fe-pnictide/chalcogenide (FePn/Ch) superconductors up to now exhibit strongly non-BCS scaling relations between $\Delta C$ {\it vs.} $T_c$ and it
deserves a theoretical understanding.

\subsubsection{Other theories}
There have been three theoretical attempts. Kogan\cite{kogan2009pair} argued that strong pair-breaking can cause $\Delta C /T_c \propto T_c ^2$. The essence of this theory is a dimensional counting. The free energy difference near $T_c$, $\Delta F = F_{s} - F_{n}$, can be expanded in powers of $\Delta^2$ ($\Delta$: the SC order parameter.
In the BCS theory, $\Delta F \propto -N(0)\frac{\Delta^4}{T_c ^2}$ \cite{abrikosov2012methods}. Using the BCS result of $\Delta^2(T)\sim T_c^2 (1 - \frac{T}{T_c})$, we get $\Delta C /T_c \propto \frac{\partial^2 \Delta F}{\partial T^2} \sim N(0)$, the well known BCS prediction.
In the case of the strong pair-breaking limit, $\Gamma_{\pi} \gg T_c$ ($\Gamma_{\pi}=$ pair-breaking rate), considered by Kogan, $\Delta F \propto - N(0)\frac{\Delta^4}{\Gamma_{\pi}^2}$ by a dimensional counting. Substituting the same BCS behavior of $\Delta^2(T) \propto T_c^2 (1 - \frac{T}{T_c})$, we immediately recover Kogan's result $\Delta C /T_c \sim N(0)\frac{T_c ^2}{\Gamma_{\pi}^2} \sim T_c ^2$. However, we believe that the self-consistent theory, in the strong pair-breaking limit $\Gamma_{\pi} \gg T_c$, should use $\Delta^2(T) \propto \Gamma_{\pi}^2 (1 - \frac{T}{T_c})$ instead of $\Delta^2(T) \propto T_c^2 (1 - \frac{T}{T_c})$\cite{bang2016anomalous}, then we would obtain $\Delta C /T_c \sim T_c ^{-2}$ instead.
The theory of Vavilov {\it et al.}\cite{vavilov2011jump} mainly studied the coexistence region with magnetic order $M$ and SC
order $\Delta$. It is a plausible theory that the coexisting magnetic order over the SC order can substantially reduce $\Delta
C$, hence develops a steep variation of $\Delta C$ {\it vs.} $T_c$. However this theory didn't reveal any generic mechanism as
to why $\Delta C $ follows the BNC scaling  $\sim T_c ^{3}$.
Finally, Zannen\cite{zaanen2009specific} attributed the origin of $\Delta C \propto T_c ^3$ to the anomalous temperature dependence of the
normal state electronic SH with the scaling form $C_{elec}^{n} \propto T^3$ due to the critical fluctuations near the quantum
critical point (QCP). A problem of this theory is that (1) this hyper-scaling argument applies to the bosonic critical fluctuations and bosonic specific heat, hence the fermionic SH jump $\Delta C_{elec}$ is irrelevant to the QC fluctuations. Furthermore, there is no experimental evidence of $C_{elec}^{n} \propto T^3$  (besides phonon SH $C_{ph}\propto T^3$)  for a wide doping range of the FePn/Ch superconductors.
All three theories mentioned above are single band theories and do not particularly utilize the unique properties of
the FePn/Ch superconductors.

\subsubsection{Theory of two band $s^{\pm}$-wave model}
The key idea of this theory is that the multi band systems should have a contribution of $\Delta C_i$ from each band $"i"$ as
\begin{equation}
\Delta C = \sum_{i=h,e} N_i(0) \Big( \frac{-d \Delta^2 _i}{d T} \Big)\Big|_{T_c}.
\label{DelC}
\end{equation}
\noindent
And the two band $s^{\pm}$-wave model described in section 2 has the interesting inverse relation between the DOS $N_i$ and the SC gap $\Delta_i$ such as $\sqrt{\frac{N_h}{N_e}} \sim \frac{\Delta_e}{\Delta_h}$ as $T \rightarrow T_c$ and $\frac{N_h}{N_e} \sim \frac{\Delta_e}{\Delta_h}$ as $T \rightarrow 0$\cite{bang2008possible}. Therefore the total SH jump can possibly have a temperature relation very different from a single band BCS prediction $\Delta C_{BCS} \sim T_c$.

Another ingredient of the model is "doping" to simulate the series of experimental data, for example, of Fig. \ref{bnc_2014}. The data of Ba$_{1-x}$Na$_x$Fe$_2$As$_2$ series in Fig. \ref{bnc_2014} is the collection of data from samples with different doping "$x$" of Na element, which introduces more "hole" carriers into the compound. In the case of the Ba(Fe$_{1-x}$Co$_x$)$_2$As$_2$ series, the doping "$x$" introduces more "electron" carriers to the compound.
To simulate this series of doping in the two band model, we first note that the undoped parent compound BaFe$_2$As$_2$ is a
compensated metal, hence has the same number of electrons and holes, i.e. $n_h = n_e$. Therefore it is a reasonable
approximation to assume $N_h \approx N_e$ at no doping and then the doping of holes (K, Na, etc.) or electrons (Co, Ni, etc.) is simulated by varying $N_h$ and $N_e$ while keeping $N_e + N_h =N_{tot} = const.$ Admittedly this modeling of doping is much too simple, but the assumption $N_{tot} = const.$ is only for convenience and can be relaxed. The real important  parameter of this model is the relative sizes between $N_e$ and $N_h$, but not the total DOS $N_{tot}$ nor an individual DOS $N_i$; those specific information are all absorbed in the plot of $\Delta C (N_{h(e)})$ vs $T_c ((N_{h(e)})$ as implicit parameters. Therefore, it is not even necessary to know the exact relation between the actual doping concentration $"x"$ of real compounds and the values of $N_{h(e)}$ in our two-band model. For more details, the readers can refer to Refs.\cite{bang2016anomalous,bang2016origin}.

\begin{figure}
\hspace{3cm}
\includegraphics[width=90mm]{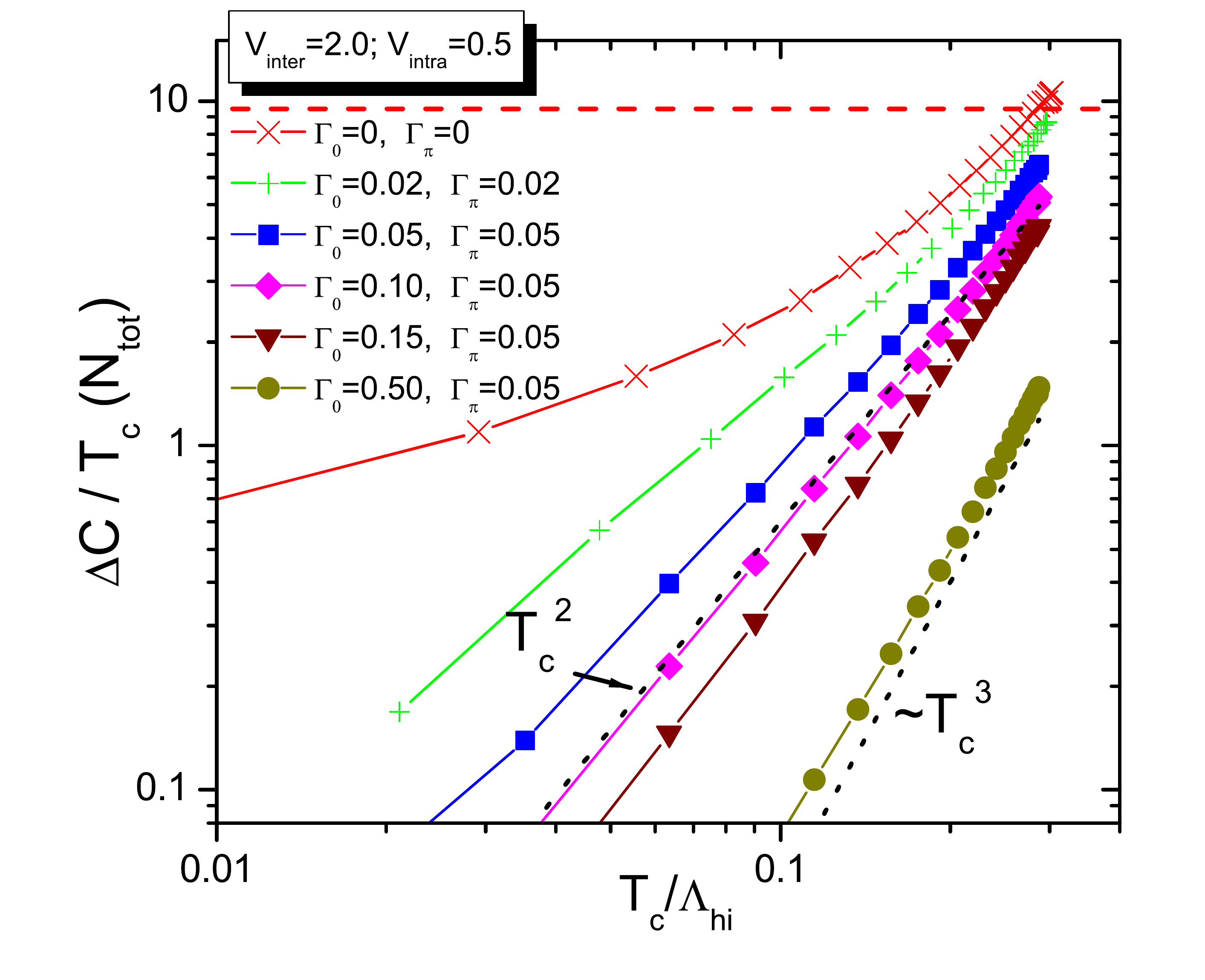}
\caption{(Color online) Numerical calculations of $\Delta C /T_c$ {\it vs.} $T_c$ with dimensionless coupling constants $\bar{V}_{inter}= 2.0$ and $\bar{V}_{intra}= 0.5$, for different impurity scattering strengths of $\Gamma_0$ and $\Gamma_{\pi}$ (in unit of $\Lambda_{hi}$). Horizontal dashed line is the BCS limit of $9.36 N_{tot}$ and the dotted lines of $\sim T_c^2$  (BNC scaling)
and $\sim T_c^3$  (super-strong scaling) are guides for the eyes. From \cite{bang2016anomalous}.\label{DelC_Tc}}
\end{figure}

\begin{figure}
\hspace{1cm}
\includegraphics[width=140mm]{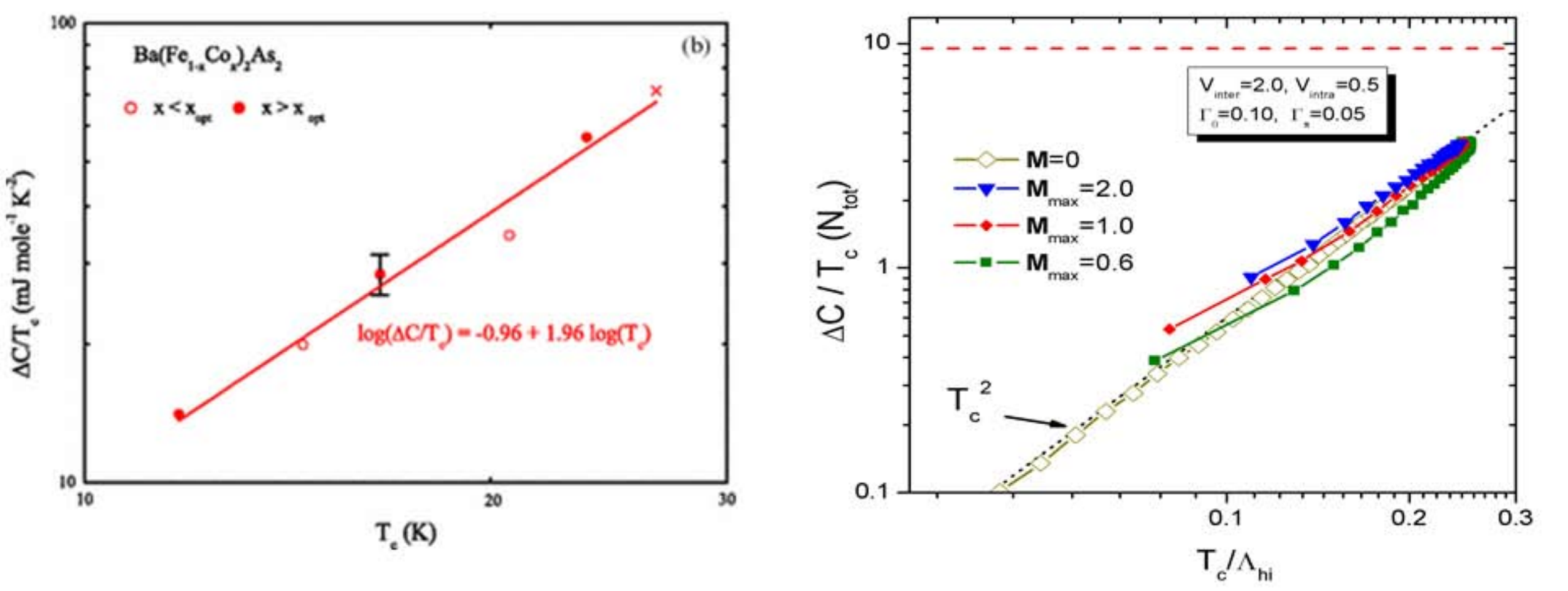}
\caption{(Color online) (Left) Normalized values of $\Delta C/T_c$ vs $T_c$ for Ba(Fe$_{1-x}$Co$_x$ )$_2$As$_2$ for annealed
compositions near (down to $T_c =11.7$ K) to the optimal $x_{opt} = 0.08$ (which composition is denoted by "X" in the graph). The data for $x < x_{opt}$ belong to the coexistence region of AFM and superconductivity. The normalized $\Delta C$ is given by $\Delta C_{measured} \times (\gamma_n)/(\gamma_n -\gamma_r)$. From \cite{kim2012specific}. (Right) The two band $s^{\pm}$-wave model calculations of $\Delta C/T_c$ vs $T_c$ with and without the AFM order ${\bf M}$ included. From \cite{bang2016anomalous}.
\label{AFM_SC}}
\end{figure}

Figure \ref{DelC_Tc} shows the results of numerical calculations of $\Delta C (N_{h(e)})$ vs $T_c ((N_{h(e)})$, which are calculated with the coupled gap equations Eqs.(5) at $T \rightarrow T_c$ and the above Eq.(\ref{DelC}). The results faithfully reproduces the anomalous BNC scaling as well as some deviations shown in experimental data of Fig.\ref{bnc_2014}. First, the results with the wide range of the non-pair-breaking impurity scattering rate $\Gamma_0 /\Lambda_{hi} = 0.05-0.15$ show the BNC scaling $\Delta C \sim T_c^3$. With decreasing the scattering rate $\Gamma_0 /\Lambda_{hi} = 0.02,$ and  $0.0$, the scaling relation continuously becomes weaker up to $\Delta C \sim T_c^{1.4}$ at $\Gamma_0 /\Lambda_{hi} =0.0$ in Fig.\ref{DelC_Tc}. Therefore, the gentler scaling power $\Delta C \sim T_c^{2}$, observed in the series of K$_{1-x}$Na$_x$Fe$_2$As$_2$ \cite{grinenko2014superconducting}, can be understood. And in the extreme clean limit with $\Gamma_{0,\pi} /\Lambda_{hi} = 0.0$, the scaling relation approaches up to $\Delta C \sim T_c^{1.4}$, which is still steeper than the observed relation $\Delta C \sim T_c$ of Ba$_{1-x}$K$_x$Fe$_2$As$_2$ for $x > 0.7$ \cite{bud2013heat}. Therefore, the data of the  Ba$_{1-x}$K$_x$Fe$_2$As$_2$ series for $x > 0.7$ appears out of scope of the minimal two band $s^{\pm}$-wave model. However, it should be noticed that the Ba$_{1-x}$K$_x$Fe$_2$As$_2$ compound has the Lifshitz transition for $x > 0.7$, where the electron band around $M$ point sinks below the Fermi surface\cite{sato2009band}, hence the simple two band model doesn't apply any more in this region of K doping. Finally, the numerical results showing the super strong scaling power law $\Delta C \sim T_c^{4}$ is only for a demonstration with an unrealistic amount of impurity scattering rate $\Gamma_0 /\Lambda_{hi} =0.5$.

In summary, Fig.\ref{DelC_Tc} shows that the BNC scaling is a generic property of the two band BCS model with a dominant interband pairing interaction as $V_{inter} > V_{intra}$, unless extreme choice of model parameters are chosen. And this two band model calculations suggest that the origin of the anomalous BNC scaling behavior is nothing but the kinematic relation of  the two band $s^{\pm}$-wave model, i.e. $\sqrt{\frac{N_h}{N_e}} \sim \frac{\Delta_e}{\Delta_h}$ as $T \rightarrow T_c$. This robustness of BNC scaling relation of FeSCs continues even in the coexistence region of AFM and superconductivity\cite{kim2012specific} as shown in the lefthand panel of Fig.\ref{AFM_SC}. The numerical results in the righthand panel of Fig.\ref{AFM_SC} show that the two band $s^{\pm}$-wave model can faithfully generate the robust BNC scaling relation with and without the AFM order ($M_{AFM}$). No other theory can possibly provide this much coherent and consistent explanation revealing a clear kinematic origin behind this anomalous scaling relation.

\subsection{Condensation Energy $\Delta E$  vs. $T_c$}

\begin{figure}[h]
\hspace{2cm}
\includegraphics[width=90mm]{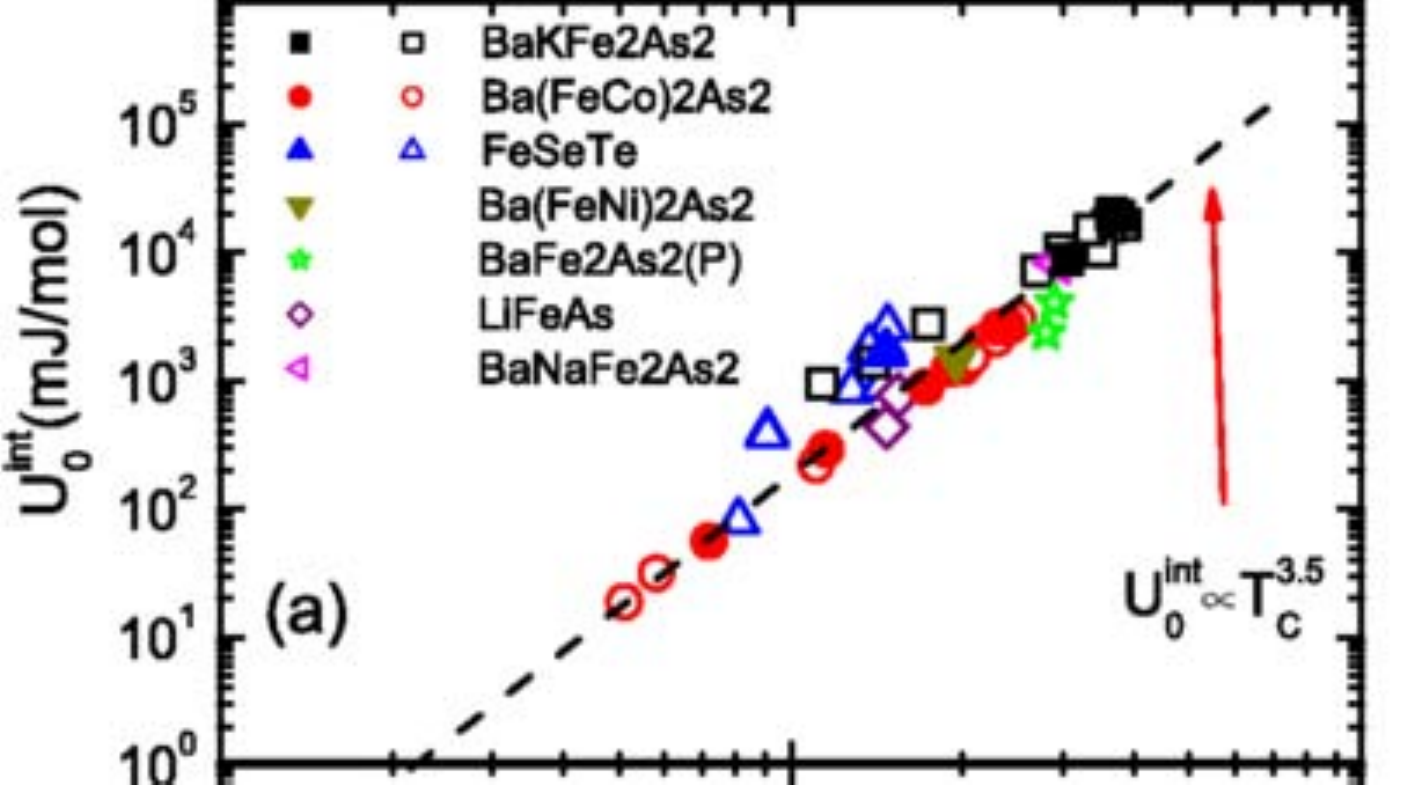}
\caption{(Color online) Collected data of the CE vs $T_c$. The CE $U_0^{int}$ are obtained by integrating the entropy in the SC state. From \cite{xing2014power}.
\label{CE_HWen}}
\end{figure}

The condensation energy (CE) $\Delta E$ of a superconductor is defined as the energy difference between the
normal state and the SC state of the same system. In general, the size of the CE of any phase transition is a measure of how much more stable the ordered state is, compared to the normal state, hence the CE is naturally related to the ordering (pairing) energetics and transition temperature $T_c$. For example, the magnetic transition with local moments such as a classic limit of Heisenberg model and Ising model has the relation $\Delta E_{mag} \propto T_c$,
while the BCS theory of the one band superconductor predicts $\Delta E_{BCS} \propto T_c ^2$\cite{BCS}.
In view of this, as show in Fig. \ref{CE_HWen}, the observation by J. Xing {\it et al.}\cite{xing2014power} and J. S. Kim {\it et al.} \cite{kim2015universal} of $\Delta E \propto T_c ^{\beta}$ ($\beta \approx
3.5$) with various FeSCs is very intriguing and should contain the crucial information about the SC pairing mechanism of the FeSCs.

\subsubsection{Other theories}
As described in the Introduction of this section, many researchers tend to connect some novel features of FeSCs to the QCP or the strong correlation effect.
According to Zaanen's hyperscaling argument\cite{zaanen2009specific}, the SH at normal state follows the scaling relation as $C \sim T^3$.
However, it was already pointed out that these critical fluctuations, which would cause $C \sim T^3$, are not fermionic fluctuations, hence no direct connection to the SH jump and the CE of the SC transition that is the fermionic reconstruction of the system.
Nevertheless, the authors of \cite{xing2014power} extrapolated the QCP relation $C \sim T^3$ to the electronic SH as $C \sim N_0 T \sim T^3$, to obtain $N_0 \sim T^2$. Combining this result with the single band BCS CE, $\Delta E =\frac{1}{2}N_0 \Delta_0 ^2$ (where again the BCS relation $\Delta_0 \sim T_c$), it leads to $\Delta E \sim T_c ^4$, close to the experimental observation $\Delta E \sim T_c ^{3.5}$. As a ballpark estimation, this result appears not bad, but as we mentioned above we believe that this QCP scenario contains several inconsistent logical loopholes.

\begin{figure}
\hspace{1cm}
\includegraphics[width=150mm]{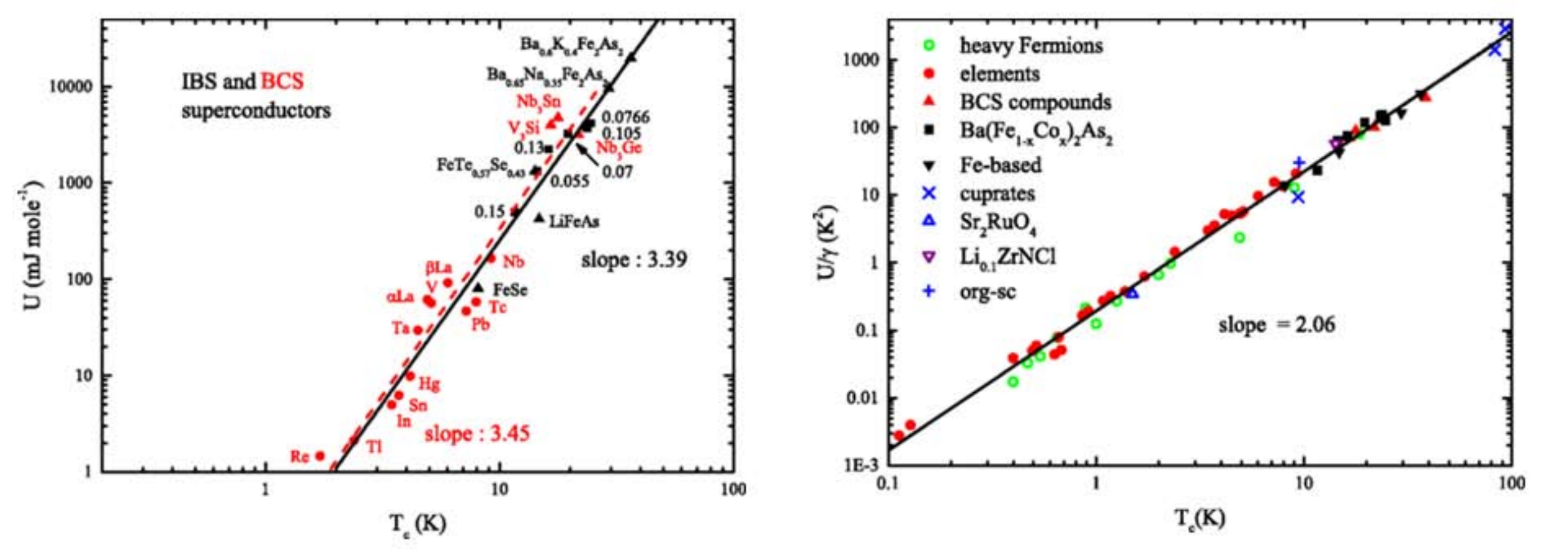}
\caption{(Color online) (Left) Collected data of the CE vs $T_c$ for BCS superconductors (with $T_c >1.4$K) and FeSCs (IBS). Both superconductors follows approximately the same scaling relation $\Delta E (U) \sim T_c ^{3.4-3.5}$;  (Right) Demonstration of the universal scaling relation $\Delta E \sim \gamma T_c ^{2}$ obeyed by wide range of superconductors.
From \cite{kim2015universal}.
\label{GS_scaling}}
\end{figure}

Another suggestion about the CE scaling was given by one of us \cite{kim2015universal}, and it was shown with extensive amount of collected data that this seemingly non-BCS scaling relation of CE, $\Delta E \sim T_c ^{3.4 - 3.5}$ (in simple BCS theory, $\Delta E_{BCS} \sim T_c ^{2}$) is actually obeyed not only by FeSCs but also by medium ($\lambda=0.46$) to
strong-coupled phonon-mediated BCS superconductors (with $T_c > 1.4$K) as shown in the left panel of Fig. \ref{GS_scaling}. However, it was also shown that many other superconductors like the phonon-mediated BCS superconductor MgB$_2$, heavy fermion and cuprate superconductors, etc fail to follow anywhere near to this scaling.
Nonetheless, the Ref.\cite{kim2015universal} found that all these superconductors -- both which do follow and which do not follow the CE$\propto T_c^{3.5}$ scaling relation -- obeyed the universal scaling relation $\Delta E / \gamma \sim T_c ^2$, as shown in the right panel of Fig.\ref{GS_scaling}. This surprisingly universal scaling relation $\Delta E / \gamma \sim T_c ^2$ has not yet a theoretical explanation, but it is suggestive of a renormalized BCS relation, namely, the BCS prediction $\Delta E =\frac{1}{2} N_0 \Delta_0 ^2$ with $\Delta_0 ^2 \sim T_c ^2$, but replacing the DOS $N_0$ by a renormalized Sommerfeld constant $\gamma \sim N_0 /(1+\lambda)$, ($\lambda=$, dimensionless coupling constant).
Although it needs more specific theory, this interpretation as a renormalized BCS relation suggests that the correlation or interaction effect could be an underlying origin of this anomalous scaling relation of the CE vs $T_c$.

\subsubsection{Theory of two band $s^{\pm}$-wave model}
Calculations of the CE for the two band $s^{\pm}$-wave model is a straightforward extension of BCS calculations to the two band model. Given parameters of $V_{inter, intra}$, and $N_{h,e}$, the coupled gap equations Eq.(5) are solved for $\Delta_{h,e}$ and using these values, the expectation value of the Hamiltonian, Eq.(1), is evaluated by a mean field theory. As before, to simulate the experimental data of a series of dopings, the values of $N_{h,e}$ continuously change while keeping $N_{tot}=N_h + N_e = const.$ More details are referred to Ref.\cite{bang2016origin}.

\begin{figure}[h]
\hspace{1cm}
\includegraphics[width=150mm]{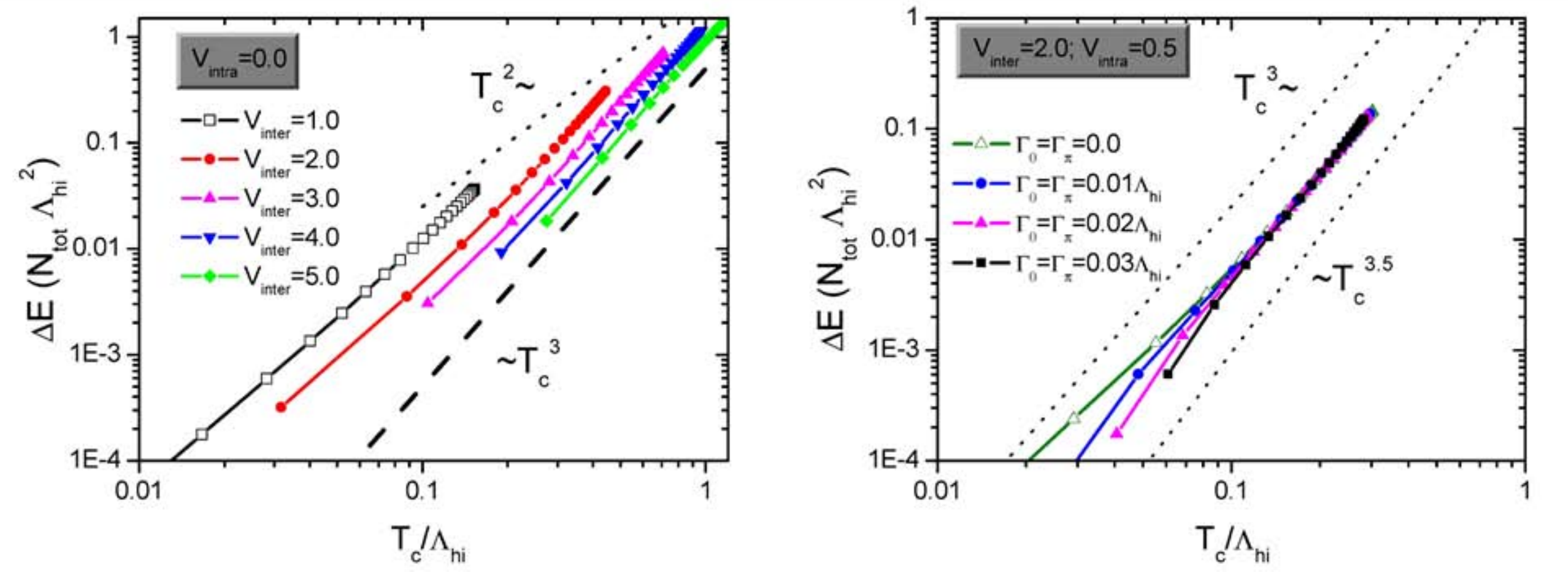}
\caption{(Color online) (Left) $\Delta E$ {\it vs.} $T_c$ calculated with the two band $s^{\pm}$-wave model with $\bar{V}_{inter}=2.0$ for $\bar{V}_{intra}=+1.0, +0.5, 0.0, -0.5$ and $-1.0$, respectively;  (Right) The same calculations of $\Delta E$ {\it vs.} $T_c$ with $\bar{V}_{inter}=2.0$ and $\bar{V}_{intra}=0.5$, and the pair-breaking impurity scattering rates $\Gamma_{\pi}/\Lambda_{hi}=0.0, 0.01, 0.02,$ and 0.03, respectively.
From \cite{bang2016origin}.
\label{bang_CE}}
\end{figure}

In Fig.\ref{bang_CE}, the left panel shows the numerical results of $\Delta E$ vs $T_c$ of the $s^{\pm}$-wave model with the interband pairing interaction only ($V_{intra} =0, V_{inter} \neq 0$). For a wide range of $V_{inter}$ ($=1.0, 2.0, 3.0, 4.0, 5.0$), it shows $\Delta E \sim T_c ^{\beta}$, with $\beta \approx 3$. Although the scaling power $\beta \approx 3$ is still weaker than the experimental power $\beta_{exp} \approx 3.5$, this is a surprisingly good result; this two band model doesn't have any tuning parameters nor any ad hoc assumptions, and it is the exactly same model which already successfully explained the BNC scaling $\Delta C \sim T_c^3$. Here again the underlying mechanism for this success which generates such a fast variation of the CE vs $T_c$ -- one order of magnitude faster than the single band BCS theory -- is the kinematic constraint ($\frac{N_h}{N_e} \sim \frac{\Delta_e}{\Delta_h}$ as $T \rightarrow 0$) of the two band BCS model with a dominant interband pairing interaction.

Adding a repulsive intraband interaction ($V_{intra} >0$) in the model increases the scaling power, but only slightly. However, adding an attractive intraband interaction ($V_{intra} <0$) reduces the scaling power $\beta$ quickly to the BCS value $\beta_{BCS} \approx 2$. All these interesting variation of the CE scaling behavior in the multiband superconductor  can be understood from the fact that the CE gain ($\Delta E <0$) in the SC transition from metallic state is a subtle balance/competition between the kinetic energy loss ($\Delta KE >0$) and the potential energy gain ($\Delta PE <0$). For more detailed discussions, we refer to Ref.\cite{bang2016origin}. In the right panel of Fig.\ref{bang_CE}, the calculation results of CE ($\Delta E$) vs $T_c$ with including impurity scattering are shown. It shows that only a tiny amount of impurity scattering ($\Gamma_{imp} /\Lambda_{hi} \approx 0.02-0.03$) is sufficient to increase the scaling power to an experimental value as $\beta \rightarrow \beta_{exp} \approx 3.5$.

\subsection{Summary}
Two anomalous scaling relations, (1) BNC scaling  $\Delta C \sim T_c^3$, and (2) the CE scaling $\Delta E \sim T_c^{3.5}$, are qualitatively different from the other previously discussed SC properties-- such as ARPES, temperature and/or field dependencies of NMR, SH/thermal conductivity, penetration depth, etc. These scaling relations do not primarily depend on the fine structure or symmetry of the SC gap function $\Delta(k)$, but mainly depend on the overall energetics of the SC transition from the norma metallic state. Therefore they should contain the generic information about the energetics of the SC transition and the pairing mechanism itself. Therefore the successful reproduction and explanation of these two seemingly non-BCS scaling behaviors with the minimal two band $s^{\pm}$-wave pairing model is an unexpected and surprising result.
Together with the previous sections which showed the extremely good compatibility of the $s^{\pm}$-wave gap with virtually all available experiments,
the successful explanation of two  anomalous scaling relations in this section strengthens the validity and consistency of the $s^{\pm}$-wave model as a correct theory for FeSCs, and leads us to speculate that the fundamental pairing mechanism of the FeSCs is basically a BCS theory; namely, strong correlation effects, abundantly observed in the normal state of the FeSCs, exist and renormalize the effective mass $m_{qp}^{*}$ of quasiparticles, DOS $N_{h,e}$, pairing interactions $V_{inter,intra}(\bf q)$, etc., but when the system enters the SC transition, the pairing mechanism and pairing energetics seem to be governed by the BCS pairing mechanism but with a non-phononic bosonic glue.

\section{Conclusions}

In this review, we have reviewed the generic SC properties of the $s^{\pm}$-wave pairing state and critically examined them in comparison with the available experiments of the Fe-based SC compounds. The generic SC properties of the $s^{\pm}$-wave pairing state are: (1) it is a $s$-wave full gap superconductor with varying degree of gap anisotropy; the gap function $\Delta(k)$ has no nodes. (2) however, the sign-changing OPs substantially modifies the usual $s$-wave coherence factor of the  large momentum exchanging processes such as INS neutron scattering, NMR $1/T_1$ spin-lattice relaxation rate, and various impurity scattering effects. (3) Combinations of (1) and (2) generate various nodal-gap-like SC features in different experimental probes. These nodal-gap-like features often cannot be distinguished from a real (kinematic) nodal gap such as the $d$-wave SC gap state with one type of experimental probe; therefore, crosschecking with different probes is important to confirm the presence of gap-nodes in the gap function $\Delta(k)$ or not. The origin of the nodal-gap-like behaviors in the $s^{\pm}$-wave pairing state can be various: it can be due to the "V"-shape DOS dynamically induced by impurity scattering, not from the kinematic constraint of the nodal gap function $\Delta(k)$ itself; it can be due to the size difference between multiple gaps, for example, $|\Delta_h| > |\Delta_e|$ or vice versa; it can be due to the inverse relation(s) $\frac{N_h}N_e \approx \frac{|\Delta_e|}{|\Delta_h|}$; and finally it can be due to the combinations of some of these.
In this review, we have explained how these various mechanisms can generate the nodal-gap-like behaviors in the $s^{\pm}$-wave pairing state and compared with the relevant experiments, side by side. As a result, we have shown that almost all nodal-gap evidences  -- initially conceived -- in the FeSCs turned out to be the supporting evidences for the $s^{\pm}$-wave pairing state.

Through the crosschecking between theory and experiment, a few Fe-based SC compounds were found indeed to have a nodal gap, for example, K-overdoped (Ba,K)Fe$_2$As$_2$ by ARPES\cite{xu2013possible,shinsik_node,okazaki2012octet}, and FeSe by STM tunneling measurement\cite{song2011direct}. However, these nodal gaps still obey $A_{1g}$ crystal symmetry, which can be continuously evolved from the $s^{\pm}$-wave pairing state. Therefore the origin of the nodal gap in these compounds are not like a $d$-wave nodal gap, and these nodes are  accidental nodes. There exist still other strong candidates for a nodal gap from the penetration depth and thermal conductivity measurements such as BaFe$_2$(As,P)$_2$\cite{hashimoto2010line,hashimoto2010evidence,hashimoto2012sharp} and LaFePO\cite{fletcher2009evidence,hicks2009evidence}, which need to be confirmed with yet different probes.
We summarize the situation as following. The absolute majority of the FeSCs have multiple $s$-wave full gaps, but often displaying nodal-gap-like behaviors in various SC properties, which is consistent with the generic $s^{\pm}$-wave paring state. A few compounds of the FeSCs were confirmed or have a strong possibility to have nodal gaps. The confirmed nodal gap structure preserve the same $A_{1g}$ symmetry as the $s^{\pm}$-wave pairing state, hence they are accidental nodes. We expect that the not-yet-confirmed ones also belong to the same category, even if these compounds indeed possess a nodal gap. Therefore, as to the pairing symmetry and pairing mechanism, finding a nodal gap or not in the FeSCs is not an essential issue.  The $s^{\pm}$-wave pairing state remains valid as the standard paradigm of the FeSCs.

As mentioned in the introduction, identifying the gap symmetry and gap function doesn't mean identifying the pairing mechanism, but only providing some constraints for the correct theory. There is seldom a direct experimental probe for the pairing mechanism because the mechanism is usually an idea and concept which cannot be seen. It can be at best agreed on only through the circumstantial evidences with the extensive consistence checks with experiments. In section 11, we discussed a possible explanation of the anomalous scaling behaviors, observed in the SH jump vs $T_c$ and the CE vs $T_c$ for about 40 to 50 Fe-based SC samples, with a generalized BCS theory. This issue is not yet closed, and other theoretical explanation based on the strong correlation might be possible.  At the moment, the BCS pairing mechanism -- with a non-phononic pairing boson -- for the FeSCs is not a very exciting idea but at least it is very much consistent with the $s^{\pm}$-wave paring model to understand these anomalous scaling behaviors.

There are several important experimental probes not covered in this review; for example, infrared (IR) spectroscopy and tunneling spectroscopy, simply because we do not have sufficient expertise and time to cover these specialized subjects with a vast amount of research papers. Nevertheless, we can say that these powerful spectroscopic tools also support the $s^{\pm}$-wave pairing state with almost all Fe-based SC compounds\cite{li2008probing,nakayama2009superconducting,hanaguri2010unconventional}, except a few, for example, FeSe\cite{song2011direct}.
We refer to the already existing excellent review articles\cite{hu2009optical,basov2011electrodynamics,charnukha2014optical,hoffman2011spectroscopic} and references therein for further discussions of these specific experimental probes.
We also didn't discuss electronic Raman spectroscopy. Although this experimental tool has played a very active role for investigating for the $d$-wave superconductivity in the high-$T_c$ cuprate superconductors\cite{devereaux2007inelastic}, it was not as much actively used with the FeSCs as in the cuprate superconductors. One reason is that the Fe-based SC compounds are multiband system having many complicated FSs, while the cuprates have a single large FS; as a result, the symmetry analysis of Raman spectra becomes more complicated and has more uncertainty\cite{scalapino2009collective,mazin2010pinpointing}. Nevertheless, there are some interesting physics uncovered uniquely with Raman spectroscopy such as new collective modes\cite{kretzschmar2013raman}, and anomalous phonon frequency shifts\cite{um2014superconductivity}, etc, in the SC phase. Again up to now Raman spectroscopy experiments also are most consistent with the $s^{\pm}$-wave pairing state.

Finally, the so-called heavily electron-doped iron selenide (HEDIS) systems, such as FeSe/SrTiO3 monolayer system ($T_c \approx$ 60 -- 100K)\cite{qing2012interface,he2013phase,ge2015superconductivity}, A$_x$Fe$_{2-y}$Se$_2$ (A=K, Rb, Cs, Tl, etc.) ($T_c \approx$ 30--40K) \cite{guo2010superconductivity,wang2011superconductivity,zhang2011nodeless}, (Li$_{1-x}$Fe$_x$OH)FeSe ($T_c \approx$ 40K)\cite{lu2015coexistence}, and pressurized bulk FeSe ($T_c \approx$37K)\cite{margadonna2009pressure} are posing a serious challenge to the standard paradigm of the $s^{\pm}$-wave pairing state for the Fe-based SC compounds, which was the only pairing state covered in this review.
These HEDIS systems share one distinct common factor totally different from the other standard Fe-based SC systems; namely, they do not have the hole pockets around $\Gamma$ point in BZ and have only the electron pockets at $M$ points. Without having the hole pockets, it is immediately clear that the standard picture of the $s^{\pm}$-wave pairing state cannot be formed (see Fig.1).
Understanding the superconductivity in the HEDIS systems is currently the hottest subject in the Fe-based superconductivity research pressing the fundamental questions: (1) why and how is $T_c$ so high, up 100K ?
(2) what is the pairing mechanism and pairing state with only electron pockets at $M$ point ?; (3) does the standard paradigm of the $s^{\pm}$-wave pairing state continue to work or not ?
We leave this extremely important subject not discussed in this review. It is a rapidly developing subject and there exist already a growing volume of research papers, we refer to Ref.\cite{hirschfeld2016using,lee2014interfacial,rademaker2016enhanced,linscheid2016high,li2016makes} and more references therein.

\ack{YKB acknowledges the NRF Grant 2016-R1A2B4-008758 funded by the National Research Foundation of Korea, and G.S acknowledges the DOE grant, Office of Basic Energy Sciences, DE-FG02-86ER45268.}

\pagebreak

\end{document}